\documentclass[a4paper,11pt]{article}
\pdfoutput=1
\usepackage{jheppub} 
\usepackage{amsmath,amssymb,amsfonts}
\usepackage{graphicx}
\usepackage{float}
\usepackage{soul}
\usepackage[export]{adjustbox}
\usepackage[utf8]{inputenc}
\usepackage{newunicodechar}
\newunicodechar{γ}{\gamma}
\newunicodechar{ε}{\varepsilon}
\usepackage{slashed}
\usepackage{bbold}
\usepackage{subfiles}
\usepackage{pifont}
\usepackage{xcolor}
\usepackage{physics}
\usepackage{enumitem}
\usepackage{ifthen}
\usepackage{mathtools}
\usepackage{centernot}
\usepackage{import}
\numberwithin{equation}{section}
\usepackage{standalone}
\usepackage{tikz}
\usepackage{comment}
\usepackage[compat=1.1.0]{tikz-feynman}
\usetikzlibrary{decorations.markings,arrows,arrows.meta,shapes}
\usetikzlibrary{decorations.text}
\usetikzlibrary{bending} 
\tikzfeynmanset{warn luatex=false}
\usepackage{slashed}
\DeclareUnicodeCharacter{2212}{-}
\usepackage[makeroom]{cancel}
\usepackage{orcidlink}
\graphicspath{{./figures/}}

\newcommand{\dmom}[1]{ \frac{d^{D-1}{\boldsymbol #1}}{(2 \pi)^{D-1} \, 2 \, |\boldsymbol #1| }}
\newcommand{\dmomM}[2]{ \frac{d^{D-1}{\boldsymbol #1}_f}{(2 \pi)^{D-1} \, 2 \, \sqrt{{\boldsymbol #1}_f^2 + {#2}^2} }}

\def\nn {\nonumber}
\def\Fcal{{\cal F}}

\author[a]{Charalampos Anastasiou\orcidlink{0000-0003-4932-5140},}
\emailAdd{babis@phys.ethz.ch}
\author[a]{Julia Karlen\orcidlink{0009-0004-7545-0863},}
\emailAdd{karlenj@phys.ethz.ch}
\author[a]{Yao Ma\orcidlink{0000-0001-5067-2508},}
\emailAdd{yaomay@phys.ethz.ch}
\author[b]{George Sterman\orcidlink{0000-0002-7408-8813}}
\emailAdd{george.sterman@stonybrook.edu}
\affiliation[a]{Institute for Theoretical Physics, ETH Z\"urich,\\8093 Z\"urich, Switzerland}
\affiliation[b]{ C.N.\ Yang Institute for Theoretical Physics and Department of Physics and Astronomy,\\Stony Brook University,\\Stony Brook NY, 11794-3840 USA}

\title{
Local finiteness for real-virtual corrections to electroweak production in partonic collisions
}

\abstract{
We present a local subtraction scheme that enables the combined integration of loop momenta and the final-state parton phase space in real-virtual NNLO QCD corrections to cross sections for hadroproduction of electroweak and other colorless states. All initial- and final-state infrared singularities are subtracted at the integrand level in momentum space, yielding a locally finite integral ready for numerical integration in four dimensions. The subtraction terms are all based on the well-understood process of single-Higgs production. 
The core of our subtraction scheme relies on achieving local factorization in all infrared limits of real and virtual momenta. This necessitates systematic modifications of the original Feynman integrand for loop amplitudes, enabling gauge symmetry cancellations before performing integrations. Our approach provides an essential step toward NNLO cross-section calculations for hadron collider processes, where both loop and phase-space integrations are carried out numerically.
}

\begin{document}
\maketitle

\section{Introduction}
\label{section-introduction}

Next-to-next-to-leading-order (NNLO) QCD corrections to cross sections for hadron-hadron processes are a critical input for precision studies at colliders. However, their calculation has not been possible for a variety of high-multiplicity processes that are important for validating the Standard Model and for enhancing the sensitivity of searches to new physics. As the complexity of these calculations at the frontier of perturbative QCD grows tremendously, it is important to develop novel approaches and computational methods.

The intricate infrared structures in different components of a cross section, discussed from different perspectives in textbooks and reviews~\cite{EdenLdshfOlvPkhn02book,Sterman:1993hfp,Collins:2011zzd,Agarwal:2021ais,BurStw13lectures,Becher:2014oda}, are a main reason for this complexity. We believe that a promising approach for developing more powerful methods for NNLO cross sections is to exploit our understanding of infrared singularities in amplitudes and cross sections, especially their universality, at the diagrammatic level. From this point of view, local infrared counterterms for exclusive NNLO amplitudes of multi-particle electroweak production were developed in refs.~\cite{Anastasiou:2018rib,Anastasiou:2020sdt,Anastasiou:2022eym,Anastasiou:2024xvk,Anastasiou:2025cvy}. In this paper, we extend this analysis to real-virtual QCD corrections to electroweak cross sections at the same order, again identifying a set of local infrared counterterms that enable in principle the numerical evaluation of dependence on the electroweak final state.

The physical cross section for an infrared-safe process includes contributions from partonic processes with additional final-state partons beyond those present in the Born approximation. At NLO, new tree-level processes with one additional final-state gluon (single-real radiation) are required, and the Born process receives a one-loop correction (single-virtual). At NNLO, we need tree-level processes with two additional final-state partons (double-real), processes with one final-state parton and a loop correction (real-virtual), and two-loop corrections to the Born process (double-virtual). Beyond leading order, all these virtual and real corrections contain infrared singularities, which cancel when combined into a physical cross section for an infrared-safe observable.

Established methods for isolating singularities in partonic cross sections and arranging their cancellations are designed to facilitate numerical integration over the phase space of real radiation~
\cite{Anastasiou:2003gr,Anastasiou:2004qd,Anastasiou:2004xq,Anastasiou:2005qj,Anastasiou:2010pw,Anastasiou:2011qx,Buhler:2012ytl,Gehrmann-DeRidder:2005btv,Daleo:2006xa,Gehrmann-DeRidder:2005alt,Gehrmann-DeRidder:2005svg,Gehrmann-DeRidder:2007nzq,Daleo:2009yj,Boughezal:2010mc,NigelGlover:2010kwr,Abelof:2011jv,Gehrmann:2011wi,Gehrmann-DeRidder:2012too,Currie:2013vh,Currie:2016ytq,Jakubcik:2024poz,NNLOJET:2025rno,Somogyi:2005xz,Somogyi:2006da,Somogyi:2006db,Somogyi:2008fc,Aglietti:2008fe,Somogyi:2009ri,Bolzoni:2009ye,Bolzoni:2010bt,DelDuca:2013kw,Somogyi:2013yk,DelDuca:2016ily,DelDuca:2016csb,Fekeshazy:2025ktp,Magnea:2018hab,Magnea:2018ebr,Magnea:2020trj,TorresBobadilla:2020ekr,Bertolotti:2022aih,Magnea:2024jqg,Caola:2017dug,Caola:2018pxp,Caola:2019pfz,Caola:2019nzf,Delto:2019asp,Asteriadis:2019dte,Bizon:2020tzr,Czakon:2010td,Czakon:2011ve,Czakon:2014oma,Dreyer:2016oyx,Han:1992hr,Brucherseifer:2014ama,Cacciari:2015jma,Frixione:2004is,Catani:2007vq,Boughezal:2011jf,Boughezal:2015aha,Gaunt:2015pea,Herzog:2018ily,Chen:2021vtu,Buonocore:2023rdw,Grazzini:2017mhc,Boughezal:2015dva,Sborlini:2016hat,Chen:2022ktf,Devoto:2023rpv,Braun-White:2023sgd,Braun-White:2023zwd,Fox:2023bma,Gehrmann:2023dxm,Fox:2024bfp,DelDuca:2024ovc,Devoto:2025kin,Devoto:2025jql,DelDuca:2025yph,Chen:2014gva,Boughezal:2015dra,Caola:2015wna,Chen:2016zka,Campbell:2019gmd,Cruz-Martinez:2018rod,Gauld:2021ule,Catani:2022mfv,Chawdhry:2019bji,Chawdhry:2021hkp,Czakon:2020coa,Gauld:2023zlv,Currie:2017eqf,Chen:2022tpk,Badger:2023mgf,Czakon:2021mjy,Czakon:2015owf,Catani:2019hip,Buonocore:2023ljm,Berger:2016oht,Campbell:2020fhf,Bronnum-Hansen:2022tmr,Buonocore:2022pqq,Alvarez:2023fhi,Armadillo:2024ncf, Devoto:2024nhl, Buonocore:2025fqs, Bonino:2025qta}.
Such a numerical approach is essential for computing cross sections of observables that reflect the complexities of realistic experimental event selection. By contrast, virtual corrections, owing to their unrestricted momentum integration domain, can in principle be treated using a wider range of analytic, semi-numerical, or numerical techniques~\cite{Soper:1999xk,Nagy:2006xy,Gong:2008ww,Becker:2010ng,Assadsolimani:2009cz,Becker:2012aqa,Becker:2011vg,Becker:2012bi,Gnendiger:2017pys,Seth:2016hmv,Capatti:2019edf,Capatti:2020ytd,Capatti:2020xjc,Capatti:2022tit,AH:2023kor,Ramirez-Uribe:2024rjg,LTD:2024yrb,Kermanschah:2021wbk,Vicini:2024ecf,Kermanschah:2024utt,Kermanschah:2025wlo,Rios-Sanchez:2024xtv,Capatti:2025khs,Anastasiou:2004vj,Smirnov:2008iw,Smirnov:2013dia,Smirnov:2014hma,Smirnov:2019qkx,Lee:2012cn,Lee:2013mka,Studerus:2009ye,vonManteuffel:2012np,Lange:2025fba,Artico:2023jrc,Chen:2024xwt,Peraro:2019svx,Klappert:2020aqs,Hidding:2020ytt,Liu:2022chg,Armadillo:2022ugh,Zhang:2024fcu,Binoth:2000ps,Anastasiou:2005pn,Anastasiou:2006ah,Lazopoulos:2007ix,Anastasiou:2007qb,Bogner:2007cr,Lazopoulos:2007bv,Heinrich:2008si,Lazopoulos:2008de,Anastasiou:2008rm,Kaneko:2009qx,Borowka:2015mxa,Borowka:2017idc,Heinrich:2023til,Jones:2025jzc,Jones:2026tkf,Smirnov:2021rhf,Borinsky:2020rqs,Borinsky:2023jdv,Smirnov:1999gc,Smirnov:1999wz,Tausk:1999vh,Smirnov:2001cm,Czakon:2005rk,Anastasiou:2005cb,Smirnov:2009up,Gluza:2010ws,Gluza:2016fwh,Dubovyk:2017cqw,Dubovyk:2018rlg,Panzer:2014caa,vonManteuffel:2014qoa}.

For Born processes with three or more particles in the final state, the analytic computation of two-loop corrections is challenging and constitutes a research frontier. The numerical evaluation of two-loop amplitudes poses its own challenges, but it has proven to be an invaluable alternative. Within the numerical integration approach, powerful techniques for loop integrations in momentum space are emerging and being applied to two-loop amplitudes with an unprecedented number of kinematic and mass scales~\cite{Kermanschah:2021wbk,Vicini:2024ecf,Kermanschah:2024utt,Kermanschah:2025wlo}. 

As numerical techniques become necessary for important components of a cross section, one may also contemplate evaluating all of them numerically.
More generally, recent work also pursues a framework where all parts of a cross section, comprising both real and virtual corrections, are combined under a common integration~\cite{Capatti:2020xjc,Capatti:2022tit,Ramirez-Uribe:2024rjg,LTD:2024yrb,AH:2023kor,Capatti:2025khs}.
In this approach, infrared singularities of these parts are arranged to cancel in a common integrand, with algorithms that directly realize the Kinoshita-Lee-Nauenberg theorem~\cite{Kinoshita:1962ur,Lee:1964is,Sterman:1978bj}. As long as the cancellation is complete, as in decay processes of neutral particles, the combined integration of real and virtual components enables the computation of rates to be carried out fully numerically.
However, for hadron collider processes or processes with identified hadrons in the final state, the cancellation of infrared singularities between real and virtual corrections is not complete. Residual initial-state collinear singularities remain, which are subtracted and replaced by parton distribution functions, and similarly, final-state singularities are absorbed into fragmentation functions.

Contributing to this broad effort, for cross sections of hadron collider processes, we aim to construct a numerical integration framework in which initial-state singularities of partonic cross sections are subtracted \emph{locally}, i.e., at the level of integrands, while final-state singularities are canceled locally in the sum of real and virtual contributions. To achieve this goal at NNLO, we would express all components of partonic cross sections as particular integrals in momentum space, for which the initial-state singularities of real and virtual corrections can be factorized at the level of integrands. This task of \emph{local factorization} has long been accomplished for double-real radiation at NNLO. The factorization of one-loop primitive amplitudes at NLO has been discussed in ref.~\cite{Assadsolimani:2009cz}. For two-loop amplitudes, the construction of locally factorizable integrands has been achieved for colorless production in electron-positron annihilation in QED~\cite{Anastasiou:2020sdt}, for quark-antiquark annihilation~\cite{Anastasiou:2022eym,Anastasiou:2025cvy}, and gluon fusion~\cite{Anastasiou:2024xvk}. A first study of this issue for three-loop quark-antiquark annihilation was carried out in ref.~\cite{Haindl:2025jte}.

Achieving local factorization poses challenges, even at the amplitude level, and requires dedicated methods to overcome several obstructions. These include, for example, non-factorizable contributions from off-shell loop momenta within a jet, which involve spurious gauge boson polarizations (termed \emph{loop polarizations}). A similar issue arises when this off-shell loop appears as a self-energy correction to a jet propagator, inducing spurious \emph{power divergences}. Another subtlety, referred to as \emph{shift mismatch}, occurs when sums of non-factorized terms cancel only after the shift of a loop momentum integration variable. While all these contributions vanish upon integration, they remain nonzero at the integrand level, thus obstructing local factorization. One of the central goals of this work is to implement the necessary additional modifications of the integrand to resolve these obstructions systematically.

In this paper, we also take an important step toward local factorization, along the lines described above, at the cross-section level, focusing on the real-virtual corrections to colorless production. Starting from the quark-antiquark channel, we construct an integrand for the real-virtual correction that is locally factorizable in all infrared limits. Our approach relies on the universality of infrared singularities. For the production of a large set of electroweak particles at short distances, the list of singular regions of integration space depends only on the choice of incoming partons that initiate the short-distance process. The list does not depend on the number of colorless particles that are produced, or on their relative locations in phase space. This universality enables us to locally factorize infrared singularities in momentum space for the real-virtual correction of colorless production from quark-antiquark annihilation, and other related channels. Additionally, it enables us to use partonic cross sections to generate subtractions from the simplest process in the class we are considering, in this case single-Higgs production. The resulting “cross-section-based” subtraction scheme factorizes all dependence on the electroweak final state from the universal singular structure, making it accessible to numerical evaluation. This feature enables the application of NNLO virtual corrections to more complex processes, such as di- and tri-photon production, as demonstrated in ref.~\cite{Kermanschah:2025wlo}.

We emphasize that this cross-section-based factorization is fully consistent with standard ``parton distribution" and ``collinear" factorizations. Once factorized from the short-distance function, a partonic single-Higgs production cross section can itself be factorized in minimal or other subtraction schemes. What we show below, however, is that it is possible to isolate the process dependence of the final states in electroweak production, independently of the details of collinear factorization. We may draw an analogy between ``cross-section-based" subtraction and the ``DIS scheme"~\cite{CTEQ:1993hwr} for structure functions in deep-inelastic scattering, an alternative collinear subtraction method that was motivated by an analogous universality of infrared structure in physical processes. General antenna subtraction methods for subtracting singularities from real radiation are founded on an analogous concept~\cite{Gehrmann-DeRidder:2005btv}.

This paper is organized as follows. In section~\ref{sec:RVamplitude}, we construct a locally finite one-loop amplitude for the production of a massive electroweak boson with a gluon via quark-antiquark annihilation, detailing our strategy for overcoming the obstructions to local factorization. Section~\ref{sec:section-finite_cross_section} introduces our conventions for partonic cross sections. For each cross section, we aim to construct subtraction terms that render the remainder locally finite. We accomplish this at order $\alpha_s$ in section~\ref{sec:NLOxsec_localfact}. In sections~\ref{sec:RVxsec_localfact} and~\ref{section-finite_RV_xsec_integrand_qg_channel}, we generalize these results to order $\alpha_s^2$ for the quark-antiquark and quark-gluon channels, respectively. A detailed analysis verifying local factorization is provided in section~\ref{sec:IRanalyze}, with the numerical checks of infrared finiteness explained in section~\ref{sec:numcheck}. We conclude with a summary and outlook in section~\ref{section-conclusions_outlook}. Additional technical details are collected in the appendices.

\section{\texorpdfstring{One-loop amplitude integrand in the $q\bar q$ channel}{One-loop amplitude integrand in the qqbar channel}}
\label{sec:RVamplitude}
We consider the production of a generic collection of massive electroweak bosons with momenta $q_1,q_2, \dots,q_n$, in association with a gluon via quark–antiquark annihilation,
\begin{align}
\label{eq:qqbar_process_definition}
    q(p_1) + \overline{q}(p_2) \longrightarrow g(p_3) + \textup{ew}(q_1,\dots,q_n) \, ,
\end{align}
where ``ew'' denotes the set of massive colorless bosons in the final state, with the corresponding particle momenta indicated in parentheses.
This process contributes beginning at leading order to the hadroproduction cross section of colorless particles in association with a jet, and beginning at NLO to the hadroproduction cross section of colorless particles including additional radiation.
For the gluon momentum $p_3$, we must therefore examine two cases relevant to the phase-space integration of the inclusive colorless production cross section: (i) a high-transverse-momentum gluon that can be resolved as a jet, and (ii) an unresolved gluon, either collinear to the incoming momenta $p_1$ or $p_2$, or soft ($p_3 \to 0$). The momenta of $q_i$ of the color-singlet final-state particles are assumed to have either large invariant masses or wide-angle separations from all other initial- and final-state particles.

The perturbative expansion of the amplitude for the process in eq.~\eqref{eq:qqbar_process_definition} reads
\begin{eqnarray}
\label{eq:Mnplus1}
 M_{q\bar q \to {\rm ew}+g}(p_1,p_2;p_3, \{q_i\}) &=& {\cal M}_{q\bar q \to {\rm ew}+g}^{(0)}(p_1,p_2; p_3, \{q_i\}) 
 \nonumber \\
&& 
\hspace{-1cm}
+ \int \frac{d^D k}{(2 \pi)^D} \, {\cal M}_{q\bar q \to {\rm ew}+g}^{(1)}(p_1, p_2;p_3,\{q_i\};k) 
+{\cal O}(g_s^5) \, . 
\end{eqnarray} 
In the notation ${\cal M}_{q\bar q \to {\rm ew}+g}^{(m)}$, the superscript $m=0,1$ denotes the loop order. From now on we will drop the color-singlet momenta $\{q_i\}$ for legibility. The Born amplitude $\mathcal{M}_{q\bar q \to {\rm ew}+g}^{(0)}$ consists of Feynman diagrams with a single fermion line connecting the incoming quark ($p_1$) and antiquark ($p_2$), an outgoing gluon ($p_3$), and outgoing colorless particles (thick dashed lines) emitted from the fermion line.
Graphically,
\begin{align}
     \mathcal{M}_{q\bar q \to {\rm ew}+g}^{(0)} &= 
    \includegraphics[scale=0.8, page=1,valign=c]{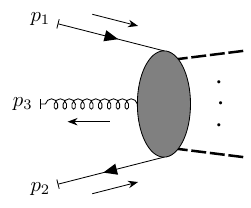}\, .
    \label{eq:M_born}
\end{align}
In this notation, the gray blob represents a tree-level subdiagram that includes all possible permutations of the final-state particles (${\rm ew}+g$). At the one-loop order, the following diagrams, classified based on their color factors and the order of gluons attached to the fermion line, are relevant:
\begin{align}
    \label{eq:M_1loop}
     \mathcal{M}_{q\bar q \to {\rm ew}+g}^{(1)} &=  
     \underset{\rm (A)}{
    \includegraphics[scale=0.8, page=5,valign=c]{figures_standalone/tikz_RVampl.pdf}} 
   + \underset{\rm (B)}{
    \includegraphics[scale=0.8, page=2,valign=c]{figures_standalone/tikz_RVampl.pdf}}
     + \underset{\rm (C)}{
    \includegraphics[scale=0.8, page=3,valign=c]{figures_standalone/tikz_RVampl.pdf}} \nonumber \\ &
      + \underset{\rm (D)}{
    \includegraphics[scale=0.8, page=4,valign=c]{figures_standalone/tikz_RVampl.pdf}}    
+ \underset{\rm (E)}{
    \includegraphics[scale=0.8,page=6,valign=c]{figures_standalone/tikz_RVampl.pdf}}
    \, .
\end{align}
The first class (A) consists of all diagrams with triple-gluon vertices. The second class (B) includes diagrams in which the final-state gluon is emitted inside the virtual gluon loop. Classes (C) and (D) contain diagrams where the final-state gluon is emitted outside the virtual gluon loop. Note that, in our notation, the ordering of the gluons attached to the gray blob is relevant. Consequently, (C) and (D) are treated as distinct configurations. Moreover, we exclude all diagrams with self-energy corrections to the incoming legs in this notation. The last class (E) comprises those with a fermion loop, to which the final-state gluon must attach. (Diagrams where the gluon attaches to the initial-state fermion line vanish due to the color structure.)
The color factors corresponding to these classes are shown below, where the superscript $a$ denotes the color of the final-state gluon (where we use the convention $T_F=1/2$):
\begin{align}
\begin{alignedat}{2}
&{\rm (A):}\  \frac{C_A T^a}{2} \, ,\qquad\qquad & {\rm (B):}\ & -\frac{T^a}{2C_A}\, ,\\
&{\rm (C)}\ \&\ {\rm (D):}\  C_F T^a \, ,      \qquad\qquad & {\rm (E):}\ & \frac{T^a}{2}\, .
\end{alignedat}
\end{align}
For each class, we shall assign a specific momentum routing in order to construct the full integrand ${\cal M}_{q\bar q \to {\rm ew}+g}^{(1)}(p_1,p_2;p_3;k)$, that is required for local factorization.

\begin{figure}[H]
    \centering
 \begin{align}
&    \includegraphics[scale=0.7, page=1,valign=c]{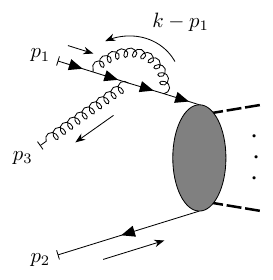}
\quad 
        \includegraphics[scale=0.7, page=5,valign=c]{figures_standalone/tikz_RV_LP.pdf}
\quad 
           \includegraphics[scale=0.7, page=3,valign=c]{figures_standalone/tikz_RV_LP.pdf}
\nonumber \\
&    \includegraphics[scale=0.7, page=2,valign=c]{figures_standalone/tikz_RV_LP.pdf}
\quad 
        \includegraphics[scale=0.7, page=6,valign=c]{figures_standalone/tikz_RV_LP.pdf}
\quad 
           \includegraphics[scale=0.7, page=4,valign=c]{figures_standalone/tikz_RV_LP.pdf}
\nonumber
\end{align}
    \caption{Diagrams that give rise to either loop polarizations (the first two columns) or power singularities (the third column) as $p_3 \parallel p_1$ (the first row) or $p_3 \parallel p_2$ (the second row).}
    \label{fig:LoopPolarizationDiagrams}
\end{figure}

Before proceeding, we note a first obstruction arising from diagrams with a one-loop correction to the quark-gluon vertex on the incoming quark or antiquark (see the first two columns of figure~\ref{fig:LoopPolarizationDiagrams}). In the limits $p_3\parallel p_1$ and $p_3\parallel p_2$ (while the virtual loop momentum is off shell), the real gluon acquires contributions where its polarization is proportional to the virtual loop momentum $k$ instead of $p_3$. Such spurious polarizations, termed ``loop polarizations''~\cite{Anastasiou:2020sdt,Anastasiou:2022eym,Anastasiou:2024xvk,Haindl:2025jte,Anastasiou:2025cvy}, can lead to logarithmic divergences in the collinear limits above and break factorization at the integrand level, although they vanish after integration over the loop momentum.
A related problem arises from diagrams with a one-loop self-energy correction on the internal fermion line (see the last column of figure~\ref{fig:LoopPolarizationDiagrams}). When the fermion momentum is on shell while the virtual loop momentum is off shell, a power-like singularity appears locally. Although it cancels after integration, special treatment is required for local subtraction.

To address these issues, we isolate the corresponding integrands and modify them into a form that preserves local factorization while yielding the same result as traditional Feynman rules after integration. Namely,
\begin{eqnarray}
\label{eq:RVpol+rest}
    {\cal M}_{q\bar q \to {\rm ew}+g}^{(1)} = {\cal M}_{q\bar q \to {\rm ew}+g}^{(1), \, {\rm lp}} +  {\cal M}_{q\bar q \to {\rm ew}+g}^{(1),\, {\rm nolp}} \,, 
\end{eqnarray}
with both ${\cal M}_{q\bar q \to {\rm ew}+g}^{(1), \, {\rm lp}}$ and ${\cal M}_{q\bar q \to {\rm ew}+g}^{(1), \, {\rm nolp}}$ having the desired property that the divergence at $p_3\parallel p_1$ or $p_3\parallel p_2$ is at most logarithmic, always with a longitudinally polarized real gluon. The loop-polarization amplitude ${\cal M}_{q\bar q \to {\rm ew}+g}^{(1), \, {\rm lp}}$ requires modification to make it locally free of power-like singularities and loop polarizations in collinear limits. The superscript ``lp'' indicates that this contribution to the amplitude stems from diagrams with these issues if one uses traditional Feynman rules.
All contributions to the term 
${\cal M}_{q\bar q \to {\rm ew}+g}^{(1), \, {\rm nolp}}$
have the desired properties without modifications. As we shall see, several diagrams contribute to both of these terms. These issues were previously addressed in ref.~\cite{Anastasiou:2022eym}, where a possible modification of the amplitude integrand was provided. Below we present a refined formulation that allows for a separation into the two terms on the right-hand side of eq.~\eqref{eq:RVpol+rest}.

Once this is established, we can further assign momentum routing to each diagram in a way that ensures local factorization. These treatments constitute the foundation for constructing local subtraction terms that render finite amplitudes and cross sections, as we will detail below.

\subsection{Loop polarizations refined}
\label{sec:LP}

The aforementioned issues of loop polarization and power-like singularity arise from diagrams depicted in figure~\ref{fig:LoopPolarizationDiagrams}, when $k$ is hard and $p_3$ is collinear to $p_1$ (the first row) or $p_2$ (the second row). To obtain an expression for ${\cal M}_{q\bar q \to {\rm ew}+g}^{(1)}$ as in eq.~\eqref{eq:RVpol+rest}, we start by replacing the original form of a self-energy correction by
\begin{align}
\label{eq:self_energy_correction_modified_expression}
    \includegraphics[scale=0.75, page=7,valign=c]{figures_standalone/tikz_RV_LP.pdf} &= \frac{i}{\slashed{p}_1-\slashed{p}_3} \, \frac{-(2-D) g_s^3 C_F T^a}{(k-p_1)^2} \frac{1}{\slashed{k}-\slashed{p}_3} \, \frac{1}{\slashed{p}_1-\slashed{p}_3}\, \gamma^\mu \, u(p_1)\nn\\
    &\xrightarrow[\phantom{aaa}]{} \frac{(2-D) g_s^3 C_F T^a}{(k-p_1)^2 (k-p_3)^2} \, \frac{i}{\slashed{p}_1-\slashed{p}_3}\,  \frac{(-\gamma^\mu)}{2} \, u(p_1) \,.
\end{align}
This removes the power-like singularity without changing the integration value. In this modified form, the full self energies will be included in ${\cal M}_{q\bar q \to {\rm ew}+g}^{(1), \, {\rm lp}}$ defined in eq.\ (\ref{eq:RVpol+rest}).

To identify the loop-polarization integrand, let us further isolate the contributions from diagrams with a triple-gluon vertex (the second column of figure~\ref{fig:LoopPolarizationDiagrams}). Specifically, we employ the following ``scalar decomposition" of the triple gluon vertex introduced in ref.~\cite{Anastasiou:2024xvk}:
\begin{align}
\label{eq:Vggg}
&\includegraphics[scale=0.8,page=1,valign=c]{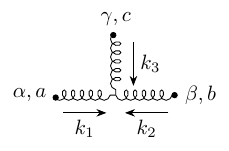}
    = \includegraphics[scale=0.8, page=2,valign=c]{figures_standalone/tikz_scalar_decomp_RV.pdf} 
    + \includegraphics[scale=0.8, page=3,valign=c]{figures_standalone/tikz_scalar_decomp_RV.pdf} + \includegraphics[scale=0.8, page=4,valign=c]{figures_standalone/tikz_scalar_decomp_RV.pdf}\,
\end{align}
with
\begin{eqnarray}
\label{eq:Vssg}
\includegraphics[scale=0.8, page=2,valign=c]{figures_standalone/tikz_scalar_decomp_RV.pdf}
&=& -g_s f_{abc} \eta^{\alpha\beta} (k_1-k_2)^{\gamma} \, . 
\end{eqnarray}
The dots on the end of the lines indicate truncation, i.e. the exclusion of propagators or polarization vectors. This construction permits a graphical decomposition of the triple-gluon vertex, as each term plays a different role in the various infrared limits. The decomposition is referred to as a ``scalar decomposition", since the vertex in eq.~\eqref{eq:Vssg} is equal to a color-octet scalar gluon vertex with an additional metric tensor. The ``scalar" lines serve only as a graphical aid and must not be interpreted as actual scalar propagators. A more in depth explanation of this decomposition can be found in ref.~\cite{Anastasiou:2024xvk}.

On the right-hand side of eq.~(\ref{eq:Vggg}), only the second term contributes to loop polarizations in figure~\ref{fig:LoopPolarizationDiagrams} when $k_1$ is taken to be $-p_3$ (with $a$ denoting the color of the outgoing gluon $p_3$). We will refer to it as the \emph{$k_1$-scalar term}, because it is precisely the vertex for a scalar fermion coupled to a gluon of momentum $k_1$, which is equal to $-p_3$ here. Anticipating the role of Ward identities below, we will find it useful to call the remaining two terms the \emph{$k_1$-ghost terms}.

Combining this QCD-like $p_3$-scalar term with the full QED-like vertex correction and the modified self-energy from eq.~\eqref{eq:self_energy_correction_modified_expression}, we obtain:
\begin{align}
\label{eq:P1jet}
&\raisebox{-1.2mm}{\includegraphics[scale=0.75, page=7,valign=c]{figures_standalone/tikz_RV_LP.pdf}}
+\includegraphics[scale=0.75, page=8,valign=c]{figures_standalone/tikz_RV_LP.pdf}
+\raisebox{-6.5mm}{\includegraphics[scale=0.75, page=9,valign=c]{figures_standalone/tikz_RV_LP.pdf} } \nonumber \\ 
&\quad = \frac{i}{\slashed{p}_1-\slashed{p}_3}\, \left[J^{a \, \mu}(k, p_1, p_3) +V_{\rm qed}^{a \mu}(k, p_1, p_3)\right]\, u(p_1)\,,
\end{align}
where all loop polarizations arise from $J^{a\mu}$, and $V_{\rm qed}^{a\mu}$ is free of loop polarizations, 
\begin{subequations}
    \begin{align}
        &V_{\rm qed}^{a \mu}(k, p_1, p_3) = -\frac 1 2 \, \frac{g_s^3 \left(C_F-\frac{C_A}{2} \right) \, T^a}{(k-p_1)^2 (k-p_3)^2 k^2} \gamma^\rho \left[ \gamma^\mu \, , \, \slashed{p}_3 \right] \slashed{k} \gamma_\rho\,, \label{eq:Vcurrent}\\
        &J^{a \mu}(k, p_1, p_3) = \frac{(2-D) \, g_s^3 \left(C_A-C_F\right) T^a}{(k-p_1)^2 (k-p_3)^2} \left[ (2 k - p_3)^\mu \frac{\slashed k}{k^2} - \frac {\gamma^\mu}{2}\right]\, . \label{eq:Jcurrent}
    \end{align}
\end{subequations}
The term $V_{\rm qed}^{a, \mu}$ has the following properties:
\begin{itemize}
    \item Only the second diagram, i.e., the QED-like vertex correction in eq.~\eqref{eq:P1jet} contributes to the infrared behavior in the limit $k\parallel p_1$, which is fully captured by $V_{\rm qed}^{a, \mu}$:
    \begin{eqnarray}
    \frac{i}{\slashed{p}_1-\slashed{p}_3}\, V_{\rm qed}^{a \mu}(k, p_1, p_3)\, u(p_1) \xrightarrow[]{k\parallel p_1} \includegraphics[scale=0.75, page=8,valign=c]{figures_standalone/tikz_RV_LP.pdf} \, .
    \label{eq:Vqed_klim}
    \end{eqnarray}
    The notation ``$\xrightarrow[]{k\parallel p_1}$'' indicates that the two expressions on either side have the same behavior in the limit $k \parallel p_1$. This motivates the notation we adopt for the current $V_{\rm qed}^{a, \mu}$.
    We will use the property of eq.~(\ref{eq:Vqed_klim}) to demonstrate local factorization in section~\ref{sec:IRdemo_RVampl}.
    \item In the limit $p_3\parallel p_1$ the current is $V_{\rm qed}^{a \, \mu} \propto p_3^\mu$, which implies that no loop polarizations are present, and all such contributions are therefore contained in $J^{a \, \mu}$.
    \item It vanishes algebraically on contracting with $p_{3\mu}$:
    \begin{equation}
    p_{3 \mu} \, V_{\rm qed}^{a \, \mu}(k, p_1, p_3) = 0 \, .
    \label{eq:p3Vqed}
    \end{equation}
    This yields a suppression when $p_3 \parallel \eta$, with $\eta$ being any lightlike vector that is not proportional to $p_1$ (e.g., $\eta=p_2$), preventing infrared divergence in this limit.
\end{itemize}

Now we shall modify the expression of $J^{a \mu}$ and construct a form that possesses local properties analogous to those satisfied by $V^{a \mu}_{\rm qed}$, specifically, being transverse to $p_3^\mu$, and the elimination of loop polarizations without modifying its integrated value, including its virtual $k\parallel p_1$ singularity. To this end, we consider the following decomposition of a generic vector $v^\mu$:
\begin{eqnarray}
\label{eq:general_vector_decomposition}
v^\mu &=& \frac{ v\cdot \eta }{ p_1 \cdot \eta} \, p_1^\mu\, + \, \frac{ v\cdot p_1 }{ p_1 \cdot \eta} \, \eta^\mu\, + \, v^\mu_\perp \, ,
\end{eqnarray}
where $\eta$ is an auxiliary lightlike vector that is not proportional to $p_1$ and $v_\perp\cdot p_1=v_\perp\cdot \eta =0$. By applying this decomposition to $\gamma^\mu$ and $\slashed{k}$ in eq.~\eqref{eq:Jcurrent}, we obtain
\begin{equation}
    J^{a \mu}(k, p_1, p_3) u (p_1) = \left[ J_\perp^{a \mu}(k, p_1, p_3, \eta) + J_\parallel^{a \mu}(k, p_1, p_3, \eta)  \right] u (p_1) \, ,
\end{equation}
with
\begin{subequations}
    \begin{align}
        &J_\perp^{a \mu}(k, p_1, p_3, \eta) = \frac{(2-D) \, g_s^3 \left(C_A-C_F\right)  T^a}{(k-p_1)^2 (k-p_3)^2} \left[ (2 k - p_3)^\mu \frac{{\slashed k}_\perp}{k^2} - \frac {\gamma^\mu_\perp}{2} \right] \, ,
        \label{eq:Jcurrent_perp}\\
        & J_\parallel^{a \mu}(k, p_1, p_3, \eta) = \frac{(2-D)}{2} \, g_s^3 \left(C_A-C_F\right)  T^a \, \nn\\
        &\hspace{4.5cm} \cdot\left[ \frac{(2 k-p_1-p_3)^\mu}{(k-p_1)^2 (k-p_3)^2}- \frac{(2 k-p_3)^\mu}{k^2 (k-p_3)^2} \right] \frac{\slashed \eta}{p_1 \cdot \eta} \, .
        \label{eq:Jcurrent_parallel}
    \end{align}
\end{subequations}
We note that the parallel component does not contribute at the integral level, since
\begin{eqnarray}
    \int d^Dk \, J_\parallel^{a \mu}(k, p_1, p_3, \eta) = 0 \, ,
\end{eqnarray}
which follows from tensor reduction.\footnote{Both terms in the square bracket of eq.~(\ref{eq:Jcurrent_parallel}) vanish upon integration. To see this, one can use the following change of variables: $k\to p_1+p_3-k$ for the first term, and $k\to p_3-k$ for the second term.} We will thus drop $J_\parallel^{a \mu}$ from the amplitude. Equivalently, we add a counterterm to the original integrand, which removes the parallel component and integrates to zero.

For the transverse part $J_\perp^{a \mu}$, we find the following properties: 
\begin{itemize}
    \item It is finite in the collinear limit  $k \parallel p_1$. 
    \item In the limit $p_3\parallel p_1$ there is always a numerator suppression, due to 
    \begin{eqnarray}
    \label{eq:Jperp_property2}
        \slashed{p}_1 \,  J_\perp^{a \mu}(k, p_1, p_3, \eta) \, u(p_1) 
        = - J_\perp^{a \mu}(k, p_1, p_3, \eta) \, \slashed{p}_1 \, u(p_1) =0 \,.
    \end{eqnarray}
    This implies that $J_\perp^{a \mu}$ is \emph{free of loop polarizations}.
    \item It does not generally vanish on contracting with $p_{3\mu}$, even in the limit $p_3\parallel \eta$:
    \begin{eqnarray}
    \label{eq:Jperp_contraction_p3_nonvanishing}
        p_{3\mu}J_\perp^{a \mu}(k, p_1, p_3, \eta)\,  \cancel{\xrightarrow{p_3 \parallel \eta}}\, 0\, .
    \end{eqnarray}
    By taking $\eta=p_2$, this inequality implies that contracting $p_2^\mu$ with the left-hand side of eq.~\eqref{eq:Jcurrent_perp} (and furthermore, eq.~(\ref{eq:P1jet})) is nonzero, which breaks the local factorization in the limit $p_3 \parallel p_2$.
\end{itemize}
Note that at the level of cross sections, the last point is resolved once the external gluon is restricted to have physical (transverse) polarization. In this case, all contributions involving $J_\perp^{a \mu}(k, p_1, p_3, \eta)$ remain finite in the limit $p_3\parallel p_2$. It is, however, permissible for this class of processes to consider an external gluon with four polarizations, in which case we need to further modify $J_\perp^{a \mu}$ in order to achieve local factorization in all infrared limits. Moreover, this refined treatment of loop polarizations and self-energy diagrams can be directly applied to the two-loop electroweak boson production amplitude of ref.~\cite{Anastasiou:2022eym}, where the momentum $p_3$ of the external gluon is instead identified with a loop momentum.

To this end, we shall present two alternative methods that further modify $J_\perp^{a \mu}$, admitting local factorization in all infrared limits. The first one symmetrizes the transverse plane with respect to $p_1$ and $\eta$. The second one applies tensor reduction to certain contributions of the current, which could be more advantageous for numerical integration.\footnote{We thank Dario Kermanschah and Matilde Vicini for the communication on this topic.}

\paragraph{Method 1: symmetrizing the transverse plane.}
We start by decomposing the loop momentum $k$ according to eq.~(\ref{eq:general_vector_decomposition}) and define $\widetilde k$ as the vector obtained by reflecting the transverse components of $k$ in this form,
\begin{align}
    k^\mu = \frac{k \cdot \eta}{p_1 \cdot \eta} \,  p_1^\mu +\frac{k \cdot p_1}{p_1 \cdot \eta} \,  \eta^\mu +k_\perp^\mu \, , \qquad \widetilde k^\mu \equiv \frac{k \cdot \eta}{p_1 \cdot \eta} \,  p_1^\mu +\frac{k \cdot p_1}{p_1 \cdot \eta} \,  \eta^\mu -k_\perp^\mu \, .
\end{align}
We then define a symmetrized current $\widetilde J_\perp$ under reflections of the loop momentum $k$ on its transverse plane:
\begin{eqnarray}
\label{eq:Jcurrent_perp_tilde}
  \widetilde J_\perp^{a \mu}(k, p_1, p_3, \eta) &\equiv& 
  \frac{J_\perp^{a \mu}(k, p_1, p_3, \eta) + J_\perp^{a \mu}(\widetilde k, p_1, p_3, \eta)}{2}
  \nonumber \\ 
  &\hspace{-4cm}=&\hspace{-2cm} \frac{(2-D) \, g_s^3 \left(C_A-C_F\right)  T^a}{(k-p_1)^2} \left\{ \left[ \frac{(2 k - p_3)^\mu}{(k-p_3)^2} - \frac{(2 \widetilde k - p_3)^\mu}{(\widetilde k-p_3)^2} \right] \frac{{\slashed k}_\perp}{ 2 \, k^2 } \right.\nonumber\\
  &&\hspace{3cm}-\left. \frac {\gamma^\mu_\perp}{4} \left[ \frac{1}{(k-p_3)^2} + \frac{1}{(\widetilde k-p_3)^2} \right] \right\} \, .
\end{eqnarray}
By construction, $J_\perp^{a \mu}$ and $\widetilde J_\perp^{a \mu}$ lead to the same value after integration over the loop momentum $k$. In contrast to eq.~\eqref{eq:Jperp_contraction_p3_nonvanishing}, $\widetilde J_\perp^{a \mu}$ further satisfies
\begin{eqnarray}
\label{eq:Jtransverse_property}
p_{3\mu} \, \widetilde J_\perp^{a \mu}(k, p_1, p_3, \eta)  \xrightarrow[]{p_3 \parallel \eta}
\frac{(2-D)\, g_s^3 \left(C_A-C_F\right)  T^a}{(k-p_1)^2 (k-p_3)^2} \left(
2 k_\perp \cdot p_3  
\frac{{\slashed k}_\perp}{k^2} - \frac {{\slashed p_3}_\perp}{2}
\right)
 \xrightarrow[]{p_3 \parallel \eta} 0 \, ,\quad
\end{eqnarray}
where both terms in the bracket vanish. Hence the obstacle of eq.~(\ref{eq:Jperp_contraction_p3_nonvanishing}) to local factorization in the limit $p_3 \parallel \eta$ is cured.

\paragraph{Method 2: tensor reduction.}
As an alternative treatment to achieve local factorization in the limit $p_3\parallel \eta$, we aim to introduce a tensor reduction from $\slashed{k}_\perp$ to $\slashed{p}_{3\perp}$. In order to permit such a tensor reduction we introduce an auxiliary vector $\rho$ and decompose $J_\perp^{a \mu}$ as follows
\begin{align}
    J_\perp^{a \mu}(k, p_1, p_3, \eta) &= \left(g^{\mu\nu} - \frac{2p_3^\nu\rho^\mu}{\left(p_3+\rho\right)^2  - \rho^2}\right)J_{\perp\nu}^{a }(k, p_1, p_3, \eta) + \frac{2p_3^\nu\rho^\mu}{\left(p_3+\rho\right)^2  - \rho^2}J_{\perp\nu}^{a}(k, p_1, p_3, \eta)\,,
    \label{eq:Jperp_split_rho}
\end{align}
where we require $\rho\cdot p_i \neq 0$ for $i=1,2,3$. Here we have retained the dependence on $p_3^2$ in the denominator, since this method is equally applicable in the two-loop amplitude, where $p_3$ appears as a loop momentum. It can be considered as an alternative for the proposal in ref.~\cite{Anastasiou:2022eym}.
The first term on the right-hand side of eq.~\eqref{eq:Jperp_split_rho} is free of infrared singularities. Especially, as $p_3\parallel\eta$ we have
\begin{align}
    \label{eq:J_perp_split_finitepart}
    p_{3\mu}\left(g^{\mu\nu} - \frac{2p_3^\nu\rho^\mu}{\left(p_3+\rho\right)^2  - \rho^2}\right)J_{\perp\nu}^{a }(k, p_1, p_3, \eta)   \xrightarrow[]{p_3 \parallel \eta} 0\, .
\end{align}
The last term in eq.~\eqref{eq:Jperp_split_rho}, which is the origin of the inequality in eq.~\eqref{eq:Jperp_contraction_p3_nonvanishing}, is 
\begin{align}
\label{eq:eq:J_perp_split_second_term}
     &\frac{2p_3^\nu\rho^\mu}{\left(p_3+\rho\right)^2  - \rho^2}J_{\perp\nu}^{a}(k, p_1, p_3, \eta) \nn\\
     &= \frac{ (2-D) \, g_s^3 \left(C_A-C_F\right)  T^a\rho^\mu}{\left(p_3+\rho\right)^2  - \rho^2} \left( \frac{2{\slashed k}_\perp - \slashed{p}_{3\perp}}{(k-p_1)^2(k-p_3)^2} -\frac{ 2{\slashed k}_\perp}{k^2(k-p_1)^2}\right)\,.
\end{align}
Using tensor reduction we can show that this term is zero due to $\slashed{p}_{1\perp} = 0$. We can hence rewrite $J_\perp$ into the following expression, which yields the same integrated value 
\begin{align}
    \label{eq:Jcurrent_perp_tilde_alt}
    \widetilde J_\perp^{a \mu}(k, p_1, p_3, \eta) &= \left(g^{\mu\nu} - \frac{2p_3^\nu\rho^\mu}{\left(p_3+\rho\right)^2  - \rho^2}\right)J_{\perp\nu}^{a }(k, p_1, p_3, \eta)  \, .
\end{align}
With this construction, $p_{3\mu}\widetilde J_\perp^{a \mu}$ vanishes in the limit $p_3\parallel \eta$.

To conclude from the above, we have cast the integrand of eq.~\eqref{eq:P1jet} into the following equivalent form:
\begin{align}
\label{eq:P1jet_good}   
&\raisebox{-1.5mm}{\includegraphics[scale=0.75, page=7,valign=c]{figures_standalone/tikz_RV_LP.pdf}}
+\includegraphics[scale=0.75, page=8,valign=c]{figures_standalone/tikz_RV_LP.pdf}
+\raisebox{-6.5mm}{\includegraphics[scale=0.75, page=9,valign=c]{figures_standalone/tikz_RV_LP.pdf} } \nonumber \\ 
& \qquad\longrightarrow \frac{i}{\slashed{p}_1-\slashed{p}_3}\, \left[ \widetilde J_\perp^{a \mu}(k, p_1, p_3, \eta_1)+
V_{\rm qed}^{a \mu}(k+p_1, p_1, p_3)\right]\, 
u(p_1).  
\end{align}
For the expression of $\widetilde J_\perp^{a \mu}$, we can choose either eq.~\eqref{eq:Jcurrent_perp_tilde} or \eqref{eq:Jcurrent_perp_tilde_alt}.
Note that we have changed the variable $k \to k+p_1$ in $V_{\rm qed}^{a, \mu}$ compared to eq.~\eqref{eq:Vcurrent}. This is required to factorize the $k \parallel p_1$ singularity once we impose the loop momentum routing in section~\ref{sec:M1L_integrand}.

Analogously, for the triangle and self-energy corrections on the antiquark incoming leg, we modify the integrand into the following form: 
\begin{eqnarray}
\label{eq:P2jet_good}
 &&   
\raisebox{1.3mm}{\includegraphics[scale=0.75, page=10,valign=c]{figures_standalone/tikz_RV_LP.pdf}}
+\includegraphics[scale=0.75, page=11,valign=c]{figures_standalone/tikz_RV_LP.pdf}
+\raisebox{6.5mm}{\includegraphics[scale=0.75, page=12,valign=c]{figures_standalone/tikz_RV_LP.pdf} }  \nonumber \\ 
&& \qquad\longrightarrow
\bar v(p_2) 
 \left[
\widetilde J_\perp^{a \mu}(k, p_2, p_3, \eta_2)
+{V}_{\rm qed}^{\dagger a \mu}(p_2-k, p_2, p_3)
\right]   
\frac{i}{\slashed{p}_3-\slashed{p}_2}\, .
\end{eqnarray}
This can be derived by following the same procedure by substituting $p_1$ with $p_2$ everywhere. 

As noted above, $V_{\rm qed}^{a \mu}$ and ${V}_{\rm qed}^{\dagger a \mu}$ are not responsible for loop polarizations, and we can define ${\cal M}_{q\bar q \to {\rm ew}+g}^{(1), \, {\rm lp}}$ in eq.~\eqref{eq:RVpol+rest} as follows:
\begin{eqnarray}
\label{eq:RVpoldefinition}
  {\cal M}_{q\bar q \to {\rm ew}+g}^{(1), \, {\rm lp}} 
  \equiv\raisebox{1.5mm}{\includegraphics[scale=0.75, page=13,valign=c]{figures_standalone/tikz_RV_LP.pdf}}
  +\raisebox{-1.5mm}{\includegraphics[scale=0.75, page=14,valign=c]{figures_standalone/tikz_RV_LP.pdf}}\, ,
\end{eqnarray}
with the $\widetilde{J}_\perp$ vertices defined as the insertion of corresponding $\widetilde{J}_\perp^{a \mu}$ functions:
\begin{eqnarray}
    \raisebox{-3.5mm}{\includegraphics[scale=0.8, page=19,valign=c]{figures_standalone/tikz_RV_LP.pdf}} &\equiv \widetilde J_\perp^{a \mu}(k, p_1, p_3, \eta_1)\, , \quad
    \raisebox{-3.5mm}{\includegraphics[scale=0.8, page=20,valign=c]{figures_standalone/tikz_RV_LP.pdf}}
    \equiv  \widetilde J_\perp^{a \mu}(k, p_2, p_3, \eta_2) \, .
\end{eqnarray}
Using the preferred method, $\widetilde{J}_\perp$ is defined either in  eq.~\eqref{eq:Jcurrent_perp_tilde} or in eq.~\eqref{eq:Jcurrent_perp_tilde_alt}. 
Similarly, we will represent the insertion of $V^{a\mu}_{\rm qed}$ or $V^{\dagger a\mu}_{\rm qed}$ by
\begin{align}
    \raisebox{-3.5mm}{\includegraphics[scale=0.8, page=21,valign=c]{figures_standalone/tikz_RV_LP.pdf}} &\equiv 
    V_{\rm qed}^{a \mu}(k+p_1
    , p_1, p_3)
    \, ,
    \quad
    \raisebox{-3.5mm}{\includegraphics[scale=0.8, page=22,valign=c]{figures_standalone/tikz_RV_LP.pdf}} \equiv
    {V}_{\rm qed}^{\dagger a \mu}(p_2-k, p_2, p_3) \, ,
    \label{eq:Vcurrent_legs}
\end{align}
where the current is defined in eq.~\eqref{eq:Vcurrent}.

\subsection{\texorpdfstring{Integrand for ${\cal M}_{q\bar q \to {\rm ew}+g}^{(1),\, {\rm nolp}}$}{Integrand for M(rest)}}
\label{sec:M1L_integrand}

With these constructions in place, we next assign loop momentum routing for the remaining diagrams and construct the integrand for ${\cal M}_{q\bar q \to {\rm ew}+g}^{(1),\, {\rm nolp}}$ in eq.~\eqref{eq:RVpol+rest} that guarantees local factorization. We further decompose ${\cal M}_{q\bar q \to {\rm ew}+g}^{(1),\, {\rm nolp}}$ into three gauge-invariant constituents, the leading-color, subleading-color and fermion-loop contributions,
\begin{eqnarray}
{\cal M}_{q\bar q \to {\rm ew}+g}^{(1),\, {\rm nolp}}  = 
{\cal M}_{q\bar q \to {\rm ew}+g}^{(1), \rm LC} 
+{\cal M}_{q\bar q \to {\rm ew}+g}^{(1),\rm SLC} \, 
+{\cal M}_{q\bar q \to {\rm ew}+g}^{(1),\rm floop} \, . 
\end{eqnarray}
The leading-color contribution is given by
\begin{align}
&\mathcal{M}_{q\bar q \to {\rm ew}+g}^{(1), \rm LC}(p_1,p_2;p_3;k)  \nn\\
&=
\includegraphics[scale=0.75, page=10,valign=c]{figures_standalone/tikz_RV_ampl_label.pdf}
-\includegraphics[scale=0.75, page=11,valign=c]{figures_standalone/tikz_RV_ampl_label.pdf}
-\includegraphics[scale=0.75, page=12,valign=c]{figures_standalone/tikz_RV_ampl_label.pdf}
\nonumber \\ 
&
+ \frac{C_A}{2 C_F}
\left\{
\includegraphics[scale=0.75, page=9,valign=c]{figures_standalone/tikz_RV_ampl_label.pdf}
-
\raisebox{-2mm}{\includegraphics[scale=0.75, page=8,valign=c]{figures_standalone/tikz_RV_ampl_label.pdf}}
\right.
\nonumber \\ 
& \hspace{2cm} +
\left.
\includegraphics[scale=0.75, page=4,valign=c]{figures_standalone/tikz_RV_ampl_label.pdf}
-
\raisebox{2mm}{\includegraphics[scale=0.75, page=5,valign=c]{figures_standalone/tikz_RV_ampl_label.pdf}}
 \right\}\, .
 \label{eq:M1LC}
\end{align}
On the first line of the right-hand side of eq.~\eqref{eq:M1LC}, all diagrams with triple-gluon vertices (class (A) in eq.~\eqref{eq:M_1loop}) are included, except for those leading to loop polarizations, which have already been taken into account by $\mathcal{M}_{q\bar q \to {\rm ew}+g}^{(1), \rm lp}$ in eq.~\eqref{eq:RVpoldefinition}.
The second and third lines include only the leading-color contribution from diagrams in which the external gluon with momentum $p_3$ is emitted outside the loop (classes (C) and (D)). To isolate their leading-color contribution, we have multiplied these diagrams with a factor $\frac{C_A}{2C_F}$. Note that we have excluded the self-energy correction next to the incoming quark and antiquark, because they are already included in $\mathcal{M}_{q\bar q \to {\rm ew}+g}^{(1), \rm lp}$. We have explicitly assigned a loop momentum routing for each diagram on the right-hand side of eq.~(\ref{eq:M1LC}). The gluon momentum that flows from $p_1$ towards $p_3$ is labeled $p_3-k$, whereas the gluon flowing from $p_2$ towards $p_3$ is labeled $k$. This momentum routing guarantees local factorization, as we will demonstrate in section~\ref{sec:IRdemo_RVampl}.

The subleading-color contribution to the amplitude is
\begin{align}
&\mathcal{M}_{q\bar q \to {\rm ew}+g}^{(1), \rm SLC}(p_1,p_2;p_3;k)  \nn\\
&=\includegraphics[scale=0.75, page=1,valign=c]{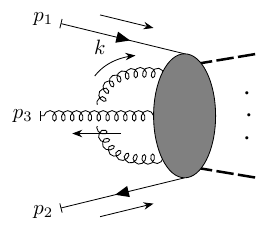}
-\raisebox{2.5mm}{\includegraphics[scale=0.75, page=2,valign=c]{figures_standalone/tikz_RV_ampl_label.pdf}}
-\raisebox{-2.5mm}{\includegraphics[scale=0.75, page=3,valign=c]{figures_standalone/tikz_RV_ampl_label.pdf}}
\nonumber \\ 
&
+ \left( 1-\frac{C_A}{2C_F} \right) 
\left\{
\includegraphics[scale=0.75, page=6,valign=c]{figures_standalone/tikz_RV_ampl_label.pdf}
-\raisebox{-2mm}{\includegraphics[scale=0.75, page=7,valign=c]{figures_standalone/tikz_RV_ampl_label.pdf}}
\right.
\nonumber \\ 
& \hspace{3cm}
+\left.
\includegraphics[scale=0.75, page=4,valign=c]{figures_standalone/tikz_RV_ampl_label.pdf}
-\raisebox{2mm}{\includegraphics[scale=0.75, page=5,valign=c]{figures_standalone/tikz_RV_ampl_label.pdf}}
\right\}
\nonumber \\ 
&
+\raisebox{0.8mm}{\includegraphics[scale=0.75, page=14,valign=c]{figures_standalone/tikz_RV_ampl_label.pdf}}
+\raisebox{-1mm}{\includegraphics[scale=0.75, page=15,valign=c]{figures_standalone/tikz_RV_ampl_label.pdf}}
\, . 
\label{eq:M1SLC}
 \end{align}
The first term on the right-hand side of eq.~\eqref{eq:M1SLC} represents most diagrams in which the external gluon $p_3$ is emitted from a quark within the loop (class (B) in eq.~\eqref{eq:M_1loop}), excluding those with triangle corrections on the incoming quark or antiquark lines. The latter are represented by the subtraction of the second and third terms, which give rise to loop polarizations, as discussed in section~\ref{sec:LP}. Terms involving ${V}^{a\mu}_{\rm qed}$, eq.\ (\ref{eq:Vcurrent}), and ${V}_{\rm qed}^\dagger$ are included on the last line to compensate for the subtractions.
Similarly, in the second and third lines we only include the subleading-color contributions from diagrams in which the external gluon with momentum $p_3$ is emitted outside the loop (classes (C) and (D)), where the self-energy corrections included in $\mathcal{M}_{q\bar q \to {\rm ew}+g}^{(1), \rm lp}$ are subtracted. Note that these diagrams are now assigned a different loop-momentum flow compared to their leading-color part.
For improved legibility in the upcoming section~\ref{sec:IRanalyze}, we summarize the subleading- and leading-color amplitudes diagrammatically using the loop momentum routing introduced above, as follows
\begin{align}
    \mathcal{M}_{q\bar q \to {\rm ew}+g}^{(1), \rm SLC}(p_1,p_2;p_3;k) &= \includegraphics[scale=0.75, page=16,valign=c]{figures_standalone/tikz_RV_ampl_label.pdf}\, ,\label{eq:M_1loop_SLC_summarized}\\
    \mathcal{M}_{q\bar q \to {\rm ew}+g}^{(1), \rm LC}(p_1,p_2;p_3;k) &= \includegraphics[scale=0.75, page=17,valign=c]{figures_standalone/tikz_RV_ampl_label.pdf} \, .\label{eq:M_1loop_LC_summarized}
\end{align}

The final class of contributions to the amplitude consists of diagrams with a fermion loop,
\begin{equation}
{\cal M}_{q\bar q \to {\rm ew}+g}^{(1), {\rm floop}}(p_1,p_2;p_3;k) = \includegraphics[scale=0.75, page=13,valign=c]{figures_standalone/tikz_RV_ampl_label.pdf}
\, .
\label{eq:M1L_floops}
\end{equation}
These diagrams are infrared finite, so explicit momentum routing is not required. For instance, in the collinear limit where two massless quark propagators adjacent to the gluon become aligned with $p_3$, the diagram acquires a suppression factor $\epsilon_3^* \cdot p_3$, where $\epsilon_3$ denotes the physical polarization vector of the gluon.\footnote{In this paper, gluon polarizations in amplitudes are taken to be transverse to their momenta. In the later discussion of cross sections, unphysical cuts are introduced and thus additional polarizations appear.} 
Consequently, no additional subtraction terms are required for these diagrams.

\subsection{Infrared approximation of the one-loop amplitude for a resolved gluon}
\label{sec:IRapprox_RVamplitude}

In the previous subsections, we have constructed a one-loop amplitude integrand that is locally factorizable in all its infrared limits,
\begin{align}
\label{eq:MR1L1_ours}
{\cal M}_{q\bar q \to {\rm ew}+g}^{(1)} &\equiv {\cal M}_{q\bar q \to {\rm ew}+g}^{(1), {\rm lp}} + {\cal M}_{q\bar q \to {\rm ew}+g}^{(1), {\rm SLC}}+{\cal M}_{q\bar q \to {\rm ew}+g}^{(1), {\rm LC}} + {\cal M}_{q\bar q \to {\rm ew}+g}^{(1), {\rm floop}}\, ,
\end{align}
where the first term on the right, the modified loop polarization term, is given by eq.~\eqref{eq:RVpoldefinition}, and the final three terms on the right-hand side together make up 
${\cal M}_{q\bar q \to {\rm ew}+g}^{(1),\, {\rm nolp}}$.
In section~\ref{sec:IRanalyze} we will verify with an analysis along the lines of refs.~\cite{Anastasiou:2022eym,Anastasiou:2024xvk}, that the integrand of eq.~\eqref{eq:MR1L1_ours} factorizes \emph{locally} in all singular infrared limits, when the loop momentum is soft or collinear to the momentum of an external parton. We can therefore introduce an overall factor depending on the loop momentum $k$, which absorbs all the infrared behavior of ${\cal M}_{q\bar q \to {\rm ew}+g}^{(1)}$. Namely,
\begin{align}
\label{eq:M1IRapprox}
    {\cal M}_{q\bar q \to {\rm ew}+g}^{(1)}(p_1,p_2;p_3;k) \overset{\textup{IR}}{\approx} \left( \mathcal{F}^{(1)}_{\rm SLC} (p_1,p_2;k) +\mathcal{F}^{(1)}_{\rm LC} (p_1,p_2;p_3;k)\right) {\cal M}_{q\bar q \to {\rm ew}+g}^{(0)}(p_1,p_2;p_3)\,,
\end{align}
where $\mathcal{F}^{(1)}_{\rm LC}$ and $\mathcal{F}^{(1)}_{\rm SLC}$ are factors for the leading- and subleading-color contributions, respectively. Later in section~\ref{sec:IRdemo_RVampl}, we will justify eq.~\eqref{eq:M1IRapprox} and derive explicit expressions for these factors:
\begin{align}
    \mathcal{F}^{(1)}_{\rm SLC} (p_1,p_2;k) &\equiv i g_s^2\, \left(C_F-\frac{C_A}{2} \right) \frac{4 p_1\cdot p_2 -4 k\cdot p_1 + 4 k\cdot p_2 }{k^2(k+p_1)^2(k-p_2)^2}\, ,
    \label{eq:FV_SLC}\\
    \mathcal{F}^{(1)}_{\rm LC} (p_1,p_2;p_3;k) &\equiv -ig_s^2 \frac{C_A}{2}\, \left( \frac{4 p_3\cdot p_2-2k\cdot p_2 -4k\cdot p_3}{k^2 (k-p_3)^2(k-p_2)^2} \right. \nn\\
    &\hspace{4cm}\left.+\frac{2p_3\cdot p_1+2k \cdot p_1 +4k\cdot p_3}{k^2(k-p_3)^2(p_1-p_3+k)^2}\right)\, .
    \label{eq:FV_LC}
\end{align}
We emphasize again that the factorization of eq.~\eqref{eq:M1IRapprox} occurs at the integrand, and it is a direct consequence of the momentum flow assigned to the diagrams in the previous section. After integrating over $k$, we obtain
\begin{align}
    &\int \frac{d^Dk}{(2\pi)^D} \, \left( \mathcal{F}^{(1)}_{\rm LC} (p_1,p_2;p_3;k) +\mathcal{F}^{(1)}_{\rm SLC} (p_1,p_2;k) \right)
    \nn\\
    &=   \frac{- g_s^2 \, \Gamma(1+\epsilon)}{(4\pi)^{D/2}}\left[ \frac{C_A}{2}\left((2p_1\cdot p_3)^{-\epsilon} + (2p_2\cdot p_3)^{-\epsilon} \right) \left(\frac{2}{\epsilon^2} + \frac{3}{\epsilon} \right) \right. \nn \\
    &\left. \hspace{3cm} + 2  \left(C_F - \frac{C_A}{2}\right)\left(-2p_1\cdot p_2\right)^{-\epsilon}\left(\frac{1}{\epsilon^2} + \frac{2}{\epsilon} \right) +\mathcal{O}(1)   \right]\, , 
\end{align}
which agrees with the result of ref.~\cite{Catani:1998bh}. This agreement becomes explicit once the poles from the self-energy corrections to the external quark, antiquark, and gluon are included. These additional corrections factorize trivially.

\section{Partonic cross sections}
\label{sec:section-finite_cross_section}

In this work, we consider the cross section for the partonic process in eq.~\eqref{eq:qqbar_process_definition}:
\begin{eqnarray}
    q(p_1) + \bar q(p_2) \to g(p_3) + \textup{ew}(q_1,\dots,q_n)\, ,
\end{eqnarray}
and another two processes related by crossing symmetry:
\begin{align}
\begin{split}
    q(p_1) + g(p_2) \to  q(p_3) +\textup{ew}(q_1,\dots,q_n) \, ,\\
    {\bar q}(p_1) + g(p_2) \to  {\bar q}(p_3) + \textup{ew}(q_1,\dots,q_n)\, .
\end{split}
\end{align}
In principle, the cross section also includes the gluon-fusion channel, but note that for the scope of this paper, the real–virtual correction of the partonic cross sections receives no contribution from this channel. Consequently, we omit it from the present analysis. In the rest of the paper, we will focus exclusively on the quark-antiquark and quark-gluon channels.

The ``bare'' partonic cross sections, before initial-state mass-factorization is carried out, are given by 
\begin{eqnarray}
\label{eq:genXsection}
    \Sigma_{ij \to {\rm ew} +l} \left[ \cal O\right] 
    && = \frac{\pi}{s} \int 
\dmom{p_3} 
 \, \left( 
 \prod_{f = 1}^{n-1} \dmomM{q}{M_f} \, 
  \right)   
  \nonumber  \\
 && \hspace{-1.5cm}
  \delta(q_n^2 - M_n^2)
\left\langle \left\vert M_{ij\to {\rm ew}+l}\left(p_1,p_2 ; p_3, q_1,\ldots , q_n\right)\right\vert^2 \right\rangle \,
{\cal O}_{{\rm ew}+1}\left( p_3, q_1,\ldots , q_n\right) \, ,
\end{eqnarray}
where $i$ and $j$ denote the initial-state partons, $l$ denotes the final-state parton, and $s\equiv (p_1+p_2)^2$ is one of the Mandelstam variables. ${\cal O}_{{\rm ew}+1}$ stands for a general infrared-safe observable~\cite{Sterman:1979uw} which is a function of the independent final-state momenta, $p_3^\mu=(|\boldsymbol{p}_3|, \boldsymbol{p}_3)$ and $q_f^\mu = \left(\sqrt{\boldsymbol{q}_f^2+M_f^2}, \boldsymbol{q}_f\right)$ for $f \neq n$. Meanwhile, the momentum $q_n^\mu$ is determined by momentum conservation,
\begin{eqnarray}
    q_n^\mu= \left( p_1^0 + p_2^0 -p_3^0 - \sum_{f =1}^{n-1} \sqrt{\boldsymbol{q}_f^2+M_f^2} \, , \,\boldsymbol{p}_1 + \boldsymbol{p}_2-\boldsymbol{p}_3 - \sum_{f =1}^{n-1} \boldsymbol{q}_f \, \right)\,.
\end{eqnarray}
The delta function puts the $n$-th colorless particle in the final state on its mass shell.

In this article, we will focus on unpolarized cross sections. For the squared amplitude, we will sum over the final-state colors and polarizations and average over the initial-state spins and colors, with the result denoted by $\left< |M|^2 \right>$ in eq.~\eqref{eq:genXsection}. For the sum over gluon polarizations, it is important to perform the sum over physical polarizations,
\begin{equation}
\label{eq:myphyscut}
\sum_\textup{pols.} \epsilon^{\mu}(p_3) \epsilon^{\nu*}(p_3) 
=-g^{\mu \nu} + \frac{p_3^\mu n^\nu+p_3^\nu n^\mu }{p_3 \cdot n }\, , \qquad p_3 \cdot n \neq 0.
\end{equation}
For this particular class of processes with only one external gluon, it is also permissible to consider an external gluon with four polarizations,
\begin{equation}
\label{eq:mynonphyscut}
\sum_\textup{pols.} \epsilon^{\mu}(p_3) \epsilon^{\nu*}(p_3) \rightarrow -g^{\mu \nu} \, . 
\end{equation}
After performing the loop integrations, the same result for the cross section is obtained using either choice in eq.~\eqref{eq:myphyscut} or~\eqref{eq:mynonphyscut}. However, for certain contributions, the integrands from these choices feature different singularity structures. In our upcoming analysis, we will mostly use eq.~\eqref{eq:mynonphyscut}, while for some specific cases (e.g., the fermion-loop contribution, as detailed in section~\ref{sec:fermloop_xsec_IRfinite}), we employ eq.~\eqref{eq:myphyscut} to simplify the local infrared structure.

The amplitudes receive loop corrections and require integration over loop momenta, as indicated in eq.~\eqref{eq:Mnplus1}. In this article, we combine integrations over loop momenta and phase-space integrations over the final-state particle momenta, and analyze the infrared singularities of the resulting integrand with respect to all these integration variables. By expanding the squared amplitude in eq.~\eqref{eq:genXsection} perturbatively,
\begin{align}
    \left\langle \left\vert M_{i j \to {\rm ew}+l}\right\vert^2 \right\rangle &= \left\langle \left\vert M^{(0)}_{ij \to {\rm ew}+l}\right\vert^2 \right\rangle + \left\langle 2 \, \mathrm{Re}\left[ M^{(0)}_{ij \to {\rm ew}+l}\left( M^{(1) }_{ij \to {\rm ew}+l}\right)^*\right]\right\rangle\,+ \mathcal{O}(\alpha_s^3)\, ,
\end{align}
the partonic cross section $\Sigma_{ij \to {\rm ew}+l} \left[ \cal O\right]$ takes the form
\begin{align}
\label{eq:genXsection_expand}
    \Sigma_{ij \to {\rm ew}+l} \left[ \cal O\right] & = \frac{\pi}{s} \int \dmom{p_3} \, \left( \prod_{f = 1}^{n-1} \dmomM{q}{M_f} \, \right) \, \frac{d^Dk}{(2\pi)^D}\nonumber \\ 
    &\hspace{1cm}
    \left[ (2\pi)^D \delta^D(k) \, \sigma_{ij\to {\rm ew}+l}^{(0)}(p_1,p_2;p_3)+ \sigma_{ij\to {\rm ew}+l}^{(1)}(p_1,p_2;p_3;k) \right]\, .
\end{align}
The integrands at orders $\alpha_s$ and $\alpha_s^2$ are given by
\begin{align}
 \label{eq:local_xsec_R}
    \sigma_{ij\to {\rm ew}+l}^{(0)}(p_1,p_2;p_3) &\equiv \left\langle \abs{\mathcal{M}^{(0)}_{ij\to {\rm ew}+l}(p_1,p_2;p_3)}^2\right\rangle \delta(q_n^2- M_n^2) \, {\cal O}_{{\rm ew}+1}\left( p_3, q_1,\ldots , q_n\right)\,,
\end{align}
and
\begin{align}
    \sigma_{ij\to {\rm ew}+l}^{(1)}(p_1,p_2;p_3;k) &
    \equiv  \left\langle 2 \, \mathrm{Re}\left[ \mathcal{M}^{(1) }_{ij\to {\rm ew}+l}(p_1,p_2;p_3;k)\left( \mathcal{M}^{(0)}_{ij\to {\rm ew}+l}(p_1,p_2;p_3)\right)^* \right]\right\rangle\, \nonumber \\
    & \hspace{1cm}\cdot \delta(q_n^2- M_n^2) \, {\cal O}_{{\rm ew}+1}\left(  {p}_3, q_1,\ldots , q_n\right) \,,
    \label{eq:local_xsec_RV}
\end{align}
respectively. Note that we use the script letter $\mathcal{M}$ to indicate an amplitude integrand. In the notation for the partonic cross-section integrands $\sigma_{ij\to {\rm ew}+l}^{(m)}$, the superscript indicates the loop order $m$ of the corresponding matrix elements. As the subscript also indicates, the process includes the radiation of a parton in the final state, and the perturbative order of these cross sections in the strong coupling is $\alpha_s^{m+1}$. The arguments of the partonic cross-section integrand functions include the momenta of the incoming and outgoing partons and the loop momentum for $m=1$. We leave their dependence on the momenta of the electroweak final state and the observable ${\cal O}_{{\rm ew}+1}$ implicit.

The cross-section integrands in eqs.~\eqref{eq:local_xsec_R} and \eqref{eq:local_xsec_RV} develop singularities as the final-state gluon momentum $p_3$ becomes soft or collinear to either $p_1$ or $p_2$, unless this is prevented by the observable ${\cal O}_{{\rm ew}+1}$ requiring a jet in the final state. At order $\alpha_s^2$, additional singularities emerge when the loop momentum $k$ becomes soft or collinear to the external partons, and when both $k$ and $p_3$ become soft or collinear to an initial-state parton.

In constructing the amplitude integrand ${\cal M}_{q \bar q \to {\rm ew} +g}^{(1)}$ in section~\ref{sec:RVamplitude} (and similarly for the other channels via parton crossing), we anticipated the presence of phase-space singularities in conjunction with singularities in the loop momentum integral. This integrand ensures that, in all infrared singular limits, the integrand of the partonic cross section factorizes into universal infrared factors and a non-singular hard function. In what follows, we will exploit this property of infrared factorization, which we have made manifest at the integrand level, to construct subtraction counterterms that isolate the finite part.

The factorized structure of infrared singularities enables the development of subtraction methods that can be applied universally to all processes with the same external partons. Here, we will provide an example of a subtraction method for the cross sections in eqs.~\eqref{eq:local_xsec_R} and \eqref{eq:local_xsec_RV} that treats the colorless electroweak final states as arbitrary. Our subtraction method will exploit the fact that the simplest process in this class, the single-Higgs production, contains \emph{all the same} infrared singularities as more general electroweak production processes. We will then use the integrands of the hadronic cross sections for single-Higgs production as universal infrared counterterms.

Concretely, it will be useful to consider the $2 \to n$ process of producing the colorless electroweak final state, without outgoing partons. Up to order $\alpha_s$ only the quark-antiquark channel contributes
\begin{eqnarray}
    q(p_1) + \bar q(p_2) \to \textup{ew}(q_1,\dots,q_n)\, . 
\end{eqnarray}
The corresponding partonic cross section is given by, 
\begin{align}
    \Sigma_{q \bar q \to {\rm ew}}(p_1, p_2) &= \int \left( \prod_{f=1}^{n-1}\, \dmomM{{q}}{M_f} \right) \, \frac{d^Dk}{(2\pi)^D} \nonumber \\
    & \hspace{1cm} \left[ (2\pi)^D \delta^D(k) \, \sigma_{q\bar q \to {\rm ew}}^{(0)}(p_1,p_2) + \, \sigma_{q\bar q \to {\rm ew}}^{(1)}(p_1,p_2 ; k) + {\cal O}(\alpha_s^2) \right], 
\end{align}
where the integrand at order $\alpha_s^0$ is given by 
\begin{align}
\label{eq:pa2tonpartonic0}
    \sigma_{q\bar q \to {\rm ew}}^{(0)}(p_1,p_2) = \left\langle \abs{\mathcal{M}^{(0)}_{q\overline{q}\to {\rm ew}}(p_1,p_2)}^2\right\rangle \delta(q_n^2- M_n^2) \, {\cal O}_{{\rm ew}} \left(q_1,\ldots, q_n\right)\,,
\end{align}
and at order $\alpha_s$ by
\begin{align}
\label{eq:pa2tonpartonic}
    \sigma_{q\bar q \to {\rm ew}}^{(1)}(p_1,p_2; k) &= \left\langle 2 \, \mathrm{Re}\left[ \mathcal{M}^{(1) }_{q\overline{q}\to {\rm ew}}(p_1,p_2;k)\left( \mathcal{M}^{(0)}_{q\overline{q}\to {\rm ew}}(p_1,p_2)\right)^* \right]
    \right\rangle \nn\\
    & \hspace{1cm}  \cdot 
    \delta(q_n^2- M_n^2) \, {\cal O}_{{\rm ew}} \left(q_1,\ldots , q_n\right)\,.
\end{align}
We define the one-loop amplitude as
 \begin{align}
     \mathcal{M}_{q\bar q \to {\rm ew}}^{(1)}(p_1,p_2;k) &= \includegraphics[scale=0.75,valign=c,page=1]{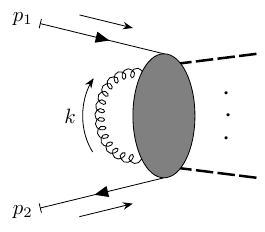}\,,
     \label{eq:amp_ew_1loop}
\end{align}
where the momentum $k$ is chosen to always flow with the charge of the quark line. 

\section{\texorpdfstring{Finite cross-section integrand at order $\alpha_s$}{Finite cross-section integrand at order alphas}}
\label{sec:NLOxsec_localfact}
In this section, we shall construct integrands for locally infrared finite partonic cross sections at order $\alpha_s$. At this perturbative order, we encounter cross sections with one parton in the final state given in eq.~\eqref{eq:local_xsec_R} for the quark-antiquark and quark-gluon channel. For these, the scattering amplitudes do not yet receive loop corrections, and the singularities arise from regions of phase-space integration. Specifically, the integrand is singular in the soft limit $p_3 \to 0$ and collinear limits $p_3 \parallel p_1$ and $p_3 \parallel p_2$. Additionally, the one-loop correction to electroweak production through quark-antiquark annihilation defined in eq.~\eqref{eq:pa2tonpartonic} contributes at order $\alpha_s$. Its integrand has singularities when the loop momentum $k$ becomes soft ($k \to 0$) or collinear to the incoming particles ($k \parallel p_1$ and $k \parallel p_2$). 

\subsection{Phase-space singularities}
It is well understood in the literature how to render the cross section for an arbitrary electroweak and a single-parton final state finite at this perturbative order, by exploiting the universality of its infrared singular limits (e.g., ref.~\cite{Catani:1996vz}). Nevertheless, it is useful to present a compact form for the infrared subtractions required for this purpose. 
First, we decompose the momentum $p_3$ of the final-state parton into components that are parallel and perpendicular to the initial-state momenta $p_1$ and $p_2$,
\begin{eqnarray}
    p_3^\mu =  (1- \xi_1) p_1^\mu + (1- \xi_2) p_2^\mu + p_{3\perp}^\mu, 
\end{eqnarray}
where
\begin{eqnarray}
\xi_1 = 1-\frac{p_3 \cdot p_2}{ p_1 \cdot p_2}\,, \quad \xi_2 = 1-\frac{p_3 \cdot p_1}{ p_1 \cdot p_2}\,,\quad p_{3\perp}^2 = -s (1-\xi_1)(1-\xi_2) \, . 
\end{eqnarray}
We further construct two lightlike momenta, $k_1$ and $k_2$, from $p_1$, $p_2$, and $p_3$, with the properties
\begin{align}
\label{eq:k1_k2}
    k_1+k_2 = p_1+p_2-p_3 =\sum_n q_n \equiv Q\,,\qquad  k_1^2=k_2^2 =0\,. 
\end{align}
Specifically, we define
\begin{align}
\begin{split}
    k_1 &= \frac{\left(1+\eta \right)}{2} \xi_ 1 p_1 + \frac{\left(1-\eta \right)}{2} \xi_ 2 p_2 - \frac{p_{3\perp}}{2}, \\
    k_2 &= \frac{\left(1-\eta \right)}{2} \xi_ 1 p_1 + \frac{\left(1+\eta \right)}{2} \xi_ 2 p_2 - \frac{p_{3\perp}}{2}, 
\end{split}
\label{eq:k1k2_param}
\end{align}
where 
\begin{align}
\label{eq:xi_eta_param}
        &\eta \equiv \sqrt{ 1 + \frac{p_{3\perp}^2}{ 2 \xi_1 \xi_2 p_1 \cdot p_2 } } =\sqrt{\frac{Q^2}{Q^2 - p_{3\perp}^2}}\,.
\end{align}
The vectors $k_1$ and $k_2$ are equivalent to the momenta of the initial-state partons after a collinear final-state parton is emitted from either of the two. 
Indeed, in the collinear limit $p_3 =  x p_1$, we have that 
\begin{eqnarray}
    k_1 
    \to (1-x) p_1 \, ,  \quad k_2 \to p_2 \, . 
\end{eqnarray}
Analogously, in the collinear limit $p_3 = x p_2$  we obtain the symmetric result, 
\begin{eqnarray}
k_1 \to p_1\, , \quad  k_2 
\to (1-x) p_2 \, .  
\end{eqnarray}
In the soft limit $p_3\to 0$ we have
\begin{align}
    k_1 \to p_1  \, , \quad k_2 \to p_2 \, .
\end{align}
Due to the property of eq.~\eqref{eq:k1_k2}, the on-shell condition for the $n$-th electroweak particle can be expressed in two forms, 
\begin{eqnarray}
    \delta \left(q_n^2 -M_n^2 \right) &=& 
    \delta\left(\left(p_1+p_2 - p_3 -\sum_{f \neq n} q_f \right)^2 - M_n^2\right) 
    \nonumber \\ 
    &=&
    \delta\left(\left(k_1+k_2 -\sum_{f \neq n} q_f \right)^2 - M_n^2\right) \, . 
\end{eqnarray}
Thus, we can interpret the phase-space measure in two equivalent ways. In the first way, we have the phase-space measure for a $2 \to n+1$ process with $p_1,p_2$ incoming momenta and a parton of momentum $p_3$ in the final state. In the second way, we have the phase-space measure for a $2 \to n$ process with $k_1,k_2$ incoming momenta and no parton in the final state.

At order $\alpha_s$, the $2 \to n$ partonic cross-section integrand in eq.~\eqref{eq:pa2tonpartonic0}, with the incoming momenta $(p_1, p_2)$ replaced by $(k_1, k_2)$, serves as the hard-function coefficient in all infrared limits. Specifically, we can show that in all these limits, the ratio of the partonic cross section for the $2 \to n+1$ process in eq.~\eqref{eq:local_xsec_R} to that of the $2 \to n$ process is approximated by the same universal infrared factor for any electroweak final state (i.e., for any value of $n$). In the quark-antiquark channel, we have
\begin{equation}
    \frac{\sigma_{q\bar q \to {\rm ew}+g}^{(0)}(p_1,p_2;p_3)}{\sigma_{q\bar q \to {\rm ew}}^{(0)}(k_1,k_2)\,}\ \overset{\textup{IR}}{\approx} \ \frac{\sigma_{q\bar q \to {\rm H}+g}^{(0)}(p_1,p_2;p_3)}{\sigma_{q\bar q \to {\rm H}}^{(0)}(k_1,k_2)}. 
\end{equation}
On the left-hand side is the ratio of partonic cross-section integrands for producing a general electroweak final state. On the right-hand side is the corresponding ratio for the simplest process in this class: the production of a single scalar color-singlet Higgs boson (H) via quark-antiquark annihilation.
Explicitly, this universal infrared factor is equal to (see eq.~(3.9) of ref.~\cite{Anastasiou:2011qx}) 
\begin{eqnarray}
\label{eq:universalityR}
    \frac{\sigma_{q\bar q \to {\rm H}+g}^{(0)}(p_1,p_2;p_3)}{\sigma_{q\bar q \to {\rm H}}^{(0)}(k_1,k_2)}  = \Fcal_{q\bar q \to {\rm H}+g}(p_1, p_2; p_3) \frac{{\cal O}_{{\rm H}+1}(p_3, q_1+ \ldots +q_n)}{{\cal O}_{\rm H}(q_1+\ldots +q_n)}\,,
\end{eqnarray}
where the form-factor $\cal F$ is given by
\begin{align}
\label{eq:Fqq2Hg}
&\Fcal_{q\bar q \to {\rm H}+g}(p_1, p_2; p_3)  =- 2 g_s^2\xi \hat{P}_{qq}\left(\xi\right) 
\left( 
\frac{1}{ 2\, p_1 \cdot p_3} 
+\frac{1}{2 \, p_2 \cdot p_3} 
\right)\nn\\
&= \frac{\includegraphics[scale=0.85,valign=c,page=1]{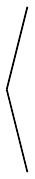}  \left\vert \includegraphics[scale=0.8,valign=c,page=1]{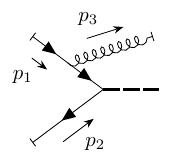}+\includegraphics[scale=0.8,valign=c,page=2]{figures_standalone/tikz_RV_ampl_LO.pdf}\right\vert^2 \includegraphics[scale=0.85,valign=c,page=2]{figures_standalone/tikz_bigangles.pdf} }{\includegraphics[scale=0.8,valign=c,page=1]{figures_standalone/tikz_bigangles.pdf} \left\vert\includegraphics[scale=0.8,valign=c,page=3]{figures_standalone/tikz_RV_ampl_LO.pdf}\right\vert^2\includegraphics[scale=0.8,valign=c,page=2]{figures_standalone/tikz_bigangles.pdf} } 
\end{align}
with 
\begin{eqnarray}
    \xi = \frac{p_1 \cdot p_2}{p_1 \cdot p_2-p_1 \cdot p_3 -p_2 \cdot p_3}\,,\qquad {\hat P}_{qq}(\xi) = C_F \left( \frac{1+\xi^2}{\xi (1-\xi)} - \epsilon \frac{1-\xi}{\xi} \right)\, . 
\end{eqnarray}
For the observable $\mathcal{O}_{{\rm H}}$ in the production process of a single Higgs, we take 
\begin{equation}
{\cal O}_{\rm H}(q_1+ \ldots +q_n)  = 1.    
\end{equation}
The observable in the Higgs and gluon production process ${\cal O}_{{\rm H}+1}$ can be taken simply to be the inclusive cross section. However, as in the soft and collinear limits the transverse momentum of the gluon vanishes, one could also impose a veto on large values,~\footnote{We thank Andrea Pelloni for a communication of a related numerical study on this choice.}
\begin{equation}
{\cal O}_{{\rm H}+1}(p_3, q_1+ \ldots +q_n)  = \Theta(p_{3\perp}^{\rm veto}-p_{3\perp}).    
\end{equation}
In the soft limit, $p_3 \to 0$, the form factor reduces to the anticipated eikonal factor,  
\begin{align}
\Fcal_{q\bar q \to {\rm H}+g}(p_1, p_2; p_3) \xrightarrow[]{p_3\to 0}  2 g_s^2  C_F \frac{p_1 \cdot p_2}{(p_1 \cdot p_3) (p_2 \cdot p_3)}\,.
\end{align}
In the collinear limit $p_3=x p_1$, it reduces to the expected splitting function,  
\begin{align}
    \Fcal_{q\bar q \to {\rm H}+g}(p_1, p_2; p_3) \xrightarrow[]{p_3=xp_1}& -\frac{2 g_s^2}{2 \, p_1 \cdot p_3} \, {\hat P}_{qq}\left(\frac{1}{1-x}\right)\cdot \frac{1}{1-x} \nonumber \\
    &= 2 g_s^2\frac{{\hat P}_{qq}(1-x)}{2 \, p_1 \cdot p_3} \, .
\end{align}
Analogously, in the other collinear limit $p_3 = x p_2$, we have
\begin{align}
\Fcal_{q\bar q \to {\rm H}+g}(p_1, p_2; p_3) &\xrightarrow[]{p_3=xp_2} 2 g_s^2
\frac{{\hat P}_{qq}(1-x)}{2 \, p_2 \cdot p_3} \, .   
\end{align}
From eq.~\eqref{eq:universalityR}, it is then natural to remove all infrared singularities via the following subtraction:
\begin{align}
\label{eq:xsec_0L_IRfinite_NLO}
{ \sigma_{q\bar q \to {\rm ew}+g}^{{(0)} \rm ,IR-finite}(p_1,p_2;p_3) = \sigma_{q\bar q \to {\rm ew}+g}^{(0)}(p_1,p_2;p_3) - \frac{\sigma_{q\bar q \to {\rm H}+g}^{(0)}(p_1,p_2;p_3)}{\sigma_{q\bar q \to {\rm H}}^{(0)}(k_1,k_2)} \sigma_{q\bar q \to {\rm ew}}^{(0)}(k_1,k_2)\,. }
\end{align}
We will demonstrate that the right-hand side of eq.~\eqref{eq:xsec_0L_IRfinite_NLO} is 
finite in section~\ref{sec:IRdemo_NLO} in more detail. 

Similarly, we can remove all infrared singularities of the quark-gluon scattering channel at order $\alpha_s$ via the following subtraction
\begin{align}
\label{eq:xsec_0L_IRfinite_NLO_qg}
    { \sigma_{qg \to {\rm ew}+q}^{{(0)} \rm ,IR-finite}(p_1,p_2;p_3) = \sigma_{qg \to {\rm ew}+q}^{(0)}(p_1,p_2;p_3) - \frac{\sigma_{qg \to {\rm H}+q}^{(0)}(p_1,p_2;p_3)}{\sigma_{q\bar q \to {\rm H}}^{(0)}(k_1,k_2)} \sigma_{q\bar q \to {\rm ew}}^{(0)}(k_1,k_2)\,. }
\end{align}
Analogously, for the antiquark-gluon channel, infrared subtraction is achieved by swapping the quark and antiquark in the equation above.

\subsection{Loop-momentum singularities}
We would like to remark that the finite remainders of eqs.~\eqref{eq:xsec_0L_IRfinite_NLO} and \eqref{eq:xsec_0L_IRfinite_NLO_qg} 
are in an analogous form as the finite remainder that can be written down for the one-loop correction to the process of electroweak production~\cite{Anastasiou:2022eym}. The one-loop virtual correction to the electroweak production integrand $\sigma_{q\bar q \to {\rm ew}}^{(1)}(p_1,p_2;k)$ is defined in eq.~\eqref{eq:pa2tonpartonic}, with the one-loop amplitude $\mathcal{M}_{q\bar q \to {\rm ew}}^{(1)}(p_1,p_2;k)$ defined in eq.~\eqref{eq:amp_ew_1loop}.
With this construction, all the infrared singular behaviors of $\sigma_{q\bar q \to {\rm ew}}^{(1)}$
are captured by
\begin{align}
\label{eq:NLO_qqbar_ew_infrared_behavior}
    \sigma_{q\bar q \to {\rm ew}}^{(1)}(p_1,p_2;k)\ \overset{\rm IR}{\approx}\ \frac{\sigma_{q\bar q \to {\rm H}}^{(1)}(p_1,p_2;k)}{\sigma_{q\bar q \to {\rm H}}^{(0)}(p_1,p_2)} \, \sigma_{q\bar q \to {\rm ew}}^{(0)}(p_1,p_2)\, ,
\end{align}
since
\begin{align}
\label{eq:FV_Higgs}
    \frac{\sigma_{q\bar q \to {\rm H}}^{(1)}(p_1,p_2;k)}{\sigma_{q\bar q \to {\rm H}}^{(0)}(p_1,p_2)}\ & \overset{\rm IR}{\approx}\  
    2 {\Re} \left[ 
    4ig_s^2 C_F \frac{p_1\cdot p_2 - k\cdot p_1 + k\cdot p_2}{\left[k^2+i0^{+}\right] \left[(k+p_1)^2 +i0^{+} \right] \left[(k-p_2)^2 +i0^{+}\right]}\right]\,  
    \nonumber \\
    &= (-2 C_AC_F) \,2 \Re \mathcal{F}^{(1)}_{\rm SLC} (p_1,p_2;k) \,  .
\end{align}
with ${\cal F}_{\rm SLC}^{(1)}$ given by
\eqref{eq:FV_SLC}.
This arises from a simple analysis of the three virtual infrared limits, $k\parallel p_1$, $k\parallel p_2$ and $k$ soft. The rationale is that the only distinction between single-Higgs and multi-particle electroweak production cross sections lies in the hard scattering, while their infrared behavior is identical. Hence the following subtraction ensures local finiteness in all virtual limits in $k$ and we can define
\begin{align}
    \sigma_{q\bar q \to {\rm ew}}^{(1),\, {\rm IR-finite}}(p_1,p_2;k)&= \sigma_{q\bar q \to {\rm ew}}^{(1)}(p_1,p_2;k) - \frac{\sigma_{q\bar q \to {\rm H}}^{(1)}(p_1,p_2;k)}{\sigma_{q\bar q \to {\rm H}}^{(0)}(p_1,p_2)} \sigma_{q\bar q \to {\rm ew}}^{(0)}(p_1,p_2) \, .
    \label{eq:xsec_virt_IRfinite}
\end{align}
As denoted by the second line of eq.~\eqref{eq:FV_Higgs}, the one-loop infrared singularities of the process $q \bar q  \to ew$, also emerge as initial state singularities of the $q \bar q  \to ew+g$ in eq.~\eqref{eq:FV_SLC} with a different color factor (subleading color).

\section{\texorpdfstring{Finite real-virtual cross-section integrand in the $q\bar q$ channel}{Finite real-virtual cross section integrand in the qqbar channel}}
\label{sec:RVxsec_localfact}

We are now ready to show how singularities from the one-loop amplitude integrand constructed in section~\ref{sec:RVamplitude} factorize locally, within a combined integration over the loop momentum and the final-state gluon momentum. 
This factorization enables us to construct a set of counterterms that separate universal but singular infrared behavior from finite dependence on the kinematics of electroweak final states. When these real-virtual results are combined with the analogous constructions for NNLO virtual corrections \cite{Anastasiou:2022eym,Anastasiou:2025cvy} and double-real corrections ~\cite{DelDuca:2025yph,Magnea:2024jqg,Devoto:2025jql,Czakon:2020coa,Braun-White:2023sgd}, this procedure will enable the numerical evaluation of arbitrary NNLO electroweak boson production processes at colliders. 
 
We will show that singularities in both ``single'' and ``double'' unresolved limits can be subtracted simultaneously using relatively simple counterterms at the integrand level. Here, single limits refer to configurations in which either the real gluon $p_3$ or the loop momentum $k$ is in an infrared limit, while double limits refer to those where both $p_3$ and $k$ are infrared.

As in the case of the amplitude $\mathcal{M}_{q\overline{q}\to {\rm ew} + g}^{(1)}$, eq.\ \eqref{eq:MR1L1_ours} we split the real-virtual correction cross section into the following four analogous parts:
\begin{align}
\label{eq:NNLO_cross_section_original}
    \sigma_{q\bar q \to {\rm ew}+g}^{(1)}(p_1,p_2;p_3;k) \equiv &\, 
    \sigma_{q\bar q \to {\rm ew}+g}^{(1), {\rm floop}}(p_1,p_2;p_3;k)+\sigma_{q\bar q \to {\rm ew}+g}^{(1), {\rm SLC}}(p_1,p_2;p_3;k) \nn\\
    & +\sigma_{q\bar q \to {\rm ew}+g}^{(1), {\rm LC}}(p_1,p_2;p_3;k) 
    +\sigma_{q\bar q \to {\rm ew}+g}^{(1), {\rm lp}}(p_1,p_2;p_3;k) \, ,
\end{align}
where each term aligns with the corresponding amplitude part in eq.~\eqref{eq:MR1L1_ours}.
Below we shall state all the necessary infrared subtractions for these terms separately. The final results will be presented in eqs.~(\ref{eq:xsec_IRfinite_fermion_loop}), (\ref{eq:xsec_IRfinite_SLC}), (\ref{eq:xsec_IRfinite_LC}), and (\ref{eq:xsec_IRfinite_pol}). 
The simplicity of the counterterms depends crucially on the local implementation of factorization, and detailed demonstrations of locality are given section~\ref{sec:IRanalyze}.

\subsection{Fermion-loop contribution to the cross section}
\label{sec:fermloop_xsec_IRfinite}
Let us begin our analysis of the terms in eq.\ \eqref{eq:NNLO_cross_section_original} with the fermion-loop contribution $\sigma_{q\bar q \to {\rm ew}+g}^{(1), {\rm floop}}(p_1,p_2;p_3;k)$, built from the  amplitudes shown in eq.~\eqref{eq:M1L_floops}. An example is given in eq.~\eqref{eq:fermloop_qqbar} with labeled momenta. The singularity structure of the fermion loop diagrams is simpler than for the others. In fact, we will see that all of its infrared singularities can be avoided without introducing additional subtraction terms.

We first note that straightforward power counting shows that a fermion loop momentum ($k$ in this case) cannot give a soft singularity. In addition, because the only on-shell lightlike external momentum of the fermion loop is $p_3$, the only collinear pinch for $k$ is in the $p_3$ direction, and hence this is the only possible source of divergences from the loop momentum $k$.

As indicated by eqs.~\eqref{eq:myphyscut} and~\eqref{eq:mynonphyscut}, there are two equivalent ways to treat external gluon polarizations in this process. These two approaches do not produce identical integrands for the fermion-loop contributions and are equivalent only after the sum over all diagrams. In particular, applying eq.~\eqref{eq:mynonphyscut} (i.e., unphysical polarizations included) generates a singular term in the integrand, which vanishes only after integration. Explicitly, we consider the following representative cut diagram,
\begin{align}
 &   \includegraphics[scale=0.8,page=2,valign=c] {figures_standalone/tikz_RV_ferm_loop.pdf} =  \ldots \frac{(\slashed{p}_3-\slashed{p}_2) \gamma^\mu \slashed{p}_2}{(p_3-p_2)^2} \ldots 
    \left( \sum_{\rm pols}
    \epsilon_\mu \epsilon_{\nu}^{*} \right) \ldots 
    \frac{\slashed k \gamma^\nu (\slashed{k}+\slashed{p_3})}{k^2 (k+p_3)^2} \ldots
    \label{eq:fermloop_qqbar}
    \nonumber \\ 
    &
\end{align}
On the right-hand side of this relation we have explicitly displayed the quark–gluon vertices connecting $p_2$ to $p_3$ and $k$ to $p_3$, along with the neighboring propagators, while omitting color factors and other constants. Other connections of the final-state gluon give similar results. The denominator structure indicates potential infrared singularities in the limits $k \parallel p_3$ and $p_2 \parallel p_3$. However, in these limits the numerators become proportional to $\epsilon^* \cdot p_3$ and $\epsilon \cdot p_3$, respectively, which vanish for physical (transverse) polarizations $\epsilon$. The double limit, $k \parallel p_3 \parallel p_2$ is doubly suppressed and also finite. Consequently, these singularities are absent when using the physical polarization sum in eq.~\eqref{eq:myphyscut}.

If instead one uses the unphysical polarization sum in eq.~\eqref{eq:mynonphyscut}, there is a singularity in the limit $k\parallel p_3$, at which the final-state gluon becomes longitudinally polarized at its connection to the fermion line in the complex conjugate diagram. The sum of these singularities vanishes locally after summing over the tree diagrams to the right of the cut. 
Similarly, for unphysical external gluon polarizations, in the other singular limit, $p_3\parallel p_2$, the final-state gluon becomes longitudinally polarized at its connection to the fermion loop. The sum of fermion-loop diagrams then yields two singular terms related by a shift in the loop momentum and opposite signs. Their sum, referred to as a ``shift mismatch" in refs.~\cite{Anastasiou:2020sdt,Anastasiou:2024xvk,Kermanschah:2025wlo}, therefore vanishes upon integration. Nevertheless, it persists at the integrand level, thereby obstructing local factorization. One can apply the strategies in these references to cure the shift mismatch at the integrand level, but for the discussion of this paper, we shall simply avoid the problem by choosing physical polarizations for $p_3$ in fermion loop contributions to the cross section. 

In summary, to simplify our discussion, we take the physical polarization sum eq.~\eqref{eq:myphyscut} for the final-state gluon in the fermion-loop contribution to $\sigma_{q\bar q \to {\rm ew}+g}^{(1)}(p_1,p_2;p_3;k)$. That is,
\begin{align}
\label{eq:xsec_IRfinite_fermion_loop}
   \sigma_{q\bar q \to {\rm ew}+g}^{(1), {\rm floop}}(p_1,p_2;p_3;k) = \left[\sigma_{q\bar q \to {\rm ew}+g}^{(1), {\rm floop}}(p_1,p_2;p_3;k)\right]_\textup{physical polarization}.
\end{align}
With this choice, fermion loop diagrams are free of infrared divergences.

\subsection{Subleading-color cross section}
\label{sec:RVxsec_SLC}
The subleading contribution to the real-virtual cross section is built from the amplitude $\mathcal{M}_{q\bar q \to {\rm ew}+g}^{(1), \rm SLC}$, defined diagrammatically in eq.\ \eqref{eq:M1SLC}. For this contribution, the locally finite integrand we construct is
\begin{align}
    &\sigma_{q\bar q \to {\rm ew}+g}^{(1), {\rm SLC, IR-finite}}(p_1,p_2;p_3;k) \nn\\
    &= \sigma_{q\bar q \to {\rm ew}+g}^{(1), {\rm SLC}}(p_1,p_2;p_3;k) - \sigma_{q\bar q \to {\rm H}+g}^{(1), {\rm SLC}}(p_1,p_2;p_3;k) \, \tau_{q\bar q \to {\rm ew}}^{(0)}(k_1,k_2) \nn\\
    & \quad- \sigma_{q\bar q \to {\rm H}}^{(1), {\rm SLC}}(p_1,p_2;k)\, \tau_{q\bar q \to {\rm ew}+g}^{(0)}(p_1,p_2;p_3) - \sigma_{q\bar q \to {\rm H}+g}^{(0), \rm SLC}(p_1,p_2;p_3)\, \tau_{q\bar q \to {\rm ew}}^{(1)}(k_1,k_2;k)  \, .
    \label{eq:xsec_IRfinite_SLC}
\end{align}
The factors $\tau$ are given by 
\begin{subequations}
\label{eq:tau_definitions}
\begin{align}
    \tau_{q\bar q \to {\rm ew}}^{(0)}(p_1, p_2) &\equiv \frac{\sigma_{q\bar q \to {\rm ew}}^{(0)}(p_1, p_2)}{\sigma_{q\bar q \to {\rm H}}^{(0)}(p_1, p_2)} \, , \label{eq:tau_tree}\\
    \tau_{q\bar q \to {\rm ew}}^{(1)}(p_1, p_2; k) &\equiv \frac{\sigma_{q\bar q \to {\rm ew}}^{(1),{\rm IR-finite}}(p_1, p_2;k)}{\sigma_{q\bar q \to {\rm H}}^{(0)}(p_1, p_2)} \, , \label{eq:tau_V}\\
    \tau_{q\bar q \to {\rm ew}+g}^{(0)}(p_1, p_2; p_3) &\equiv \frac{\sigma_{q\bar q \to {\rm ew}+g}^{(0),{\rm IR-finite}}(p_1, p_2;p_3)}{\sigma_{q\bar q \to {\rm H}}^{(0)}(p_1, p_2)}\, ,
    \label{eq:tau_R}
\end{align}
\end{subequations}
where $\sigma_{q\bar q \to {\rm ew}+g}^{(0),{\rm IR-finite}}$ and $\sigma_{q\bar q \to {\rm ew}}^{(1),{\rm IR-finite}}$ are defined in eqs.~\eqref{eq:NLO_qqbar_ew_infrared_behavior} and \eqref{eq:xsec_virt_IRfinite}, respectively. The subscript ``H" in the subtraction terms indicates single-Higgs production processes, and ``ew" a process with the full electroweak final state under consideration. Each subtraction term in eq.\ \eqref{eq:xsec_IRfinite_SLC} is then the product of an infrared-singular factor labeled $\sigma$, a cross section with a single Higgs or Higgs plus gluon in the final state, and an infrared-finite factor labeled $\tau$, which contains all information on the electroweak final state.

The momenta $k_1,k_2$ in eq.\ \eqref{eq:xsec_IRfinite_SLC} depend on the external momenta $p_1,p_2$ and $p_3$ through eq.~\eqref{eq:k1k2_param}. The subleading-color contributions of the virtual and real correction to single-Higgs production are obtained by isolating only the contribution proportional to $-\frac{1}{2C_A}$. For example, for the one-loop $\sigma$ factor in the third term on the right-hand side of eq.~\eqref{eq:xsec_IRfinite_SLC}, whose diagrams all carry the color factor $C_F$, we define
\begin{align}
    \sigma_{q\bar q \to {\rm H}}^{(1), {\rm SLC}}(p_1,p_2;k) & \equiv \frac{-1}{2C_A C_F} \, \sigma_{q\bar q \to {\rm H}}^{(1)}(p_1,p_2;k)\, 
    \nonumber \\[2mm]
    & \overset{\rm IR}{\approx} \sigma_{q\bar q \to {\rm H}}^{(0)}(p_1,p_2) \, 2 \Re  \mathcal{F}^{(1)}_{\rm SLC} (p_1, p_2 ; k) \, .
    \label{eq:xsec_1L_H_SLC}
\end{align}
As shown in the second relation, the infrared limits of this factor ($k$ soft or collinear to $p_1$ or $p_2$) are given by $\mathcal{F}^{(1)}_{\rm SLC}$ defined, including the same color factor, by eq.\ \eqref{eq:FV_SLC}.

Let us make some general remarks on the subtraction structure in eq.~\eqref{eq:xsec_IRfinite_SLC}.
On the right-hand side, the first term is the integrand for the subleading-color part of a generic electroweak production, which is singular in both ``single'' and ``double'' infrared limits, corresponding to one or both of the momenta $k$ and $p_3$ in a soft or collinear configuration. By substituting the definitions, eqs.\ (\ref{eq:tau_tree}), (\ref{eq:tau_V}) and (\ref{eq:tau_R}) for the factors $\tau$, it becomes clear that each of the remaining terms removes a specific class of singularities. 
Each subtraction term corresponds either to a single approximation targeting a specific divergence, or to repetitive approximations addressing two singularities nested within one another. This structure is characteristic of \emph{nested subtraction}. For a Feynman integral with nested and overlapping singularities, the complete set of subtraction terms can be constructed via a BPHZ-like scheme~\cite{Bogoliubov:1957gp,Hepp:1966eg,Zimmermann:1969jj}: each term involves a combination of approximations acting only on nested singularities~\cite{Collins:1981uk,Collins:2011zzd,Erdogan:2014gha,Anastasiou:2018rib,Ma:2019hjq}. Such a construction is free of double counting by design. 

In detail, the second term of the right-hand side of eq.~\eqref{eq:xsec_IRfinite_SLC} removes all double singularities (where $k$ and $p_3$ are either soft, collinear to $p_1$, or to $p_2$) as well as single-real singularities ($p_3$ soft, collinear to $p_1$ or to $p_2$) of the $V_{\rm qed}$ terms, defined in eq.~\eqref{eq:Vcurrent_legs}. The third term removes the single-virtual singularities, which for the subleading-color case involve only a gluon exchange between the incoming quark pair. As indicated in eq.~\eqref{eq:xsec_1L_H_SLC}, this counterterm yields the form-factor $\mathcal{F}^{(1)}_{\rm SLC} (p_1, p_2 ; k)$ in eq.~\eqref{eq:FV_SLC} of the one-loop amplitude in the single infrared regions of the loop momentum $k$. The final term removes the single-real singularities in the terms not involving $V_{\rm qed}$. 

As mentioned above, the counterterms are structured to subtract singularities only once. 
In particular, the factor $\tau_{q\bar q \to {\rm ew}+g}^{(0)}(p_1,p_2;p_3)$ is finite in the limits where the external gluon momentum $p_3$ becomes soft or collinear to initial-state partons and the corresponding counterterm (third term of the right-hand side) subtracts only singularities originating from the loop that do not overlap with the subtractions of ``double'' singularities in the first term of the right-hand side. 
Analogously, $\tau_{q\bar q \to {\rm ew}}^{(1)}(k_1,k_2;k)$ is finite in the limits where the loop momentum $k$ becomes soft or collinear to external partons and thus the corresponding counterterm (last term in the right-hand side) subtracts only non-overlapping single-real singularities.

\subsection{Leading-color cross section}
\label{sec:RVxsec_LC}
The initial-state and final-state infrared singularities of the leading-color cross section can be removed in a similar way as in eq.~(\ref{eq:xsec_IRfinite_SLC}) for the subleading-color case. The diagrammatic structure of the leading color contribution to the real-virtual cross section, based on the amplitude $\mathcal{M}_{q\bar q \to {\rm ew}+g}^{(1), \rm LC}$, is given in eq.\ \eqref{eq:M1LC}. We render finite the leading-color cross-section integrand 
with nested subtractions of double and single singularities as follows:
\begin{align}
        &\hspace{-1cm}\sigma_{q\bar q \to {\rm ew}+g}^{(1), {\rm LC, IR-finite}}(p_1,p_2;p_3;k)\nn\\
        =& \sigma_{q\bar q \to {\rm ew}+g}^{(1), {\rm LC}}(p_1,p_2;p_3;k) - \sigma_{q\bar q \to {\rm H}+g}^{(1), {\rm LC}}(p_1,p_2;p_3;k)\, \tau_{q\bar q \to {\rm ew}}^{(0)}(k_1,k_2) \nn\\
        &- \frac{\sigma_{q\bar q \to {\rm H}+g}^{(1), {\rm LC}}(p_1,p_2;p_3;k)}{\sigma_{q\bar q \to {\rm H}+g}^{(0)}(p_1,p_2;p_3)} 
        \sigma_{q\bar q \to {\rm H}}^{(0)}(p_1,p_2)\, \tau_{q\bar q \to {\rm ew}+g}^{(0)}(p_1,p_2;p_3) 
        \nn\\ 
        &- \mathcal{F}_{\rm R}^{(0)\rm ,LC}(p_1,p_2;p_3) \sigma_{q\bar q \to {\rm H}}^{(0)}(k_1,k_2) \, \tau_{q\bar q \to {\rm ew}}^{(1)}(k_1,k_2;k) \frac{\mathcal{O}_{{\rm H}+1}}{\mathcal{O}_{{\rm H}}} \nn \\
        &- \mathcal{F}_{\rm R}^{(0)\rm ,LC}(p_2,p_1;p_3) \sigma_{q\bar q \to {\rm H}}^{(0)}(k_1,k_2) \, \tau_{q\bar q \to {\rm ew}}^{(1)}(k_1,k_2; k-p_3)\frac{\mathcal{O}_{{\rm H}+1}}{\mathcal{O}_{{\rm H}}} \, ,
        \label{eq:xsec_IRfinite_LC}
\end{align}
where the infrared-finite $\tau$ functions are the same as for the subleading color integrand, and are defined in eqs.\ (\ref{eq:tau_tree}), (\ref{eq:tau_V}) and (\ref{eq:tau_R}).

In the following, we discuss terms on the right-hand side of eq.\ \eqref{eq:xsec_IRfinite_LC} in order.
The first term, $\sigma_{q\bar q \to {\rm ew}+g}^{(1), {\rm LC}}$, is the full integrand for the leading color cross-section. The second term (the first subtraction) removes all the double singularities. It also removes singularities in single-real limits of a number of terms in which the virtual loop also factorizes from the electroweak subdiagram. This is the case for the $p_3$-ghost terms, identified above in the discussion of the three-gluon vertex in eq.~\eqref{eq:Vggg}. For vertex corrections on the external fermions in the single-real limit, the $p_3$-ghost terms are precisely those contributions that were not included in the loop polarization analysis following eq.\ \eqref{eq:Vggg}. As noted there, the term $p_i$-ghost refers to the role these terms play in the Ward identity that is relevant in the single-real collinear region for the gluon of momentum $p_i$. This is discussed in more detail in section~\ref{sec:Collapprox_Wardids}.
The third line of eq.\ \eqref{eq:xsec_IRfinite_LC} subtracts the single-virtual singularities. In all the virtual single-infrared limits ($k$ soft or $\parallel p_1$, $p_2$ or $p_3$), the singularities are captured by the factor 
\begin{eqnarray}
\frac{\sigma_{q\bar q \to {\rm H}+g}^{(1), {\rm LC}}(p_1,p_2;p_3;k)}{\sigma_{q\bar q \to {\rm H}+g}^{(0)}(p_1,p_2;p_3)}\ 
\overset{\rm IR}{\approx} \
2 {\rm Re} \mathcal{F}^{(1)}_{\rm LC} (p_1,p_2;p_3;k) \, .
\label{eq:p3-co-limits}
\end{eqnarray}
This counterterm, matches the corresponding color component of the 
infrared form-factor of eq.~\eqref{eq:FV_LC} for the infrared behavior of the one-loop amplitude.  

The last two terms of eq.\ \eqref{eq:xsec_IRfinite_LC} remove the single-real ($p_3$) singularities, aside from those already accounted for in the second term (the $p_3$-ghost terms).
There are two terms in this single-real subtraction, in contrast to the subleading-color case, where one subtraction term is sufficient. This is necessary because in the 
two $p_3$ collinear limits, $p_3 \parallel p_1$ and $p_3\parallel p_2$,  the leading color radiative singularities factor from the same one-loop non-radiative cross section, but evaluated at different loop momentum $k$.   This
is confirmed in sec.\ \ref{sec:single-real-qqbar-LC}. To be explicit, given the momentum flow defined for leading color diagrams in eq.\ \eqref{eq:M1LC}, we show, in eqs.\ \eqref{eq:p1-p3-LC-2} and \eqref{eq:p3-to-xp2-ew}, that
\begin{align}
\begin{split}
    \sigma_{q\bar q \to {\rm ew}+g}^{(1), {\rm LC}}(p_1,p_2;p_3;k) &\xrightarrow[]{p_3\parallel p_1} \mathcal{F}_{\rm R}^{(0)\rm ,LC}(p_1,p_2;p_3)\sigma_{q\bar q \to {\rm ew}}^{(1)}(k_1,k_2;k) \, ,\\[2mm]
    \sigma_{q\bar q \to {\rm ew}+g}^{(1), {\rm LC}}(p_1,p_2;p_3;k) &\xrightarrow[]{p_3\parallel p_2} \mathcal{F}_{\rm R}^{(0)\rm ,LC}(p_2,p_1;p_3)\sigma_{q\bar q \to {\rm ew}}^{(1)}(k_1,k_2;k-p_3) \, ,
\end{split}
\label{eq:eq:xsec_LC_single_limits}
\end{align}
for a specific function ${\cal F}^{(0),{\rm LC}}_{\rm R}$ whose infrared behavior is defined by these relations. Note that the arguments $p_1$ and $p_2$ are exchanged on the right-hand side of the second relation. Since the right-hand sides in eq.~(\ref{eq:eq:xsec_LC_single_limits}) depend on distinct momentum routings, one cannot subtract both the $p_3 \parallel p_1$ and $p_3\parallel p_2$ singularities with a single subtraction term. In particular, the difference in the arguments of $\sigma_{q\bar q \to {\rm ew}}^{(1)}$ explains the need for two single-real subtractions, i.e., the last two lines in eq.~(\ref{eq:xsec_IRfinite_LC}). In the remaining infrared limit of $p_3$, the single-soft limit, $p_3\to 0$, the sum of these two subtraction terms removes the singularity of $\sigma_{q\bar q \to {\rm H}+g}^{(0), \rm LC}$ locally.

We can now extend the scalar function $\mathcal{F}_{\rm R}^{(0)\rm ,LC}$ to all momenta $p_3$, without introducing further singularities, in terms of cross section integrands by defining
\begin{align}
     &\mathcal{F}_{\rm R}^{(0)\rm ,LC}(p_1,p_2;p_3) \equiv \frac{C_A}{2} g_s^2  \left[ \frac{p_1\cdot p_2}{\left(p_1\cdot p_3 \right)\left(p_2\cdot p_3\right)} + \frac{p_1\cdot \chi}{\left(p_1\cdot p_3 \right)\left(\chi\cdot p_3\right)}  \right. \nn\\
     &\hspace{3cm}\left. - \frac{p_2\cdot \chi}{\left(p_2\cdot p_3 \right)\left(\chi\cdot p_3\right)} -\frac{2-D}{2} \frac{p_2\cdot p_3 + p_1\cdot p_3}{\left(p_1\cdot p_3 \right)(p_1\cdot p_2-p_1\cdot p_3-p_2\cdot p_3)} \right] \, ,
    \label{eq:FR_def}
\end{align}
where $\chi$ is an auxiliary vector satisfying $p_i\cdot \chi \neq 0$ for $i=1,2,3$.
The function of eq.~\eqref{eq:FR_def} can be thought of as a partial fraction of the single-real radiation form-factor ${\cal F}_{q\bar q \to {\rm H}+g}$ defined in eq.\ \eqref{eq:Fqq2Hg}, designed to separate the singularities in the collinear limits, $p_3\parallel p_1$ and $p_3 \parallel p_2$. Defined in this fashion, the form factor ${\cal F}^{(0),{\rm LC}}_{\rm R}$ is related to the universal infrared factor by
\begin{eqnarray}
\label{eq:FRparfrac}
 \frac{C_A}{2C_F}\,\Fcal_{q\bar q \to {\rm H}+g}(p_1, p_2; p_3) = \mathcal{F}_{\rm R}^{(0)\rm ,LC}(p_1,p_2;p_3) 
 + \mathcal{F}_{\rm R}^{(0)\rm ,LC}(p_2,p_1;p_3) \, , 
\end{eqnarray}
where again, in the second term on the right-hand side arguments $p_1$ and $p_2$ are exchanged. Note that the dependence on the fixed vector $\chi$ then cancels trivially in the sum. For the terms in eq.\ \eqref{eq:xsec_IRfinite_LC} containing ${\cal F}$, we have indicated phase space limitations on $p_3$ with the abbreviated notation,
\begin{align}
    \frac{\mathcal{O}_{{\rm H}+1}}{\mathcal{O}_{{\rm H}}} \equiv  \frac{\mathcal{O}_{{\rm H}+1}(p_3,q_1+\ldots + q_n)}{\mathcal{O}_{{\rm H}}(q_1+\ldots + q_n)}\, .
    \label{eq:OH1/OH}
\end{align}

Finally, we note that at the integral level, one can combine the two single-real subtraction terms (last two subtractions in eq.~\eqref{eq:xsec_IRfinite_LC}) by shifting $k-p_3 \to k$ in the last term of eq.~\eqref{eq:xsec_IRfinite_LC} to obtain:
\begin{align}
    &\int \frac{\mathrm{d}^Dk}{(2\pi)^D} \left[
    \mathcal{F}_{\rm R}^{(0)\rm ,LC}(p_1,p_2;p_3) 
    + \mathcal{F}_{\rm R}^{(0)\rm ,LC}(p_2,p_1;p_3) \right] \tau_{q\bar q \to {\rm ew}}^{(1)}(k_1,k_2;k)  \sigma_{q\bar q \to {\rm H}}^{(0)}(k_1,k_2)\frac{\mathcal{O}_{{\rm H}+1}}{\mathcal{O}_{{\rm H}}}  \nn\\
    &= \int \frac{\mathrm{d}^Dk}{(2\pi)^D} \sigma_{q\bar q \to {\rm H}+g}^{(0), \rm LC}(p_1,p_2;p_3) \tau_{q\bar q \to {\rm ew}}^{(1)}(k_1,k_2;k) \, ,
    \label{eq:FR_integrated}
\end{align}
whose integrand simplifies to the product of the lower-order cross sections, $\sigma_{q\bar q \to {\rm H}+g}^{(0), \rm LC}$ and $\tau_{q\bar q \to {\rm ew}}^{(1)}(k_1,k_2;k)$. This can be implemented by the addition of a shift counterterm, similar to those discussed in ref.\ \cite{Anastasiou:2024xvk}, but we shall not do so here.

\subsection{Loop-polarization cross section}
\label{sec:RVxsec_LP}
From eq.~(\ref{eq:Jcurrent_perp_tilde}), the loop-polarization contribution $\sigma_{q\bar q \to {\rm ew}+g}^{(1), {\rm lp}}$ has the following logarithmic singularities: (1) $k$ hard, $p_3$ collinear to $p_1$ or $p_2$; (2) $k$ and $p_3$ both collinear to $p_1$ or $p_2$.\footnote{Note that each individual diagram involving $\widetilde J_\perp^{a \mu}$ features a divergence at $k\parallel p_3$, but these divergences cancel in the sum due to the Ward identity.} The following subtraction is sufficient to remove them both simultaneously and render it locally finite:
\begin{align}
        &\sigma_{q\bar q \to {\rm ew}+g}^{(1), {\rm lp, IR-finite}}(p_1,p_2;p_3;k)\nn \\
        &\qquad\qquad= \sigma_{q\bar q \to {\rm ew}+g}^{(1), {\rm lp}}(p_1,p_2;p_3;k) - \sigma_{q\bar q \to {\rm H}+g}^{(1), {\rm lp}}(p_1,p_2;p_3;k) \, \tau_{q\bar q \to {\rm ew}}^{(0)}(k_1,k_2) \, .
        \label{eq:xsec_IRfinite_pol}
\end{align}

\subsection{Comment on subtractions}

In eqs.~\eqref{eq:xsec_IRfinite_SLC}, \eqref{eq:xsec_IRfinite_LC}, and \eqref{eq:xsec_IRfinite_pol}, we derive an infrared-finite part of the real-virtual cross section, capitalizing on the fact that the one-loop amplitude integrand constructed in section~\ref{sec:RVamplitude}, factorizes locally in all infrared limits for the loop momentum and the momentum of the final-state gluon simultaneously. All dependence on final-state electroweak kinematics is contained in the factors $\tau^{(m)}$ for $m=0,1$, defined in eqs.~\eqref{eq:tau_tree}--\eqref{eq:tau_R}, which are integrable by construction. All infrared-singular dependence is contained in the components of $\sigma_{q\bar q\to H}^{(1)}$ and $\sigma_{q\bar q\to H+g}^{(m)}$ for $m=0,1$. The real-virtual correction cross section to single-Higgs production is known. They can be further factorized into parton distributions once the final-state singularities cancel in a suitably inclusive sum over final states, when combined with the NNLO ``double-real" diagrams.

\section{\texorpdfstring{Finite real-virtual cross-section integrand in the $qg$ channel}{Finite real-virtual cross-section integrand in the qg channel}}
\label{section-finite_RV_xsec_integrand_qg_channel}

In this section, we present the infrared subtractions for the production of massive electroweak bosons in association with a quark via quark-gluon scattering:
\begin{align}
    q(p_1) + g(p_2) \longrightarrow q(p_3) + \textup{ew}(q_1,\dots,q_n) \, ,
\end{align}
where ``ew'' denotes the set of massive colorless bosons in the final state, with the corresponding momenta indicated in parentheses. This process is related to the process we have studied so far, i.e., the production of massive electroweak bosons in association with a gluon via quark-antiquark annihilation, by crossing the antiquark and the gluon.
We similarly split the one-loop amplitude of this process into four classes, namely,
\begin{align}
\label{eq:Mqg_split}
{\cal M}_{qg\to {\rm ew}+q}^{(1)} &\equiv   
{\cal M}_{qg\to {\rm ew}+q}^{(1), {\rm floop}}
+{\cal M}_{qg\to {\rm ew}+q}^{(1), {\rm SLC}}+{\cal M}_{qg\to {\rm ew}+q}^{(1), {\rm LC}} +
 {\cal M}_{qg\to {\rm ew}+q}^{(1), {\rm lp}}\, .
\end{align}
They are connected to the $q\overline{q}$ channel, eq.\ \eqref{eq:MR1L1_ours} via the following crossing:
\begin{subequations}
\begin{align}
    {\cal M}_{qg\to {\rm ew}+q}^{(1), {\rm floop}}(p_1,p_2;p_3;k) &= {\cal M}_{q\bar q \to {\rm ew}+g}^{(1), {\rm floop}}(p_1,-p_3;-p_2;k) \vert_{\bar{q}\leftrightarrow g} \, , \\
    {\cal M}_{qg\to {\rm ew}+q}^{(1), {\rm SLC}}(p_1,p_2;p_3;k) &= {\cal M}_{q\bar q \to {\rm ew}+g}^{(1), {\rm SLC}}(p_1,-p_3;-p_2;k) \vert_{\bar{q}\leftrightarrow g} \, , \\
    {\cal M}_{qg\to {\rm ew}+q}^{(1), {\rm LC}}(p_1,p_2;p_3;k) &= {\cal M}_{q\bar q \to {\rm ew}+g}^{(1), {\rm LC}}(p_1,-p_3;-p_2;k-p_2) \vert_{\bar{q}\leftrightarrow g} \, , \label{eq:Mqg_LC}\\
    {\cal M}_{qg\to {\rm ew}+q}^{(1), {\rm lp}}(p_1,p_2;p_3;k) &= {\cal M}_{q\bar q \to {\rm ew}+g}^{(1), {\rm lp}}(p_1,-p_3;-p_2;k) \vert_{\bar{q}\leftrightarrow g} \, , 
\end{align}
\end{subequations}
where we shift the routing of the loop momentum $k$ in the leading-color contribution, to simplify the construction of the single-real infrared counterterm.

The virtual infrared structure and the corresponding subtractions follow in the same way as in the quark–antiquark channel discussed above. We can then approximate the one-loop amplitude in all the virtual infrared limits by
\begin{align}
\label{eq:M1Lqg_IRapprox}
    &{\cal M}_{qg \to {\rm ew}+q}^{(1)}(p_1,p_2;p_3;k) \nn\\
    &\quad  \overset{\textup{IR}}{\approx}
    \left(
    \mathcal{F}^{(1)}_{\rm LC} (p_1,-p_3;-p_2;k-p_2)
    +\mathcal{F}^{(1)}_{\rm SLC} (p_1,-p_3;k) 
    \right) {\cal M}_{qg \to {\rm ew}+q}^{(0)}(p_1,p_2;p_3)\,,
\end{align}
where the scalar functions are defined in eqs.~\eqref{eq:FV_SLC} and \eqref{eq:FV_LC}, and the tree-level amplitude is connected to the quark–antiquark channel amplitude via
\begin{align}
     {\cal M}_{qg\to {\rm ew}+q}^{(0)}(p_1,p_2;p_3) &= {\cal M}_{q\bar q \to {\rm ew}+g}^{(0)}(p_1,-p_3;-p_2) \vert_{\bar{q}\leftrightarrow g} \, .
\end{align}

The subtraction of the single-real singularities and the real-virtual singularities at the cross-section level proceeds similarly to the quark–antiquark channel. However, there are several notable differences that require a further discussion. 

As in the case of the quark–antiquark channel in eqs.\ \eqref{eq:local_xsec_R} and \eqref{eq:local_xsec_RV}, the integrands of the tree-level and one-loop-level quark-gluon channel cross sections are given by
\begin{align}
    \sigma_{qg\to {\rm ew}+q}^{(0)}(p_1,p_2;p_3) &= \left\langle \abs{\mathcal{M}^{(0)}_{qg\to {\rm ew}+q}(p_1,p_2;p_3)}^2\right\rangle \delta(q_n^2- M_n^2) \, {\cal O}_{{\rm ew}+1}\left( p_3, q_1,\ldots , q_n\right)\,,\label{eq:local_xsec_R_qg}\\ 
    \sigma_{qg\to {\rm ew}+q}^{(1)}(p_1,p_2;p_3;k) &=  \left\langle 2 \, \mathrm{Re}\left[ \mathcal{M}^{(1) }_{qg\to {\rm ew}+q}(p_1,p_2;p_3;k)\left(\mathcal{M}^{(0)}_{qg\to {\rm ew}+q}(p_1,p_2;p_3) \right)^*\right]\right\rangle \nn\\
    &\hspace{1cm}\cdot \delta(q_n^2- M_n^2) \, {\cal O}_{{\rm ew}+1}\left( p_3, q_1,\ldots , q_n\right)\, .
    \label{eq:local_xsec_RV_qg}
\end{align}
At order $\alpha_s^2$, the cross section inherits the division of the amplitude 
in eq.~\eqref{eq:Mqg_split}, 
\begin{align}
\label{eq:qgNNLO_cross_section_original}
\sigma_{qg \to {\rm ew}+q}^{(1)}(p_1,p_2;p_3;k) \equiv &\, \sigma_{qg \to {\rm ew}+q}^{(1), {\rm floop}}(p_1,p_2;p_3;k)
 +\sigma_{qg \to {\rm ew}+q}^{(1), {\rm SLC}}(p_1,p_2;p_3;k) \nn\\
 & +\sigma_{qg \to {\rm ew}+q}^{(1), {\rm LC}}(p_1,p_2;p_3;k) 
+\sigma_{qg \to {\rm ew}+q}^{(1), {\rm lp}}(p_1,p_2;p_3;k) \, ,
\end{align}
as in eq.\ \eqref{eq:NNLO_cross_section_original} for the quark-antiquark channel.

The singularity structure for quark-gluon channel in the single-real limits is similar to that of the  quark-antiquark channel with a transversely polarized external gluon. As we show below, this follows from the fact that, in the  quark-gluon channel, the momentum $p_2$ is an incoming physical gluon and thus restricted to transverse polarizations. This results in a suppression of the real singularities,  $p_3 \parallel p_1$, which appears in a subset of the cross-section terms, and $p_3 \to 0$, which appears in all. 

Only the fermion-loop contribution exhibits singularities in the limit $p_3 \parallel p_1$. As an example, consider the following cut diagram containing this singularity:
\begin{align} 
    \includegraphics[scale=1,valign=c,page=1]{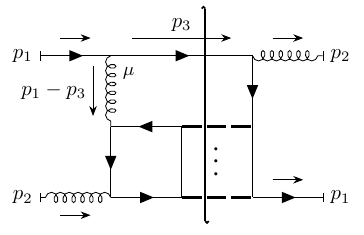}
    = \ldots \slashed{p}_3 \gamma^\mu \slashed{p}_1 \ldots\,,
    \label{eq:ferm_loop}
\end{align}
where we explicitly display the quark-gluon vertex from $p_1$ to $p_3$, appearing on the left-hand side of the cut, on the right-hand side of the equation. In the limit $p_3 \parallel p_1$, this produces a longitudinally polarized gluon $p_1 - p_3$ entering the fermion loop. After summing over all cut diagrams, the longitudinally polarized gluon attaches to every side of the fermion loop. Applying Ward identities, this sum yields a local shift mismatch, as in the discussion of eq.\ \eqref{eq:fermloop_qqbar}, between two terms of opposite sign that differ only by a shift of loop momentum, and hence which cancel after integration.
These terms can be removed via shift counterterms as constructed in refs.~\cite{Anastasiou:2020sdt,Anastasiou:2024xvk,Kermanschah:2025wlo}. Specifically, the $\hat{\gamma}$-prescription avoids these singularities altogether by modifying the quark-gluon vertex adjacent to $p_1$ shown in eq.~\eqref{eq:ferm_loop} as follows:
\begin{align}
\label{eq:gamma_hat_prescription}
    \gamma^\mu \to \hat{\gamma}^\mu = \gamma^\mu - (p_1-p_3)^\mu \frac{2\slashed{\eta}}{(p_1-p_3+\eta)^2-\eta^2}\,.
\end{align}
The vector $\eta$ is chosen with $\eta^2\neq 0$ and such that it produces no new pinches from the new denominator. In the limit we then immediately have
\begin{align}
    \slashed{p}_3 \hat\gamma^\mu \slashed{p}_1 \xrightarrow[]{p_3\parallel p_1} 0\,.
\end{align}
The modification in eq.~\eqref{eq:gamma_hat_prescription} subtracts a term proportional to $(p_1-p_3)^\mu$ which leads again to a shift mismatch and cancels after integration. Consequently, this modification renders the fermion-loop cross-section integrand infrared finite in this limit, without changing the value of the cross section. Note that the single-virtual limit, where the fermion loop momentum next to the external gluon $p_2$ becomes collinear to the gluon, is suppressed in the $qg$ channel, since the gluon $p_2$ is a transversely polarized physical gluon. Additionally, the fermion-loop contribution is finite in the single-real limit where the outgoing quark is collinear to the incoming gluon $p_3\parallel p_2$, again due to the transversality of the incoming gluon.

Hence, after the modification -- either via $\hat{\gamma}$ or via shift counterterms, collectively denoted as the subtraction $\Delta \sigma_{qg \to {\rm ew}+q}^{(1), {\rm floop}}(p_1,p_2;p_3;k)$ -- the shift mismatch of the unmodified integral is eliminated and the fermion-loop contribution becomes free of all infrared singularities:
\begin{align}
    \sigma_{qg \to {\rm ew}+q}^{(1), {\rm floop, IR-finite}}(p_1,p_2;p_3;k) &= \sigma_{qg \to {\rm ew}+q}^{(1), {\rm floop}}(p_1,p_2;p_3;k) - \Delta \sigma_{qg \to {\rm ew}+q}^{(1), {\rm floop}}(p_1,p_2;p_3;k)\,.
    \label{eq:qgxsec_IRfinite_floops}
\end{align}

The leading- and the subleading-color contributions are rendered locally finite by the following subtractions: 
\begin{align}
    &\sigma_{qg \to {\rm ew}+q}^{(1), {\rm SLC + LC, IR-finite}}(p_1,p_2;p_3;k) \nn\\
    &= \sigma_{qg \to {\rm ew}+q}^{(1), {\rm SLC + LC}}(p_1,p_2;p_3;k) - \sigma_{qg \to {\rm H}+q}^{(1), {\rm SLC + LC}}(p_1,p_2;p_3;k)\, \tau_{q\bar q \to {\rm ew}}^{(0)}(k_1,k_2) \nn\\
    &\qquad- \frac{\sigma_{qg \to {\rm H}+q}^{(1), {\rm SLC + LC}}(p_1,p_2;p_3;k)}{\sigma_{qg \to {\rm H}+q}^{(0)}(p_1,p_2;p_3)}\sigma_{qg \to {\rm ew}+q}^{(0) {\rm, finite}}(p_1,p_2;p_3) 
    \nn\\ 
    &\qquad- \sigma_{qg \to {\rm H}+q}^{(0)}(p_1,p_2;p_3)\, \tau_{q\bar q \to {\rm ew}}^{(1)}(k_1,k_2;k)
    \label{eq:qgxsec_IRfinite_SLCLC}
\end{align}
where we have employed the infrared-finite cross sections in eqs.~\eqref{eq:xsec_0L_IRfinite_NLO_qg},\eqref{eq:tau_tree} and \eqref{eq:tau_V}. All infrared singularities are canceled in a pattern similar to that of eqs.~(\ref{eq:xsec_IRfinite_SLC}) and (\ref{eq:xsec_IRfinite_LC}). The second term on the right-hand side removes all double singularities, as well as the single-real singularities of the $p_2$-ghost terms (introduced in section~\ref{sec:Collapprox_Wardids}) from the leading-color contributions and the $V_{\rm qed}$ terms from the subleading-color contributions. The third line subtracts the single-virtual singularities, while the last term removes the remaining single-real singularities. Compared with the last two lines of eq.~\eqref{eq:xsec_IRfinite_LC}, only one subtraction is needed for the leading-color cross section in this channel, as the only infrared singularity arises from the $p_3 \parallel p_2$ limit. Note that without the momentum shift in $k$ in eq.~\eqref{eq:Mqg_LC}, the leading-color part of the finite cross section $\tau$ on the last line would depend on $k+p_2$ rather than $k$.

The loop-polarization contribution has a single-real logarithmic singularity where $k$ is hard and $p_3$ is collinear to $p_2$, and a real-virtual singularity when $k$ and $p_3$ are collinear to $p_2$. Meanwhile, it is finite in all single-virtual limits. All infrared singularities therefore are subtracted in the
expression
\begin{align}
    &\sigma_{qg \to {\rm ew}+q}^{(1), {\rm lp, IR-finite}}(p_1,p_2;p_3;k) = \sigma_{qg \to {\rm ew}+q}^{(1), {\rm lp}}(p_1,p_2;p_3;k) - \sigma_{qg \to {\rm H}+q}^{(1), {\rm lp}}(p_1,p_2;p_3;k)\,\tau_{q\bar q \to {\rm ew}}^{(0)}(k_1,k_2) .
    \label{eq:qgxsec_IRfinite_pol}
\end{align}

In the following section, we provide a detailed demonstration of the subtraction procedure for the quark–antiquark annihilation channel. The local factorization for the quark-gluon scattering channel presented in this section proceeds analogously and is therefore not discussed further.

\section{\texorpdfstring{Demonstration of local infrared factorization in the $q\bar q$ channel}{Demonstration of local infrared factorization in the qqbar channel}}
\label{sec:IRanalyze}

In this section, we demonstrate that the subtraction terms introduced in sections~\ref{sec:IRapprox_RVamplitude}, \ref{sec:NLOxsec_localfact}, and \ref{sec:RVxsec_SLC}--\ref{sec:RVxsec_LP} suffice to remove all infrared singularities at the integrand level. Our presentation is structured as follows: first, we show how each infrared singularity can be locally factorized in the relevant limit, then we explain how the subtraction term reproduces the same infrared behavior in each case.

In the demonstration below, we use external gluons with four polarizations as in~\eqref{eq:mynonphyscut}, rather than restricting to the transverse polarizations of eq.~\eqref{eq:myphyscut}. This choice connects the factorization of the cross-section integrand more directly to the factorization of the two-loop amplitude integrand in ref.~\cite{Anastasiou:2022eym}.

Our analysis relies primarily on the Ward identity together with the graphical notation introduced in section~\ref{sec:Collapprox_Wardids}. In section~\ref{sec:IRdemo_RVampl} we demonstrate the local factorization of single-virtual singularities for the one-loop amplitude and derive the subtraction factors $\mathcal{F}^{(1)}_{\rm SLC}$ and $\mathcal{F}^{(1)}_{\rm LC}$ defined in eqs.~\eqref{eq:FV_SLC} and \eqref{eq:FV_LC}. We then turn to the cross sections, where in section~\ref{sec:IRdemo_NLO}, we first establish the local factorization of electroweak boson production at order $\alpha_s$, stated in eq.~\eqref{eq:xsec_0L_IRfinite_NLO}. Afterwards we extend the single-virtual treatment of the amplitude to the cross-section level in section~\ref{sec:IRdemo_RVcrsec_singleV}, and finally, in section~\ref{sec:IRdemo_RVcrsec_singleR}, we address the subtraction of single-real singularities for the real-virtual correction cross section.

\subsection{Graphical notations}
\label{sec:Collapprox_Wardids}

Due to the specific construction of the loop amplitude integrands in section~\ref{sec:RVamplitude}, the infrared singularities in the combined loop and phase-space integrals of the real-virtual cross-sections are logarithmic. This is shown in more detail as a power-counting analysis for some examples in the appendix~\ref{appendix-power_counting}. To isolate the singularity in a given infrared limit, it therefore suffices to retain only the leading contribution, where polarizations of certain gauge bosons are constrained. In particular, when a collinear quark–gluon pair enters the hard subdiagram $\mathcal{V}$, merely the \emph{longitudinal} polarization of the gluon contributes to the infrared singularity. Only showing the relevant terms, the limit $k\parallel p$ of a general amplitude is 
\begin{align}
\label{eq:collinear_approximation}
    \includegraphics[scale=0.9, page=28,valign=c]{figures_standalone/tikz_RV_Ward.pdf}&= \mathcal{V}^\alpha\left(k,\dots\right) \frac{1}{k^2}\frac{\slashed{p}-\slashed{k}}{(p-k)^2}  \gamma_\alpha \,u(p)\nonumber\\
    &\hspace{-1em}\xrightarrow{k\parallel p} k_\alpha \mathcal{V}^\alpha\left(k,\dots\right) \frac{1}{k^2}\frac{\slashed{p}-\slashed{k}}{(p-k)^2} \frac{\slashed{\chi}}{k\cdot\chi} \,u(p) \equiv \includegraphics[scale=0.9, page=24,valign=c]{figures_standalone/tikz_RV_Ward.pdf}\,,
\end{align}
where $\chi$ is an auxiliary lightlike vector that is not in the direction of $p$.
The approximation in eq.~(\ref{eq:collinear_approximation}), referred to as the \emph{hard-collinear approximation} in more general cases~\cite{Collins:1989gx}, demonstrates that only the longitudinal polarization of the gluon $k^\alpha$ contributes to the collinear singularity. Equivalently, this amounts to replacing the gluon propagator Feynman rule by,
\begin{align}
    \label{eq:coll_approx}
    \frac{-i\delta^{ab}}{k^2}g^{\alpha\beta} \xrightarrow{\quad } \frac{-i\delta^{ab}}{k^2}\frac{\chi^\alpha k^\beta}{k\cdot \chi}\equiv \raisebox{2mm}{\includegraphics[width=0.35\textwidth, page=1,valign=c]{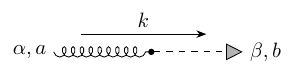} },
\end{align}
in the collinear limit $k\parallel p$. The diagram on the right-hand side of eq.~\eqref{eq:coll_approx} denotes this approximation. In the limit the longitudinal gluon $k^\alpha$ contracts with the rest of the diagram, yielding the product $k_\alpha\mathcal{V}^\alpha$ in eq.~(\ref{eq:collinear_approximation}), allowing us to apply a diagrammatic Ward identity to this unphysically-polarized gluon. Since we are concerned with the local behavior of integrands at a particular (NNLO) order, our diagrammatic Ward identities are algebraic in nature, the lowest-order tree expressions of the elementary Ward identities of QCD. 
Naturally, the identity depends on the vertex to which the longitudinally polarized gluon is attached, and in our NNLO cut diagrams we encounter only two cases.

The first case is when a longitudinally polarized gluon is attached to a fermion propagator. In this case the identity reads, pictorially, 
\begin{align}
    \includegraphics[scale=0.85, page=2,valign=c]{figures_standalone/tikz_RV_Ward.pdf} &=\raisebox{0.2mm}{\includegraphics[scale=0.85, page=3,valign=c]{figures_standalone/tikz_RV_Ward.pdf}} \hspace{-1cm} + \hspace{-1cm}\includegraphics[scale=0.85, page=4,valign=c]{figures_standalone/tikz_RV_Ward.pdf}= g_s T^c \frac{i}{\slashed{r}} - g_s T^c \frac{i}{\slashed{l}+\slashed{r}}\, , 
    \label{eq:Wardid_qqg_eq}
\end{align}
where the symbol ``$\times$'' denotes the deletion of the corresponding propagator. Note that the second term on the right-hand side has a minus sign, which has been absorbed into the vertex, defined as
\begin{align}
    \raisebox{0.5cm}{\includegraphics[scale=0.85,  page=5,valign=c]{figures_standalone/tikz_RV_Ward.pdf}} = g_sT^c\, ,\qquad \raisebox{0.5cm}{\includegraphics[scale=0.85, page=6,valign=c]{figures_standalone/tikz_RV_Ward.pdf}} = -g_sT^c\,.
    \label{eq:WI_qqg_vertdef}
\end{align}

The other example that we encounter at NNLO is when the longitudinally polarized gluon is attached to a gluon propagator via a triple-gluon vertex, we first decompose the vertex according to eq.~(\ref{eq:Vggg}), after which the Ward identity can be represented as
\begin{align}
    &\includegraphics[width=0.24\textwidth, page=23,valign=c]{figures_standalone/tikz_RV_Ward.pdf} \nn\\
    &= \includegraphics[width=0.24\textwidth, page=7,valign=c]{figures_standalone/tikz_RV_Ward.pdf}+ \includegraphics[width=0.24\textwidth, page=15,valign=c]{figures_standalone/tikz_RV_Ward.pdf}+\includegraphics[width=0.24\textwidth, page=16,valign=c]{figures_standalone/tikz_RV_Ward.pdf}\nn\\
    &=\ \  g_sf^{abc}\left[ \left( \frac{g^{\alpha\beta}}{r^2} - \frac{g^{\alpha\beta}}{(r+l)^2}\right) +\frac{r^\alpha r^\beta}{r^2(r+l)^2} - \frac{(l+r)^\alpha(l+r)^\beta}{r^2(r+l)^2} \right]\nn\\
    &= \raisebox{0.75mm}{
    \includegraphics[width=0.2\textwidth, page=9,valign=c]{figures_standalone/tikz_RV_Ward.pdf} }\hspace{-0.3cm}+ \hspace{-0.3cm} \includegraphics[width=0.2\textwidth, page=8,valign=c]{figures_standalone/tikz_RV_Ward.pdf}+ \includegraphics[width=0.24\textwidth, page=17,valign=c]{figures_standalone/tikz_RV_Ward.pdf}+ \includegraphics[width=0.24\textwidth, page=18,valign=c]{figures_standalone/tikz_RV_Ward.pdf}\, .
    \label{eq:scalar_to_ghost}
\end{align}
The first equality is exactly the representation of eq.\ \eqref{eq:Vggg}, of the three-gluon vertex, with an arrow where factor $l^\gamma$ is contracted where the line of momentum $l$ enters the vertex. The first term on the right is what we have called the $l$-scalar term for this vertex after eq.\ \eqref{eq:Vggg}, where we also mentioned that the remaining two terms will be described as $l$-ghost terms. The next two equalities show the motivation for this terminology.

The second equality in eq.\ \eqref{eq:scalar_to_ghost} shows the explicit algebraic result found after the contraction of the full vertex with $l^\mu$. In this expression, the first two terms are from the $l$-scalar vertex of the first equality contracted with vector $l^\gamma$. In each of these two terms, one propagator adjacent to the vertex is canceled, in much the same way as for the fermion in eq.\ \eqref{eq:Wardid_qqg_eq}. These two terms are illustrated diagrammatically on the final line, which is defined, term-by-term, by the algebraic expression of the second equality.
The remaining two terms are from the $l$-ghost terms contracted with $l^\gamma$ in the first equality, whose sum gives the result shown. This result corresponds to third and fourth diagrams of the final equality, where the dashed lines with small black arrows are identified with QCD ghosts, coupled to a gluons by standard ghost-gluon vertices. When eq.~\eqref{eq:scalar_to_ghost} is considered as a part of a diagram, the ghost lines terminate at vectors (index $\alpha$ or $\beta$) which contract with vertices in the rest of the diagram. This requires us to apply similar algebraic reasoning to these vertices. As we shall see below, the ghost terms lead to the factorization of two-loop subdiagrams, even when one of the loops is not infrared. More detailed examples and explanations of identities following from the rules above can be found in refs.~\cite{Anastasiou:2022eym,Anastasiou:2024xvk}.
In concluding this discussion, we remark that the diagrammatic notation above is introduced to verify the local factorization of infrared behavior from the details of electroweak production. They are not needed for the definition or construction of the infrared counterterms introduced in sections \ref{sec:RVxsec_localfact} and \ref{section-finite_RV_xsec_integrand_qg_channel}. They enable us to develop arguments that confirm that these universal counterterms render the relevant integrands integrable.

In the following subsections, we apply the graphical notations introduced above to the cross sections at order $\alpha_s$, the real-virtual correction cross section as well as in the one-loop amplitude to show their local factorization properties in all infrared limits.

\subsection{One-loop amplitude}
\label{sec:IRdemo_RVampl}

For the local factorization of $\mathcal{M}^{(1)}_{q\overline{q}\to {\rm ew}+g}$, presented in eqs.~\eqref{eq:M1IRapprox}–\eqref{eq:FV_LC}, the loop polarization terms ${\cal M}_{q\bar q \to {\rm ew}+g}^{(1), {\rm lp}}$, eq.\ \eqref{eq:RVpoldefinition} and fermion loop terms, ${\cal M}_{q\bar q \to {\rm ew}+g}^{(1), {\rm floop}}$, eq.\ \eqref{eq:M1L_floops}, require no discussion, because they contain \emph{no} virtual singularities as explained in sections~\ref{sec:LP} and \ref{sec:M1L_integrand}, respectively. In what follows, we focus on the infrared limits of $\mathcal{M}_{q\overline{q}\to {\rm ew}+g}^{(1),{\rm LC}}$ and $\mathcal{M}_{q\overline{q}\to {\rm ew}+g}^{(1),{\rm SLC}}$ separately.

\subsubsection{Infrared limits of $\mathcal{M}_{q\overline{q}\to {\rm ew}+g}^{(1) \rm,SLC}$}
\label{sec:demo_IR_SLC_ampl}
The subleading-color amplitude $\mathcal{M}_{q\overline{q}\to {\rm ew}+g}^{(1) \rm,SLC}$ is defined in eq.~\eqref{eq:M1SLC}, which is free of power-like singularities and loop-polarization terms. From the original triangle corrections to the incoming quark and antiquark, the amplitude includes only the ${V}_{\rm qed}$ and ${V}^\dagger_{\rm qed}$ vertices. It is worth noting that this replacement does not change the infrared behavior, as shown in eq.~\eqref{eq:Vqed_klim}. Therefore, the $k\parallel p_1$ limit of $\mathcal{M}_{q\overline{q}\to {\rm ew}+g}^{(1) \rm,SLC}$ is equal to that of the original subleading-color amplitude. We thus have:
\begin{align}
    \mathcal{M}_{q\overline{q}\to {\rm ew}+g}^{(1) \rm,SLC}(p_1,p_2;p_3;k)\xrightarrow[]{k\parallel p_1}&
     \left( 1-\frac{C_A}{2C_F} \right) \includegraphics[scale=0.8,page=1,valign=c]{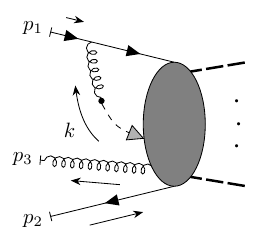} 
     + \includegraphics[scale=0.8,page=2,valign=c]{figures_standalone/tikz_RV_ampl_limits.pdf}  
     \, \nn\\
     =&  \left( 1-\frac{C_A}{2C_F} \right)\, \includegraphics[scale=0.8,page=3,valign=c]{figures_standalone/tikz_RV_ampl_limits.pdf} 
    + \includegraphics[scale=0.8,page=4,valign=c]{figures_standalone/tikz_RV_ampl_limits.pdf} \, \nn\\
    =& \frac{ig_s^2 }{2C_A}\, \frac{2(k+p_1)\cdot (-\chi)}{(-\chi\cdot k)\,k^2(k+p_1)^2} \,\mathcal{M}_{q\overline{q}\to {\rm ew}+g}^{(0)}(p_1,p_2;p_3)
    \label{eq:M_SLC_collp1_extleg}\,.
\end{align}
On the right-hand side of the first line, we have used the collinear approximation in eq.~\eqref{eq:coll_approx}, where the longitudinally polarized gluon attaches to all possible positions on the fermion line represented as a gray blob. Note that in the first diagram, we have excluded all diagrams containing self-energy corrections to the incoming leg. The Ward identity is then applied to obtain the second line, which is already in factorized form.\footnote{In the first diagram on the second line we have singled out an electroweak boson closest to the $p_1$ leg. We note that for single electroweak boson production in this diagram the gray blob should be thought of as a fermion line with only the external gluon $p_3$ attached.}
The auxiliary vector $\chi$ used for the collinear approximation appearing on the last line of eq.~\eqref{eq:M_SLC_collp1_extleg} can be chosen to take the specific value $p_2$, such that
\begin{align}
    \mathcal{M}_{q\overline{q}\to {\rm ew}+g}^{(1) \rm,SLC}(p_1,p_2;p_3;k) \xrightarrow[]{k\parallel p_1}   \frac{ig_s^2 }{2C_A}\, \frac{4(k+p_1)\cdot (-p_2)}{k^2(k+p_1)^2(k-p_2)^2} \,\mathcal{M}_{q\overline{q}\to {\rm ew}+g}^{(0)}(p_1,p_2;p_3)
    \label{eq:M_SLC_collp1_rewritten}\,.
\end{align}
Analogously, in the other collinear limit $k\parallel p_2$ (for which we choose $\chi=p_1$) and the soft limit, we find:
\begin{subequations}
    \begin{align}
        \mathcal{M}_{q\overline{q}\to {\rm ew}+g}^{(1) \rm,SLC}(p_1,p_2;p_3;k) &\xrightarrow[]{k\parallel p_2} \frac{i g_s^2 }{2C_A}\, \frac{4(k-p_2)\cdot p_1}{k^2(k+p_1)^2(k-p_2)^2} \,\mathcal{M}_{q\overline{q}\to {\rm ew}+g}^{(0)}(p_1,p_2;p_3) \,,
        \label{eq:M_SLC_collp2_extleg}\\
        \mathcal{M}_{q\overline{q}\to {\rm ew}+g}^{(1) \rm,SLC}(p_1,p_2;p_3;k) &\xrightarrow[]{k\to 0} \frac{i g_s^2 }{2C_A}\, \frac{4p_1\cdot(-p_2)}{k^2(k+p_1)^2(k-p_2)^2} \,\mathcal{M}_{q\overline{q}\to {\rm ew}+g}^{(0)}(p_1,p_2;p_3) \,.
        \label{eq:M_SLC_soft}    
    \end{align}
\end{subequations}
It is straightforward to verify that the subtraction $\mathcal{F}^{(1)}_{\rm SLC} (k)$, as defined in eq.~\eqref{eq:FV_SLC}, approximates all the aforementioned singularities in the corresponding infrared limits. This is a natural result, since in each limit, the singular part is identical to that of a one-loop form factor for $q\overline{q}\to {\rm H}$, whose expression (ignoring the color factor) is precisely eq.~\eqref{eq:FV_SLC}.

\subsubsection{Infrared limits of $\mathcal{M}_{q\overline{q}\to {\rm ew}+g}^{(1) \rm,LC}$}
\label{sec:coll_lim_M_LC}

The leading-color amplitude $\mathcal{M}_{q\overline{q}\to {\rm ew}+g}^{(1) \rm,LC}$, defined in eq.~\eqref{eq:M1LC}, has both initial-state singularities ($k-p_3\parallel p_1$ and $k\parallel p_2$) and final-state singularities ($k\parallel p_3$, $k\to 0$ and $k-p_3\to 0$). Below, we demonstrate how $\mathcal{M}_{q\overline{q}\to {\rm ew}+g}^{(1) \rm,LC}$ can be locally factorized in these infrared limits, with the infrared factor ${\cal F}^{(1)}_{\rm LC}$, as shown in eqs.\ \eqref{eq:M1IRapprox} and \eqref{eq:FV_LC}.

Let us start from the limit $k-p_3\parallel p_1$, where graphically we obtain
\begin{align}
\label{eq:LC_amplitude_collinear1_divergent_graphs}
    \mathcal{M}_{q\overline{q}\to {\rm ew}+g}^{(1) \rm,LC}(p_1,p_2;p_3;k) \xrightarrow[]{k-p_3\parallel p_1}&
    \includegraphics[scale=0.8,page=5,valign=c]{figures_standalone/tikz_RV_ampl_limits.pdf} +
    \includegraphics[scale=0.8,page=6,valign=c]{figures_standalone/tikz_RV_ampl_limits.pdf} \nn\\
    +& \includegraphics[scale=0.8,page=7,valign=c]{figures_standalone/tikz_RV_ampl_limits.pdf} + \frac{C_A}{2C_F}\includegraphics[scale=0.8,page=8,valign=c]{figures_standalone/tikz_RV_ampl_limits.pdf} \, ,
\end{align}
using the collinear approximation in eq.~\eqref{eq:coll_approx}.
The first three terms represent all the singular diagrams with triple-gluon vertices, where the scalar decomposition in eq.~(\ref{eq:Vggg}) has been applied to each of them. Note that in the second diagram (a set of $(p_3-k)$-ghost terms) there is always an external colorless particle adjacent to the gluon $p_3-k$, since we have excluded diagrams with external vertex functions that generate loop polarizations, in this case the $p_3$-scalar diagrams. (These are the second and third diagrams on the right-hand side of eq.~\eqref{eq:M1LC} from ${\cal M}_{q\bar q \to {\rm ew}+g}^{(1),{\rm LC}}$.)
The second and third terms above are both $(p_3-k)$-ghost terms, as explained in eq.~\eqref{eq:scalar_to_ghost}. 
Note that the third set of diagrams in the equation above includes the $(p_3-k)$-ghost diagram where the two virtual gluons attach next to each other on the incoming quark leg, forming a vertex correction. The other $(p_3-k)$-ghost diagram of this vertex correction is not present in the leading-color amplitude, since it is accounted for in the loop-polarization part of the amplitude as a $p_3$-scalar diagram (and is zero in this limit). We can however show that in the limit $k-p_3=-x p_1$ the single diagram from the third set still gives us both ghost terms, due to
\begin{align}
    \includegraphics[scale=0.8,page=25,valign=c]{figures_standalone/tikz_RV_Ward.pdf} 
    &\xrightarrow[]{k-p_3\parallel p_1}
    ig_s^3\frac{C_A}{2} T^a \, \frac{2(1-x)}{x}  \frac{\epsilon_{3\alpha}^*\left(p_3^\mu p_3^\alpha - k^\mu k^\alpha\right)(\slashed{p}_1-\slashed{p}_3)\gamma_\mu u(p_1)}{k^2(p_3-k)^2(p_1-p_3+k)(p_1-p_3)^2}   \nn\\
     &= \includegraphics[scale=0.8,page=26,valign=c]{figures_standalone/tikz_RV_Ward.pdf} + \includegraphics[scale=0.8,page=27,valign=c]{figures_standalone/tikz_RV_Ward.pdf} \, ,
     \label{eq:ghost_term}
\end{align}
by making use of $\slashed{p}_1 u(p_1)=0$ and $k-p_3=-x p_1$. All the ghost diagrams where the longitudinal ghost points into the external gluon, as in the first term on the last line of eq.~\eqref{eq:ghost_term}, do not contribute to this limit due to the vanishing factor $\epsilon_3^* \cdot p_3$. Therefore, only the term with the longitudinal ghost entering the fermion line remains. Summing up all these diagrams and using the fermionic diagrammatic Ward identity in eq.\ \eqref{eq:Wardid_qqg_eq}, we obtain:
\begin{align}
    &  \hspace{-1cm}\mathcal{M}_{q\overline{q}\to {\rm ew}+g}^{(1) \rm,LC}(p_1,p_2;p_3;k) \nn\\
    \xrightarrow[]{k
    -p_3\parallel p_1} &
    \includegraphics[scale=0.8,page=5,valign=c]{figures_standalone/tikz_RV_ampl_limits.pdf}
    + \includegraphics[scale=0.8,page=9,valign=c]{figures_standalone/tikz_RV_ampl_limits.pdf} + \frac{C_A}{2C_F}\, \includegraphics[scale=0.8,page=8,valign=c]{figures_standalone/tikz_RV_ampl_limits.pdf} \nn\\
    =&
    \includegraphics[scale=0.8,page=10,valign=c]{figures_standalone/tikz_RV_ampl_limits.pdf}
    +\includegraphics[scale=0.8,page=12,valign=c]{figures_standalone/tikz_RV_ampl_limits.pdf}
    +\includegraphics[scale=0.8,page=13,valign=c]{figures_standalone/tikz_RV_ampl_limits.pdf}
   \nn\\
    & + 
    \frac{C_A}{2C_F}\,\left\{\includegraphics[scale=0.8,page=11,valign=c]{figures_standalone/tikz_RV_ampl_limits.pdf}\ +\includegraphics[scale=0.8,page=14,valign=c]{figures_standalone/tikz_RV_ampl_limits.pdf}\right\}\, .
    \label{eq:M_LC_collp1}
\end{align}
Here, in the second relation we explicitly split the contribution from first diagram of the first relation into two components on the right-hand side of the equation, displaying explicitly  the contribution from the vertex diagram.
The third term on the right of the equality, which arises from the $p_3-k$-ghost contributions, is finite in the limit $k-p_3\parallel p_1$, since there are not enough propagators to pinch this singularity. The first and fourth terms have opposite signs to each other (note they have the same color factor) thus cancel algebraically, leaving only the second and fifth terms in this infrared limit, which are locally factorized:
\begin{align}
    &\mathcal{M}_{q\overline{q}\to {\rm ew}+g}^{(1) \rm,LC}(p_1,p_2;p_3;k) \nn\\
    &\xrightarrow[]{k-p_3\parallel p_1} \includegraphics[scale=0.8,page=12,valign=c]{figures_standalone/tikz_RV_ampl_limits.pdf} +
    \frac{C_A}{2C_F}\,\includegraphics[scale=0.8,page=14,valign=c]{figures_standalone/tikz_RV_ampl_limits.pdf}\,\nn\\
    =&\, \frac{ig_s^2\, C_A }{2}\, \frac{2(p_1-p_3+k)\cdot \chi}{(\chi\cdot (p_3-k))\,(p_3-k)^2(p_1-p_3+k)^2} \,\mathcal{M}_{q\overline{q}\to {\rm ew}+g}^{(0)}(p_1,p_2;p_3)\nn\\
    =& -ig_s^2\, C_A \, \frac{2\,(p_1+k)\cdot p_3}{k^2 (p_3-k)^2(p_1-p_3+k)^2} \,\mathcal{M}_{q\overline{q}\to {\rm ew}+g}^{(0)}(p_1,p_2;p_3)
    \label{eq:M_LC_collp1_rewritten}\, ,
\end{align}
where in the last equality we have chosen the auxiliary vector of the collinear approximation, eq.\ \eqref{eq:coll_approx}, $\chi$ to be $p_3$. The resulting expression gives precisely the same infrared limit as in ${\cal F}_{\rm LC}^{(1)}$, eq.\ \eqref{eq:FV_LC}  when $k-p_3$ becomes parallel to $p_1$. 

Analogously, in the limit $k\parallel p_2$ we obtain
\begin{align}
    \mathcal{M}_{q\overline{q}\to {\rm ew}+g}^{(1) \rm,LC}(p_1,p_2;p_3;k) &\xrightarrow[]{k\parallel p_2} - ig_s^2\, C_A\, \frac{2(p_2-k)\cdot p_3}{k^2(p_3-k)^2(k-p_2)^2} \,\mathcal{M}_{q\overline{q}\to {\rm ew}+g}^{(0)}(p_1,p_2;p_3)
    \label{eq:M_LC_collp2_extleg}\, .
\end{align}
Again, this expression gives the same infrared limit as found from the form factor, eq.\ \eqref{eq:FV_LC} for $k$ parallel to $p_2$.

Next we consider the final-state singularities of $\mathcal{M}_{q\overline{q}\to {\rm ew}+g}^{(1) \rm,LC}$, starting with the limit $k\parallel p_3$. The diagrams that are singular in this limit are
\begin{align}
    &\hspace{-1cm}\mathcal{M}_{q\overline{q}\to {\rm ew}+g}^{(1) \rm,LC}(p_1,p_2;p_3;k)\nn\\
      \xrightarrow[]{k\parallel p_3} &
     \includegraphics[scale=0.7,page=15,valign=c]{figures_standalone/tikz_RV_ampl_limits.pdf}  
    + \includegraphics[scale=0.7,page=16,valign=c]{figures_standalone/tikz_RV_ampl_limits.pdf}  
    =  \includegraphics[scale=0.7,page=17,valign=c]{figures_standalone/tikz_RV_ampl_limits.pdf}  
    + \includegraphics[scale=0.7,page=18,valign=c]{figures_standalone/tikz_RV_ampl_limits.pdf} \nonumber\\
    =& \frac{iC_A\, g_s^2}{2} \left( \frac{(2p_3-k)\cdot \chi_1}{k\cdot \chi_1}+\frac{(p_3+k)\cdot \chi_2}{(p_3-k)\cdot \chi_2}\right)\frac{1}{k^2(p_3-k)^2} \mathcal{M}_{q\overline{q}\to {\rm ew}+g}^{(0)}(p_1,p_2;p_3)
    \label{eq:M_LC_collp3}\,.
\end{align}
The first equality results from the tree-level Ward identity, from which the gluons collinear to $p_3$ can be clearly factorized from the remaining hard scattering. The second equality manifests this local factorization. By choosing $\chi_1=p_2$ and $\chi_2=p_1$, we have
\begin{align} 
     &\mathcal{M}_{q\overline{q}\to {\rm ew}+g}^{(1) \rm,LC}(p_1,p_2;p_3;k) \nn\\
     &\xrightarrow[]{k\parallel p_3}-iC_A\, g_s^2\, \left( \frac{(2p_3-k)\cdot p_2}{(k-p_2)^2}+\frac{(p_3+k)\cdot p_1}{(p_1-p_3+k)^2}\right)\frac{1}{k^2(p_3-k)^2} \mathcal{M}_{q\overline{q}\to {\rm ew}+g}^{(0)}(p_1,p_2;p_3)
    \label{eq:M_LC_collp3_rewritten}\, ,
\end{align}
again matching the infrared behavior of the form factor \eqref{eq:FV_LC} in this limit.

Finally, we can check the soft limits:
\begin{subequations}
    \begin{align}
        \mathcal{M}_{q\overline{q}\to {\rm ew}+g}^{(1) \rm,LC}(p_1,p_2;p_3;k) &\xrightarrow[]{k\to 0} \frac{-iC_A\, g_s^2}{2} \frac{4p_2\cdot p_3}{k^2(p_3-k)^2(k-p_2)^2} \mathcal{M}_{q\overline{q}\to {\rm ew}+g}^{(0)}(p_1,p_2;p_3)\,,
        \label{eq:M_LC_softk} \\
        \mathcal{M}_{q\overline{q}\to {\rm ew}+g}^{(1) \rm,LC}(p_1,p_2;p_3;k) &\xrightarrow[]{k-p_3\to 0} \frac{-iC_A\, g_s^2}{2} \frac{4p_1\cdot p_3}{k^2(p_3-k)^2(p_1-p_3+k)^2} \mathcal{M}_{q\overline{q}\to {\rm ew}+g}^{(0)}(p_1,p_2;p_3)\, .
        \label{eq:M_LC_softkprime}
    \end{align}
\end{subequations}
Note that the results above can also be obtained by taking the soft limits on top of their ``nested'' collinear limits. This elegant infrared structure allows us to employ a single subtraction term for all the infrared singularities of $\mathcal{M}_{q\overline{q}\to {\rm ew}+g}^{(1),\rm LC}$, precisely the factor $\mathcal{F}^{(1)}_{\rm LC}$ defined in eq.~\eqref{eq:FV_LC}, as we have verified above directly in all the single-virtual infrared limits.

\subsection{\texorpdfstring{Cross section at order $\alpha_s$}{Cross section at order alphas}}
\label{sec:IRdemo_NLO}

In this section we aim to give a more detailed demonstration of the local factorization of the cross section at order $\alpha_s$ already described in sec.\ \ref{sec:NLOxsec_localfact}. We shall examine the single-real singularities of $\sigma_{q\bar q \to {\rm ew}+g}^{(0)}(p_1,p_2;p_3)$ diagrammatically and see how they can be factorized and subtracted at the integrand level. For the local factorization of the one-loop amplitude contributing to the virtual correction cross section we refer to ref. \cite{Anastasiou:2022eym}.

A single-real singularity arises when $p_3$ is either soft or collinear to $p_1$ or $p_2$. To demonstrate its local factorization in these limits, we consider the limit of $p_3\parallel p_1$, where we have
\begin{align}
    &\sigma_{q\bar q \to {\rm ew}+g}^{(0)}(p_1,p_2;p_3) = \includegraphics[scale=0.75,valign=c,page=1]{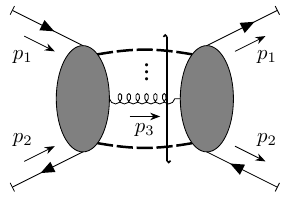}\nn\\
    &\xrightarrow[]{p_3=xp_1}\includegraphics[scale=0.75,valign=c,page=2]{figures_standalone/tikz_RV_xsces_NLO.pdf} 
    +\includegraphics[scale=0.75,valign=c,page=3]{figures_standalone/tikz_RV_xsces_NLO.pdf}
    +\includegraphics[scale=0.75,valign=c,page=4]{figures_standalone/tikz_RV_xsces_NLO.pdf}\nn\\
    &\qquad = \includegraphics[scale=0.75,valign=c,page=2]{figures_standalone/tikz_RV_xsces_NLO.pdf} + 
    \includegraphics[scale=0.75,valign=c,page=5]{figures_standalone/tikz_RV_xsces_NLO.pdf} 
    + \includegraphics[scale=0.75,valign=c,page=6]{figures_standalone/tikz_RV_xsces_NLO.pdf}\nn\\
    &\qquad = -C_F\, g_s^2 \frac{1}{2p_1\cdot p_3} \left( \frac{2(1-\epsilon)x}{x-1} -\frac{4}{x}\right) \sigma_{q\bar q \to {\rm ew}}^{(0)}\left((1-x)p_1,p_2\right) \, .    
    \label{eq:xsec_0L_Sg_p3p1limit}
\end{align}
In the pictorial representation, we sum over all possible attachments of the colorless particles, that is, all permutations and crossings. Note that we include the infrared-safe observable $\mathcal{O}$ and the on-shell condition for the $n$-th electroweak boson, present in the definition of the local cross section in eq.~\eqref{eq:local_xsec_R}, implicitly in this cut diagram notation. We adopt this convention for the remainder of the paper. In the graphical depiction above, we have explicitly distinguished the first term on the second and third lines, in which the gluon attaches directly to the incoming quark line on both sides of the cut. The two quark propagators $\frac{1}{p_1\cdot p_3}$ seem to produce a power-like singularity in this limit, that gets suppressed by the numerator, such that the singularity is in fact only logarithmic, as explained in more detail in appendix~\ref{appendix-power_counting}. Moreover, this term is already factorized in the limit $p_3\parallel p_1$ and does not participate in the Ward identity cancellations with the remaining terms. For these we have applied the collinear approximation defined in eq.~\eqref{eq:coll_approx} with the arrow attached to all possible positions on the fermion line (gray blob). Note that the second diagram in the second line of eq.~\eqref{eq:xsec_0L_Sg_p3p1limit} also exists for single electroweak boson production, for which the blob on the right hand side is just a fermion line with the longitudinal gluon $p_3$ attached to it. Similarly for the other diagrams where one electroweak boson is singled out and attached next to the incoming quark $p_1$.
The Ward identity can then be employed to obtain the third line, which is already in a factorized form. Specifically, $\sigma_{q\bar q \to {\rm ew}+g}^{(0)}$ can be expressed as the product of a scalar function independent of the electroweak particles and the lower-order cross section $\sigma_{q\bar q \to {\rm ew}}^{(0)}$. Note that $\sigma_{q\bar q \to {\rm ew}}^{(0)}$ depends on the variables $(1-x)p_1$ and $p_2$, which correspond to $k_1$ and $k_2$, defined in eq.~\eqref{eq:k1k2_param}, in the limit $p_1\parallel p_3$. By using the Ward identity, one can show that the cross section of single-Higgs production in association with an external gluon, leads to the same infrared singular structure as in eq.~\eqref{eq:xsec_0L_Sg_p3p1limit}, thus differing from $\sigma_{q\bar q \to {\rm ew}+g}^{(0)}(p_1,p_2;p_3)$ only by a hard factor. With this observation, we can isolate the singularities in the factor 
\begin{align}
    &\frac{\sigma_{q\bar q \to {\rm H}+g}^{(0)}(p_1,p_2;p_3)}{\sigma_{q\bar q \to {\rm H}}^{(0)}(k_1,k_2)}\nn\\ 
    &= 
    \frac{
    \includegraphics[scale=0.7,valign=c,page=1]{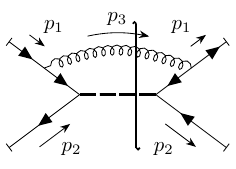} 
    + \includegraphics[scale=0.7,valign=c,page=2]{figures_standalone/tikz_RV_xsces_LO.pdf} 
    + \includegraphics[scale=0.7,valign=c,page=3]{figures_standalone/tikz_RV_xsces_LO.pdf}
    + \includegraphics[scale=0.7,valign=c,page=4]{figures_standalone/tikz_RV_xsces_LO.pdf}
    }{
    \includegraphics[scale=0.7,valign=c,page=5]{figures_standalone/tikz_RV_xsces_LO.pdf} }\,.
\label{eq:FR}
\end{align}
A direct evaluation yields 
\begin{align}
\label{eq:FR_direct_evaluation}
    \frac{\sigma_{q\bar q \to {\rm H}+g}^{(0)}(p_1,p_2;p_3)}{\sigma_{q\bar q \to {\rm H}}^{(0)}(k_1,k_2)} &= \frac{C_F\, g_s^2\, \xi}{p_1\cdot p_2} \left( (1-\epsilon)\, \left(\frac{p_2\cdot p_3}{p_1\cdot p_3} + \frac{p_1\cdot p_3}{p_2\cdot p_3} + 2\right) + \frac{2 \left(p_1\cdot p_2 \right)^2}{\xi\,  p_1\cdot p_3\,  p_2\cdot p_3} \right)\frac{{\cal O}_{{\rm H}+1}}{{\cal O}_H} \, ,
\end{align}
where
\begin{align}
    \xi \equiv \frac{p_1 \cdot p_2}{p_1 \cdot p_2-p_1 \cdot p_3 -p_2 \cdot p_3}\, .
\end{align}
In the limit $p_3\parallel p_1$ it takes the form
\begin{align}
    \frac{\sigma_{q\bar q \to {\rm H}+g}^{(0)}(p_1,p_2;p_3)}{\sigma_{q\bar q \to {\rm H}}^{(0)}(k_1,k_2)}\xrightarrow[]{p_3\parallel p_1} -
    C_F\, g_s^2\,\frac{1}{2p_1\cdot p_3} \left(  \frac{2(1-\epsilon)x}{x-1}  - \frac{4}{x}\right) \, ,
\end{align}
and thus matches the singular part on the last line of eq.~\eqref{eq:xsec_0L_Sg_p3p1limit}. Similarly, all other infrared singularities can be approximated by this factor. Therefore, the following expression
\begin{align}
    \sigma_{q\bar q \to {\rm ew}+g}^{(0)}(p_1,p_2;p_3) - \frac{\sigma_{q\bar q \to {\rm H}+g}^{(0)}(p_1,p_2;p_3)}{\sigma_{q\bar q \to {\rm H}}^{(0)}(k_1,k_2)} \sigma_{q\bar q \to {\rm ew}}^{(0)}(k_1,k_2)\, 
\end{align}
is locally finite in all infrared limits and corresponds exactly to the right-hand side of eq.~\eqref{eq:xsec_0L_IRfinite_NLO}, thereby defining $\sigma_{q\bar q \to {\rm ew}+g}^{{(0)} \rm ,IR-finite}(p_1,p_2;p_3)$.

\subsection{\texorpdfstring{Cross section at order $\alpha_s^2$: single-virtual singularities}{Cross section at order alphaS2: single-virtual singularities}}
\label{sec:IRdemo_RVcrsec_singleV}

As stated in section~\ref{sec:IRapprox_RVamplitude} and demonstrated in section~\ref{sec:demo_IR_SLC_ampl}, the subleading-color amplitude integrand for the production of an electroweak multi-particle state ``ew" and a gluon can be approximated in all virtual infrared limits by 
\begin{align}
    {\cal M}_{q\bar q \to {\rm ew}+g}^{(1), {\rm SLC}}(p_1,p_2;p_3;k) \overset{\textup{IR}}{\approx} \mathcal{F}^{(1)}_{\rm SLC} (p_1,p_2;k) {\cal M}_{q\bar q \to {\rm ew}+g}^{(0)}(p_1,p_2;p_3)\, .
\end{align}
This also holds at the cross-section level after interference with the tree-level amplitude ${\cal M}_{q\bar q \to {\rm ew}+g}^{(0)}$:
\begin{align}
    \sigma_{q\bar q \to {\rm ew}+g}^{(1), {\rm SLC}}(p_1,p_2;p_3;k) \overset{\textup{IR}}{\approx} 2 \text{Re}\left(\mathcal{F}^{(1)}_{\rm SLC} (p_1,p_2;k)\right)
    \,\sigma_{q\bar q \to {\rm ew}+g}^{(0)}(p_1,p_2;p_3)\, .
\end{align}
Since this is a statement for an arbitrary number of colorless particles, single-Higgs production exhibits the same infrared behavior. Moreover, the singular part of the subleading-color amplitude is proportional to that of a one-loop form factor for $q\bar q \to {\rm H}$, when changing the color factor $C_F$ to $- \frac{1}{2C_A}$:
\begin{align}
    2 \text{Re}\left(\mathcal{F}^{(1)}_{\rm SLC} (p_1,p_2;k)\right) \overset{\textup{IR}}{\approx}  
     \frac{\sigma_{q\bar q \to {\rm H}+g}^{(1), {\rm SLC}}(p_1,p_2;p_3;k) }{\sigma_{q\bar q \to {\rm H}+g}^{(0)}(p_1,p_2;p_3)\, }\overset{\textup{IR}}{\approx}
    \frac{\sigma_{q\bar q \to {\rm H}}^{(1), {\rm SLC}}(p_1,p_2;k) }{\sigma_{q\bar q \to {\rm H}}^{(0)}(p_1,p_2)}
    \,,
\end{align}
where  $\sigma_{q\bar q \to {\rm H}}^{(1), {\rm SLC}}(p_1,p_2;k)$ is defined in eq. \eqref{eq:xsec_1L_H_SLC}.
Therefore, the subtraction
\begin{align}
\label{eq:NNLO_cross_section_single_virtual_subtraction_SLC}
    \sigma_{q\bar q \to {\rm ew}+g}^{(1), {\rm SLC}}(p_1,p_2;p_3;k) -  \frac{\sigma_{q\bar q \to {\rm H}}^{(1), {\rm SLC}}(p_1,p_2;k)}{\sigma_{q\bar q \to {\rm H}}^{(0)}(p_1,p_2)} \sigma_{q\bar q \to {\rm ew}+g}^{(0)}(p_1,p_2;p_3)
\end{align}
removes the single-virtual singularities of the cross section completely, as presented in eq.~\eqref{eq:xsec_IRfinite_SLC}. 

Analogously, we can remove the virtual singularities in $k$ from the leading-color contribution using the single-Higgs production in association with a gluon in the final state:
\begin{align}
\label{eq:NNLO_cross_section_single_virtual_subtraction_LC}
    \sigma_{q\bar q \to {\rm ew}+g}^{(1), {\rm LC}}(p_1,p_2;p_3;k) -  \frac{\sigma_{q\bar q \to {\rm H}+g}^{(1), {\rm LC}}(p_1,p_2;p_3;k)}{\sigma_{q\bar q \to {\rm H}+g}^{(0)}(p_1,p_2;p_3)} \sigma_{q\bar q \to {\rm ew}+g}^{(0)}(p_1,p_2;p_3)\,.
\end{align}
This subtraction not only removes the initial-state but also the final-state singularities, as presented in eq.~\eqref{eq:xsec_IRfinite_LC}.

In contrast to the amplitude-level subtraction demonstrated in section~\ref{sec:IRdemo_RVampl}, a key difference arises at the cross-section level.
In $\mathcal{M}_{q\overline{q}\to {\rm ew}+g}^{}$, the final-state gluon is restricted to physical polarizations only, i.e., those transverse to its momentum. Here, instead, we allow for unphysical polarizations of the external gluon. This difference introduces an additional subtlety in the local factorization analysis. Recall that below eq.~(\ref{eq:LC_amplitude_collinear1_divergent_graphs}) we showed that the $(p_3-k)$-ghost term with the ghost line entering the external gluon is suppressed in the $k-p_3\parallel p_1$ limit by the factor $\epsilon_3\cdot p_3$. At the cross-section level, however, this factor can be $\mathcal{O}(1)$, since $\epsilon_3$ is not restricted to be physical; thus, these ghost terms do not vanish individually. Nevertheless, an $\mathcal{O}(1)$ value of $\epsilon_3\cdot p_3$ implies that the $p_3$ gluon is longitudinally polarized and contracts with the tree-level. Consequently the sum over all terms locally cancel by virtue of the tree-level Ward identity:
\begin{align}
    \includegraphics[scale=0.8,page=1,valign=c]{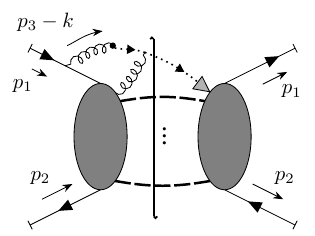}
    +\includegraphics[scale=0.8,page=2,valign=c]{figures_standalone/tikz_RV_xsecs_NNLO_LC.pdf} 
   = 0\, ,
   \label{eq:xsec_ghost_virtlim}
\end{align}
where all possible attachments of the arrow to the fermion line are summed over. As a result, eq.~\eqref{eq:NNLO_cross_section_single_virtual_subtraction_LC} suffices to remove all the single-virtual singularities for the leading-color cross section.

An analogous subtlety arises in $\sigma_{q\overline{q}\to {\rm ew}+g}^{(1), \rm lp}$. From eq.~\eqref{eq:Jcurrent_perp_tilde} or eq.~\eqref{eq:Jcurrent_perp_tilde_alt}, the function $\widetilde{J}_\perp^{\mu,a}$ is free of infrared singularities in the limits $k\parallel p_1$, $k\parallel p_2$, and $k\to 0$. In the limit $k\parallel p_3$, however, each individual cut diagram features a logarithmic divergence, where $\widetilde{J}_\perp^{\mu,a}$ is proportional to $p_3^\mu$, rendering the final-state gluon longitudinally polarized. Summing over all possible attachments of $p_3$ on the tree-level side of the cut, these singularities cancel due to the Ward identity. Pictorially,
\begin{align}
  \sigma_{q\overline{q}\to {\rm ew}+g}^{(1) \rm, lp}(p_1,p_2;p_3;k) \xrightarrow[]{k\parallel p_3}&  \includegraphics[width = 0.3\textwidth,valign=c,page=1]{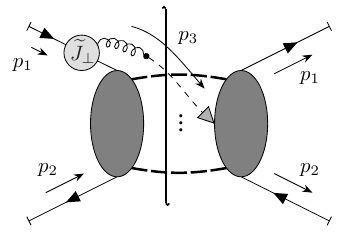}+ \includegraphics[width = 0.3\textwidth,valign=c,page=2]{figures_standalone/tikz_RV_xsec_pol.pdf} \nn\\
   &+\includegraphics[width = 0.3\textwidth,valign=c,page=3]{figures_standalone/tikz_RV_xsec_pol.pdf}
   +\includegraphics[width = 0.3\textwidth,valign=c,page=4]{figures_standalone/tikz_RV_xsec_pol.pdf} \nn\\
   &= 0\, .
   \label{eq:xsec_Jperp_virtlim}
\end{align}
Hence, no single-virtual subtractions are required for $\sigma_{q\overline{q}\to {\rm ew}+g}^{(1) \rm, lp}$.

\subsection{\texorpdfstring{Cross section at order $\alpha_s^2$: single-real singularities}{Cross section at order alphaS2: single-real singularities}}
\label{sec:IRdemo_RVcrsec_singleR}

In this section we aim to demonstrate the local factorization of the single-real singularities in the subleading-color, leading-color, and loop-polarization cross sections respectively, and how the corresponding local subtractions, as shown in eqs.~\eqref{eq:xsec_IRfinite_SLC}, \eqref{eq:xsec_IRfinite_LC}, and \eqref{eq:xsec_IRfinite_pol} suffice to remove them.

\subsubsection{Single-real singularities of $\boldsymbol{\sigma_{q\bar q \to {\rm ew}+g}^{(1), {\rm SLC}}}$}

The subleading-color real–virtual cross section $\sigma_{q\bar q \to {\rm ew}+g}^{(1), {\rm SLC}}$ is defined as the subleading-color contribution to the full local cross section $\sigma_{q\bar q \to {\rm ew}+g}^{(1)}$ presented in eq.~\eqref{eq:local_xsec_RV}, with the proper modification due to the self-energy and vertex correction adjacent to the incoming lines.

In what follows, we detail the local factorization of $\sigma_{q\bar q \to {\rm ew}+g}^{(1), {\rm SLC}}$ in the limit $p_3\parallel p_1$. The remaining limits can be treated analogously. The cut diagrams contributing to the infrared singularity in this limit are:
\begin{align}
    &\sigma_{q\bar q \to {\rm ew}+g}^{(1), {\rm SLC}}(p_1,p_2;p_3;k) \xrightarrow[]{p_3\parallel p_1}\nn\\
    &\left( 1-\frac{C_A}{2C_F} \right) \left(\includegraphics[scale = 0.7,page=1,valign=c]{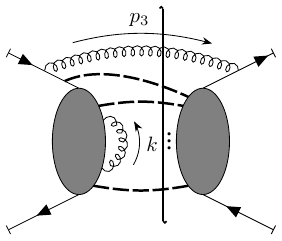}
    + \includegraphics[scale = 0.7,page=2,valign=c]{figures_standalone/tikz_RV_xsecs_NNLO_SLC.pdf} \right)
    + \includegraphics[scale = 0.7,page=3,valign=c]{figures_standalone/tikz_RV_xsecs_NNLO_SLC.pdf}\nn\\
    &+\left( 1-\frac{C_A}{2C_F} \right) \left(
     \includegraphics[scale = 0.7,page=4,valign=c]{figures_standalone/tikz_RV_xsecs_NNLO_SLC.pdf}+ 
      \includegraphics[scale = 0.7,page=5,valign=c]{figures_standalone/tikz_RV_xsecs_NNLO_SLC.pdf}\right) +  \includegraphics[scale = 0.7,page=6,valign=c]{figures_standalone/tikz_RV_xsecs_NNLO_SLC.pdf}\nn\\
       &+\includegraphics[scale = 0.7,page=26,valign=c]{figures_standalone/tikz_RV_xsecs_NNLO_SLC.pdf}
    \quad+ \text{c.c.}
    \label{eq:xsec_SLC_limitp3p1}
\end{align}
Each cut diagram displayed is of the form $\mathcal{M}^{(1),\rm SLC}_{q\overline{q}\to {\rm ew}+ g} \cdot \left( \mathcal{M}^{(0),\rm SLC}_{q\overline{q}\to {\rm ew}+ g}\right)^*$, with the complex conjugate implied by the notation ``c.c.''. We omit the incoming particle momenta as well as the loop momentum label $k$ in the subleading-color one-loop amplitude, defined in eq.~\eqref{eq:M_1loop_SLC_summarized}, for legibility. The first line contains cut diagrams where the gluon $p_3$ acts as an internal jet-1 propagator, i.e. it does not attach to any hard vertex, and hence are locally factorized. Moreover they are logarithmically divergent, due to a numerator suppression explained in more detail in appendix~\ref{appendix-power_counting}. The second line contains cut diagrams in which $p_3$ attaches to a hard vertex on the right-hand side of the cut. The third line summarizes all cut diagrams in which the longitudinally polarized gluon $p_3$ instead attaches to the hard subleading-color one-loop amplitude on the left-hand side of the cut. The collinear approximation in eq.~\eqref{eq:coll_approx} applies to all cut diagrams in the second and third lines.

We now turn to the demonstration of local factorization of $\sigma_{q\bar q \to {\rm ew}+g}^{(1), {\rm SLC}}$. In the first line, the real gluon factorizes trivially from the hard process. In the second line, the tree-level Ward identity can be applied to the fermion line on the right-hand-side, where all possible attachments of $p_3$ are summed over, yielding
\begin{align}
    \left( 1-\frac{C_A}{2C_F} \right) \left(
    \includegraphics[scale = 0.7,page=17,valign=c]{figures_standalone/tikz_RV_xsecs_NNLO_SLC.pdf}
     + \includegraphics[scale = 0.7,page=18,valign=c]{figures_standalone/tikz_RV_xsecs_NNLO_SLC.pdf} \right)
     + \includegraphics[scale = 0.7,page=19,valign=c]{figures_standalone/tikz_RV_xsecs_NNLO_SLC.pdf}\, ,
\end{align}
which are locally factorized. Note that, for double or single electroweak boson production, one or two of the thick dashed lines exchanged between the gray blobs may be absent. This convention applies to all subsequent diagrams in this article as well. In the case of single electroweak boson production the first cut diagram is absent.

The remaining diagrams summarized in the cut diagram on the last line also factorize via the Ward identity, though the situation is slightly more subtle. In general, applying the Ward identity directly to a loop diagram produces a ``shift mismatch''~\cite{Anastasiou:2020sdt,Anastasiou:2022eym,Anastasiou:2024xvk} as we have mentioned in section~\ref{sec:RVxsec_localfact}: this vanishes after integration but persists at the integrand level, thereby obstructing local factorization. At first sight, this seems to be exactly the case here, since the fermion line on the left-hand-side, where all possible attachments of $p_3$ should be summed over, is a one-loop amplitude. However, the propagators to which $p_3$ can attach, while preserving the subleading-color structure, form a tree rather than a loop.
This can be seen in the definition of the subleading-color amplitude in eq. \eqref{eq:M1SLC}. Consequently, the Ward identity remains effectively at tree level in this context, and local factorization is still preserved. Let us see this through an explicit example for two external colorless particles:
\begin{align}
\label{eq:Ward_identity_equivalent_tree_SLC}
    \sum_{\circ}\ \raisebox{0.4cm}{\includegraphics[scale=0.8,page=1,valign=c]{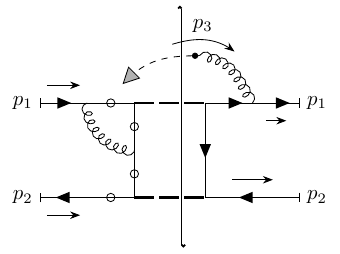}}=\raisebox{0.4cm}{\includegraphics[scale=0.8,page=3,valign=c]{figures_standalone/tikz_RV_xsec_WI.pdf}}\,.
\end{align}
Here we sum over different subleading-color contributions in which the longitudinally polarized gluon $p_3$ attaches to the left-hand side of the cut, with the attachments denoted by ``$\circ$''. The cancellations within this sum clearly follow from the tree-level Ward identity rather than the loop-level one, thus local factorization is guaranteed. A more detailed analysis for the cut diagram on the last line in eq. \eqref{eq:xsec_SLC_limitp3p1} is provided in appendix~\ref{appendix-Ward_identity_cancellations_single_real_SLC}. We report our findings here directly:
\begin{align}
    \includegraphics[scale = 0.7,page=26,valign=c]{figures_standalone/tikz_RV_xsecs_NNLO_SLC.pdf} & = \left( 1-\frac{C_A}{2C_F} \right) \left(
    +\includegraphics[scale = 0.7,page=25,valign=c]{figures_standalone/tikz_RV_xsecs_NNLO_SLC.pdf}
    \right)
    +\includegraphics[scale = 0.7,page=20,valign=c]{figures_standalone/tikz_RV_xsecs_NNLO_SLC.pdf}\, .
    \label{eq:M1L_SLC_p3lim_general_factorized}
\end{align}
With the analysis above, $\sigma_{q\bar q \to {\rm ew}+g}^{(1), {\rm SLC}}$ can be rewritten into the following locally factorized form:
\begin{align}
\label{eq:xsec_SLC_limitp3p_final_factorization}
     &\sigma_{q\bar q \to {\rm ew}+g}^{(1), {\rm SLC}}(p_1,p_2;p_3;k) \xrightarrow[]{p_3\parallel p_1}\nn\\
    &\left( 1-\frac{C_A}{2C_F} \right) \left(
    \includegraphics[scale = 0.7,page=1,valign=c]{figures_standalone/tikz_RV_xsecs_NNLO_SLC.pdf}
    + \includegraphics[scale = 0.7,page=2,valign=c]{figures_standalone/tikz_RV_xsecs_NNLO_SLC.pdf}
    + \includegraphics[scale = 0.7,page=17,valign=c]{figures_standalone/tikz_RV_xsecs_NNLO_SLC.pdf}\right.\nn\\
    &\hspace{2.75cm}\left.
    +\includegraphics[scale = 0.7,page=18,valign=c]{figures_standalone/tikz_RV_xsecs_NNLO_SLC.pdf}
    +\includegraphics[scale = 0.7,page=25,valign=c]{figures_standalone/tikz_RV_xsecs_NNLO_SLC.pdf}
    \right)
    +\includegraphics[scale = 0.7,page=20,valign=c]{figures_standalone/tikz_RV_xsecs_NNLO_SLC.pdf}
   \nn\\
    &\hspace{0.5cm}
    + \includegraphics[scale = 0.7,page=3,valign=c]{figures_standalone/tikz_RV_xsecs_NNLO_SLC.pdf}
    + \includegraphics[scale = 0.7,page=19,valign=c]{figures_standalone/tikz_RV_xsecs_NNLO_SLC.pdf}
    \quad + \text{c.c.}
\end{align}
By using $p_3=xp_1$ to parameterize this limit as detailed in eq.~\eqref{eq:collapprox_l}, we can extract the infrared singularity as an $x$-dependent factor of the hard scattering. Namely, 
\begin{align}
&\sigma_{q\bar q \to {\rm ew}+g}^{(1), {\rm SLC}}(p_1,p_2;p_3;k)\xrightarrow[]{p_3=xp_1} \nn\\
    & -\left(C_F -\frac{C_A}{2} \right) \,g_s^2 \frac{1}{2p_1\cdot p_3} \left( \frac{2(1-\epsilon)x}{x-1}  -\frac{4}{x}\right) \sigma_{q\bar q \to {\rm ew}}^{(1)}\left((1-x)p_1,p_2;k\right) 
    \nn\\
    &\hspace{1cm}
    + \frac{\left( \includegraphics[scale = 0.7,page=22,valign=c]{figures_standalone/tikz_RV_xsecs_NNLO_SLC.pdf} + \includegraphics[scale = 0.7,page=23,valign=c]{figures_standalone/tikz_RV_xsecs_NNLO_SLC.pdf} + \text{c.c.}\right)}{\includegraphics[scale = 0.7,page=24,valign=c]{figures_standalone/tikz_RV_xsecs_NNLO_SLC.pdf}}
    \includegraphics[scale = 0.7,page=21,valign=c]{figures_standalone/tikz_RV_xsecs_NNLO_SLC.pdf}\, .
    \label{eq:xsec_SLC_limitp3p1_simplified} 
\end{align}
The expression above can be separated into two parts. The first part on the second line corresponds to the sub-leading color part of the cross section $\sigma^{(0)}_{q\bar{q} \to {\rm H}+g}$ divided by the leading order $\sigma^{(0)}_{q\bar{q} \to {\rm H}}$(see eq.~\eqref{eq:FR}), times the virtual correction to electroweak multi-particle production via quark-antiquark annihilation, $\sigma^{(1)}_{q\bar q \to {\rm ew}}$. The virtual correction has initial-state quark momenta $k_1,k_2$, defined in eq.~\eqref{eq:k1k2_param}, which reduce to $(1-x)p_1$ and $p_2$, respectively, in this limit. The $p_3\parallel p_1$ divergence from this contribution is removed by the last term in eq.~\eqref{eq:xsec_IRfinite_SLC}.
The second part consists of the sum over the virtual $q\bar{q}\to {\rm H}g$ cut diagrams containing a single $V_{\rm qed}$ (or $V^\dagger_{\rm qed}$) vertex, defined in eq.~\eqref{eq:Vcurrent_legs}, multiplied by the leading-order electroweak production cross section $\sigma_{q\bar{q} \to {\rm ew}}^{(0)}$. The $p_3\parallel p_1$ singularity from this contribution factorizes at the order $\alpha_s^2$ and is removed by the first subtraction on the right-hand side of eq.~\eqref{eq:xsec_IRfinite_SLC}.

Analogously, one can show that the counterterms in eq.~~\eqref{eq:xsec_IRfinite_SLC} remove the singularities in the other collinear limit $p_3 \parallel p_2$, as well as in the soft limit $p_3 \to 0$.

Note that, as in the fermion-loop cross section, some cut diagrams in the subleading-color cross section exhibit singularities when the quark loop momentum $k+Q$, where $Q$ is some combinations of external particle momenta, becomes collinear to the adjacent $p_3$ gluon momentum. In these limits, the numerator produces a longitudinal $p_3^\mu$ pointing into the tree-level side of the cut diagram. Summing over all the cut diagrams singular in these limits, results in the attachments of $p_3^\mu$ everywhere on the tree. Due to Ward identity, similar as in eqs.~\eqref{eq:xsec_ghost_virtlim} and \eqref{eq:xsec_Jperp_virtlim}, the singularities vanish at the cross-section level and require no further subtraction.

We have therefore demonstrated the local factorization of $\sigma_{q\bar q \to {\rm ew}+g}^{(1), {\rm SLC}}$ in all single-real limits. The proper subtraction terms have been constructed and have been justified to remove all the single-real singularities. 

\subsubsection{Single-real singularities of $\boldsymbol{\sigma_{q\bar q \to {\rm ew}+g}^{(1), {\rm LC}}}$}
\label{sec:single-real-qqbar-LC}

The leading-color cross section, $\sigma_{q\bar q \to {\rm ew}+g}^{(1), {\rm LC}}$, can be obtained by extending $\mathcal{M}_{q\bar q \to {\rm ew}+g}^{(1), {\rm LC}}$, defined in eq.~\eqref{eq:M1LC}, to the cut diagram level, by interfering with the tree amplitude including the infrared finite observable $\mathcal{O}_{{\rm ew}+1}$ and delta distribution, as defined for the general cross section in eq.~\eqref{eq:local_xsec_RV}.
Our goal is to show the factorization of single-real singularities in the leading-color cross section and that these singularities are of the form of the subtractions in eq.\ \eqref{eq:xsec_IRfinite_LC}.

Again, we focus on the limit $p_3\parallel p_1$, for which the cross section is
\begin{align}
    &\sigma_{q\bar q \to {\rm ew}+g}^{(1), {\rm LC}}(p_1,p_2;p_3;k)\xrightarrow[]{p_3=xp_1} \nn\\
    & \frac{C_A}{2C_F} \left(\includegraphics[scale = 0.65,page=1,valign=c]{figures_standalone/tikz_RV_xsecs_NNLO_SLC.pdf}
    + \includegraphics[scale = 0.65,page=2,valign=c]{figures_standalone/tikz_RV_xsecs_NNLO_SLC.pdf}
    +\includegraphics[scale = 0.65,page=4,valign=c]{figures_standalone/tikz_RV_xsecs_NNLO_SLC.pdf} 
    + \includegraphics[scale = 0.65,page=5,valign=c]{figures_standalone/tikz_RV_xsecs_NNLO_SLC.pdf}\right)\nn\\
     &+\includegraphics[scale = 0.65,page=3,valign=c]{figures_standalone/tikz_RV_xsecs_NNLO_LC.pdf}
    +\includegraphics[scale = 0.65,page=4,valign=c]{figures_standalone/tikz_RV_xsecs_NNLO_LC.pdf}
    + \includegraphics[scale = 0.65,page=5,valign=c]{figures_standalone/tikz_RV_xsecs_NNLO_LC.pdf}
    + \includegraphics[scale = 0.65,page=6,valign=c]{figures_standalone/tikz_RV_xsecs_NNLO_LC.pdf} \nn\\
    &    + \includegraphics[scale = 0.65,page=25,valign=c]{figures_standalone/tikz_RV_xsecs_NNLO_LC.pdf} 
    \quad + \text{c.c.}
    \label{eq:xsec_LC_limitp3p1}
\end{align}
On the right-hand side we have listed all the diagrams that are singular in this limit in the form of $\mathcal{M}^{(1),\rm LC}_{q\overline{q}\to {\rm ew}+ g} \cdot \left( \mathcal{M}^{(0),\rm LC}_{q\overline{q}\to {\rm ew}+ g}\right)^*$. The scalar decomposition and collinear approximation have been employed.
We again omit the incoming particle momenta as well as the loop momentum label of the leading-color amplitude, defined in eq.~\eqref{eq:M_1loop_LC_summarized}, for legibility. 
Cut diagrams in which $p_3$ is internal to the jet $p_1$ are automatically factorized. Those with $p_3$ attached to a hard vertex on the right-hand side of the cut can be factorized using the tree-level Ward identity, following the same pattern as in the subleading-color case. Note that the cut diagrams on the second line have no loop polarizations from the vertex correction in this limit. The vertex correction of the third and fourth diagrams on the second line become proportional to $p_3^\mu$ and the collinear approximation is recovered.

On the third line we summarize all the cut diagrams in which the longitudinally polarized $p_3$ attaches to the hard leading-color one-loop amplitude on the left-hand side of the cut. This contribution comprises numerous terms and is analyzed in detail in appendix \ref{appendix-Ward_identity_cancellations_single_real_LC}. It can be divided in two subsets. The first subset contains all $p_3$-ghost contributions, which arise from the attachment of the longitudinal gluon $p_3$ to a triple gluon vertex, i.e. the last two diagrams in eq.~\eqref{eq:scalar_to_ghost}. The second subset consists of all remaining diagrams of the leading-color amplitude contracted with a longitudinal gluon $p_3$, including the $p_3$-scalar terms for diagrams other than the external vertex diagram, for which the $p_3$-scalar term is incorporated into ${\cal M}^{(1),{\rm lp}}_{q\bar q \to {\rm ew}+g}$. The $p_3$-ghost terms factorize into a form involving a ghost self-energy correction , which we demonstrate in more detail in appendix \ref{appendix-Ward_identity_cancellations_single_real_LC} (see also section 5.3 of ref.~\cite{Anastasiou:2022eym}). The remaining cut diagrams factorize in a manner analogous to the subleading-color case. All diagrams carry the same overall color factor, and although the fermion line on the left-hand side contains a loop, the possible attachments of $p_3$ form a tree subdiagram. As a consequence, the tree-level Ward identity remains applicable, and, as in the subleading color case, no shift-mismatch terms arise. Using the previous example in Eq.~\eqref{eq:Ward_identity_equivalent_tree_SLC}, we obtain here 
\begin{align}
\label{eq:Ward_identity_equivalent_tree_LC}
    {\sum_{\circ}}\ \raisebox{0.4cm}{\includegraphics[scale=0.8,page=2,valign=c]{figures_standalone/tikz_RV_xsec_WI.pdf}}=\raisebox{0.4cm}{\includegraphics[scale=0.8,page=4,valign=c]{figures_standalone/tikz_RV_xsec_WI.pdf}} \, .
\end{align}
Again, the sum is over those leading-color contributions where the longitudinally polarized gluon $p_3$ attaches to the left-hand side of the cut with the attachments ``$\circ$''. The cancellations within this sum follow from the tree-level (rather than loop-level) Ward identity to guarantee local factorization. 

Here, we only report the factorized form of the last cut diagram in eq.~\eqref{eq:xsec_LC_limitp3p1}, deferring the detailed derivation to appendix~\ref{appendix-Ward_identity_cancellations_single_real_LC}:
\begin{align}
    \includegraphics[scale = 0.65,page=25,valign=c]{figures_standalone/tikz_RV_xsecs_NNLO_LC.pdf} =& 
    \includegraphics[scale = 0.65,page=19,valign=c]{figures_standalone/tikz_RV_xsecs_NNLO_LC.pdf}+ \frac{C_A}{2C_F} \includegraphics[scale = 0.65,page=25,valign=c]{figures_standalone/tikz_RV_xsecs_NNLO_SLC.pdf} \nn\\
    & +\frac{1}{2}\left(\includegraphics[scale = 0.65,page=22,valign=c]{figures_standalone/tikz_RV_xsecs_NNLO_LC.pdf}
    +\includegraphics[scale = 0.65,page=23,valign=c]{figures_standalone/tikz_RV_xsecs_NNLO_LC.pdf}
     \right) + \includegraphics[scale = 0.65,page=24,valign=c]{figures_standalone/tikz_RV_xsecs_NNLO_LC.pdf}
     \label{eq:M1L_cutgrpah_p3p1_limit_summary}\, ,
\end{align}
where the last line originates from $p_3$-ghost terms.
In total, the cross section $\sigma_{q\bar q \to {\rm ew}+g}^{(1), {\rm LC}}$ completely factorizes at the integrand level in the single-real limit $p_3\parallel p_1$ as
\begin{align}
    &\sigma_{q\bar q \to {\rm ew}+g}^{(1), {\rm LC}}(p_1,p_2;p_3;k)\xrightarrow[]{p_3\parallel p_1} \nn\\
    & \frac{C_A}{2C_F} \left(
    \includegraphics[scale = 0.65,page=1,valign=c]{figures_standalone/tikz_RV_xsecs_NNLO_SLC.pdf}
    + \includegraphics[scale = 0.65,page=2,valign=c]{figures_standalone/tikz_RV_xsecs_NNLO_SLC.pdf}
    + \includegraphics[scale = 0.65,page=17,valign=c]{figures_standalone/tikz_RV_xsecs_NNLO_SLC.pdf}
    +\includegraphics[scale = 0.65,page=18,valign=c]{figures_standalone/tikz_RV_xsecs_NNLO_SLC.pdf} \right)\nn\\
    &+\includegraphics[scale = 0.65,page=19,valign=c]{figures_standalone/tikz_RV_xsecs_NNLO_LC.pdf}+ \frac{C_A}{2C_F} \includegraphics[scale = 0.65,page=25,valign=c]{figures_standalone/tikz_RV_xsecs_NNLO_SLC.pdf} \nn\\
    &+\includegraphics[scale = 0.65,page=3,valign=c]{figures_standalone/tikz_RV_xsecs_NNLO_LC.pdf}
    +\includegraphics[scale = 0.65,page=4,valign=c]{figures_standalone/tikz_RV_xsecs_NNLO_LC.pdf}
    +\includegraphics[scale = 0.65,page=20,valign=c]{figures_standalone/tikz_RV_xsecs_NNLO_LC.pdf}
    +\includegraphics[scale = 0.65,page=21,valign=c]{figures_standalone/tikz_RV_xsecs_NNLO_LC.pdf}
    \nn\\
    & +\frac{1}{2}\left(\includegraphics[scale = 0.65,page=22,valign=c]{figures_standalone/tikz_RV_xsecs_NNLO_LC.pdf}
    +\includegraphics[scale = 0.65,page=23,valign=c]{figures_standalone/tikz_RV_xsecs_NNLO_LC.pdf}
     \right) + \includegraphics[scale = 0.65,page=24,valign=c]{figures_standalone/tikz_RV_xsecs_NNLO_LC.pdf}
    +\, \text{c.c.}
    \label{eq:p1-p3-LC}
\end{align}
The first two lines are factorized in terms of a real gluon and a lower order virtual correction with loop momentum $k$. These terms are subtracted by the third line on the right-hand side of eq.~\eqref{eq:xsec_IRfinite_LC}. The second to last line factorizes in the form of vertex corrections to the $p_1$ line times $\sigma_{q\bar{q} \to {\rm ew}}^{(0)}$. The last line similarly factorizes in terms of the real-virtual correction $p_3$-ghost terms times $\sigma_{q\bar{q} \to {\rm ew}}^{(0)}$. That is, the collinear approximation for the $p_3$-ghost terms factorizes at order $\alpha_s^2$, independently of the value of loop momentum $k$, in the same manner for all electroweak production processes with one gluon in the final state, including single-Higgs plus a gluon. All terms in the last two lines of eq.~\eqref{eq:p1-p3-LC}  are therefore removed by $p_3$-ghost terms in the first subtraction (second term on the right)
in eq.~\eqref{eq:xsec_IRfinite_LC}. By setting $p_3=xp_1$, we can then write
\begin{align}
    &\sigma_{q\bar q \to {\rm ew}+g}^{(1), {\rm LC}}(p_1,p_2;p_3;k)\nn\\
    & \xrightarrow[]{p_3=x p_1} -\frac{C_A}{2} \,g_s^2 \frac{1}{2p_1\cdot p_3} \left( \frac{2(1-\epsilon)x}{x-1}  -\frac{4}{x}\right) \sigma_{q\bar q \to {\rm ew}}^{(1)}\left((1-x)p_1,p_2;k\right) 
    \nn\\
    &\hspace{0.5cm} + \frac{1}{
    \includegraphics[scale=0.6,page=5,valign=c]{figures_standalone/tikz_RV_xsces_LO.pdf}} \left(
    \includegraphics[scale=0.6,page=1,valign=c]{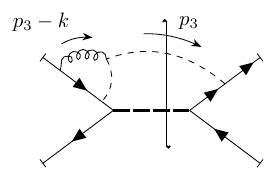}
    +\includegraphics[scale=0.6,page=2,valign=c]{figures_standalone/tikz_RV_xsces_NNLO_H.pdf}
    +\includegraphics[scale=0.6,page=3,valign=c]{figures_standalone/tikz_RV_xsces_NNLO_H.pdf}
    +\includegraphics[scale=0.6,page=4,valign=c]{figures_standalone/tikz_RV_xsces_NNLO_H.pdf}
    \right. \nn\\
    &\left. \hspace{3cm}    
    +\includegraphics[scale=0.6,page=5,valign=c]{figures_standalone/tikz_RV_xsces_NNLO_H.pdf}
    +\includegraphics[scale=0.6,page=6,valign=c]{figures_standalone/tikz_RV_xsces_NNLO_H.pdf}
    +\includegraphics[scale=0.6,page=7,valign=c]{figures_standalone/tikz_RV_xsces_NNLO_H.pdf}
    +\includegraphics[scale=0.6,page=8,valign=c]{figures_standalone/tikz_RV_xsces_NNLO_H.pdf}\right. \nn\\
    &\left. \hspace{6cm} +\, \text{c.c.} \vphantom{\begin{array}{c} a \\ b \\ c \\ \end{array}} \right) \cdot \includegraphics[scale = 0.65,page=21,valign=c]{figures_standalone/tikz_RV_xsecs_NNLO_SLC.pdf}\, 
    \, ,
    \label{eq:p1-p3-LC-2}
\end{align}
where we have explicitly calculated the first two lines  for terms that factor to order $\alpha_s$ and diagrammatically depicted the last two lines for the $p_3$-ghost terms that factorize to order $\alpha_s^2$. All singularities associated with the latter terms are already canceled by the first subtraction in eq.~\eqref{eq:xsec_IRfinite_LC} and do not require a separate single-collinear subtraction.

To construct the subtraction term in eq.~\eqref{eq:xsec_IRfinite_LC} for the single-real singularities that factor to order $\alpha_s$, it is natural to take guidance from the subleading-color cross section, where all single-real limits were removed by one term in eq.~\eqref{eq:xsec_IRfinite_SLC} (except for the $V_{\rm qed}$ terms). 
For the leading-color cross section however, the singularities are not removable locally by just one subtraction. This is due to the fact that the lower order cross section $\sigma_{q\bar{q}\to {\rm ew}}^{(1)}$ requires different arguments in the two collinear single-real limits. Compared to the limit $p_3\parallel p_1$, where the cross section $\sigma_{q\bar q \to {\rm ew}}^{(1)}$, in eqs.~\eqref{eq:p1-p3-LC} and \eqref{eq:p1-p3-LC-2}, depends on $k$, in the collinear limit $p_3\parallel p_2$, we have
\begin{align}
    &\sigma_{q\bar q \to {\rm ew}+g}^{(1), {\rm LC}}(p_1,p_2;p_3;k) \nn\\
    & \xrightarrow[]{p_3=x p_2} -\frac{C_A}{2} \,g_s^2 \frac{1}{2p_2\cdot p_3} \left( \frac{2(1-\epsilon)x}{x-1} -\frac{4}{x}\right) \sigma_{q\bar q \to {\rm ew}}^{(1)}\left(p_1,(1-x)p_2;k-p_3\right)
    \nn\\
    &\hspace{0.5cm}+ \frac{1}{
    \includegraphics[scale=0.6,page=5,valign=c]{figures_standalone/tikz_RV_xsces_LO.pdf}} \left(
    \includegraphics[scale=0.6,page=11,valign=c]{figures_standalone/tikz_RV_xsces_NNLO_H.pdf} 
    + \includegraphics[scale=0.6,page=12,valign=c]{figures_standalone/tikz_RV_xsces_NNLO_H.pdf} 
    +\includegraphics[scale=0.6,page=7,valign=c]{figures_standalone/tikz_RV_xsces_NNLO_H.pdf}
    +\includegraphics[scale=0.6,page=8,valign=c]{figures_standalone/tikz_RV_xsces_NNLO_H.pdf} \right. \nn\\
    &\left. \hspace{3cm} 
    +\includegraphics[scale=0.6,page=9,valign=c]{figures_standalone/tikz_RV_xsces_NNLO_H.pdf}
    +\includegraphics[scale=0.6,page=10,valign=c]{figures_standalone/tikz_RV_xsces_NNLO_H.pdf}
    +\includegraphics[scale=0.6,page=3,valign=c]{figures_standalone/tikz_RV_xsces_NNLO_H.pdf}
    +\includegraphics[scale=0.6,page=4,valign=c]{figures_standalone/tikz_RV_xsces_NNLO_H.pdf}\right.  \nn\\
    &\left. \hspace{6cm} +\, \text{c.c.}  \vphantom{\begin{array}{c} a \\ b \\ c \\ \end{array}}\right)\cdot
    \includegraphics[scale = 0.65,page=21,valign=c]{figures_standalone/tikz_RV_xsecs_NNLO_SLC.pdf}\,
    \, .
    \label{eq:p3-to-xp2-ew}
\end{align}
Note that the cross section $\sigma_{q\bar q \to {\rm ew}}^{(1)}$ depends on $k-p_3$. Therefore it is impossible for a single term to locally subtract the singularities in the limits $p_3\parallel p_1$ and $p_3\parallel p_2$ simultaneously. As a result we introduced two single-real subtraction terms in eq.~\eqref{eq:xsec_IRfinite_LC}, removing the aforementioned collinear divergences separately:
\begin{align}
    &\mathcal{F}_{\rm R}^{(0)\rm ,LC}(p_1,p_2;p_3)\sigma_{q\bar q \to {\rm ew}}^{(1),\textup{IR-finite}}(k_1,k_2;k)  \cdot \frac{\mathcal{O}_{{\rm H}+1}}{\mathcal{O}_{{\rm H}}} \nn\\
    &\quad+ \mathcal{F}_{\rm R}^{(0)\rm ,LC}(p_2,p_1;p_3)\sigma_{q\bar q \to {\rm ew}}^{(1),\textup{IR-finite}}(k_1,k_2;k-p_3)  \cdot \frac{\mathcal{O}_{{\rm H}+1}}{\mathcal{O}_{{\rm H}}} \, ,
    \label{eq:real_sub_LC}
\end{align}
with $\mathcal{F}_{\rm R}^{(0)\rm ,LC}$ defined in eq.~\eqref{eq:FR_def}. The subscript R indicates that the scalar function is singular in a single-real limit, exchanging the roles of $p_1$ and $p_2$.
Note the change in variables of the real function $\mathcal{F}_{\rm R}^{(0)\rm ,LC}$ in the two subtraction terms in eq.~\eqref{eq:real_sub_LC}. To see that this subtraction works, we examine the behavior of the two $\mathcal{F}_{\rm R}$ functions in the single-collinear limits. When $p_3\parallel p_1$,
\begin{align}
    \mathcal{F}_{\rm R}^{(0)\rm ,LC}(p_1,p_2;p_3) &\xrightarrow[]{p_3=xp_1}  - \frac{C_A}{2} \,g_s^2 \frac{1}{2p_1\cdot p_3} \left( \frac{2(1-\epsilon)x}{x-1}-\frac{4}{x}\right)\,, \nn\\
    \mathcal{F}_{\rm R}^{(0)\rm ,LC}(p_2,p_1;p_3) &\xrightarrow[]{p_3=xp_1} 0  
    \, .
\end{align}
Meanwhile, as $p_3\parallel p_2$,
\begin{align}
    \mathcal{F}_{\rm R}^{(0)\rm ,LC}(p_1,p_2;p_3) &\xrightarrow[]{p_3=xp_1} 0 \,, \nn \\
    \mathcal{F}_{\rm R}^{(0)\rm ,LC}(p_2,p_1;p_3) &\xrightarrow[]{p_3=xp_1}  - \frac{C_A}{2} \,g_s^2 \frac{1}{2p_2\cdot p_3} \left( \frac{2(1-\epsilon)x}{x-1} -\frac{4}{x}\right) 
    \, .
\end{align}
Therefore, in each single-collinear limit, exactly one term in eq.~\eqref{eq:real_sub_LC} is singular and captures the infrared behavior of $\sigma_{q\bar q \to {\rm ew}+g}^{(1), {\rm LC}}$. In the single-soft limit $p_3\to 0$, the cross section behaves as
\begin{align}
    & \sigma_{q\bar q \to {\rm ew}+g}^{(1), {\rm LC}}(p_1,p_2;p_3;k) \nn\\
    & \xrightarrow[]{p_3\to 0} \frac{C_A}{2} \,g_s^2 \frac{2\ p_1\cdot p_2}{p_1\cdot p_3\; p_2\cdot p_3}  \sigma_{q\bar q \to {\rm ew}}^{(1)}\left(p_1,p_2;k\right)
    \nn\\
    &\hspace{1cm}+ 
   \frac{1}{
    \includegraphics[scale=0.6,page=5,valign=c]{figures_standalone/tikz_RV_xsces_LO.pdf}} \left(
    \includegraphics[scale=0.6,page=7,valign=c]{figures_standalone/tikz_RV_xsces_NNLO_H.pdf}
    +\includegraphics[scale=0.6,page=8,valign=c]{figures_standalone/tikz_RV_xsces_NNLO_H.pdf} \right. \nn\\
    &\left. \hspace{3.5cm} 
    +\includegraphics[scale=0.6,page=3,valign=c]{figures_standalone/tikz_RV_xsces_NNLO_H.pdf}
    +\includegraphics[scale=0.6,page=4,valign=c]{figures_standalone/tikz_RV_xsces_NNLO_H.pdf}
    +\, \text{c.c.}  \vphantom{\begin{array}{c} a \\ b \\ c \\ \end{array}}\right)\cdot
    \includegraphics[scale = 0.65,page=21,valign=c]{figures_standalone/tikz_RV_xsecs_NNLO_SLC.pdf}\,
    \label{eq:p3-to-soft-ew}
    \, ,
\end{align}
where we again write the vertex correction terms as their subtraction from the real-virtual counterterm diagrammatically. Importantly, the first line is now indeed removed by the two single-real subtraction in terms of $\mathcal{F}_{\rm R}$. The combination of the two subtraction involving $\mathcal{F}_{\textup{R}}$ in eq.~\eqref{eq:real_sub_LC} combines due to $k-p_3\xrightarrow[]{p_3\to 0} k$, to exactly subtract the soft limit:
\begin{align}
    &\mathcal{F}_{\rm R}^{(0)\rm ,LC}(p_1,p_2;p_3)\sigma_{q\bar q \to {\rm ew}}^{(1)}(k_1,k_2;k) + \mathcal{F}_{\rm R}^{(0)\rm ,LC}(p_2,p_1;p_3)\sigma_{q\bar q \to {\rm ew}}^{(1)}(k_1,k_2;k-p_3) \nn\\
    & \quad \xrightarrow[]{p_3\to 0} \frac{C_A}{2} \,g_s^2 \frac{2\ p_1\cdot p_2}{p_1\cdot p_3\; p_2\cdot p_3}  \sigma_{q\bar q \to {\rm ew}}^{(1)}\left(p_1,p_2;k \right).
\end{align}

To summarize, we have demonstrated the local factorization of $\sigma_{q\bar q \to {\rm ew}+g}^{(1), {\rm LC}}$ in the single-real limits. Based on the factorized form, proper subtraction terms have been constructed and justified to remove all the single-real singularities, as in eq.\ \eqref{eq:xsec_IRfinite_LC}.

\subsubsection{Single-real singularities of $\boldsymbol{\sigma_{q\bar q \to {\rm ew}+g}^{(1), {\rm lp}}}$}
The loop-polarization part of the cross section, $ \sigma_{q\bar q \to {\rm ew}+g}^{(1), {\rm lp}}(p_1,p_2;p_3;k)$, exhibits single-real collinear singularities, and some double collinear singularities where both the real and virtual momenta are collinear to the same initial state parton.
Pictorially, the cross section can be represented through cut diagrams as, 
\begin{align}
    \sigma_{q\bar q \to {\rm ew}+g}^{(1), {\rm lp}}(p_1,p_2;p_3;k) 
    =  \includegraphics[scale=0.8,page=5,valign=c]{figures_standalone/tikz_RV_xsec_pol.pdf}
    + \includegraphics[scale=0.8,page=6,valign=c]{figures_standalone/tikz_RV_xsec_pol.pdf}+ \text{c.c.}
\end{align}
The vertex $\widetilde{J}_\perp$, introduced in section~\ref{sec:LP}, leads to a numerator suppression in the single-real collinear limits and therefore renders the cut diagrams free of loop polarizations. In the limit $p_3\parallel p_1$ we have 
\begin{align}
    \sigma_{q\bar q \to {\rm ew}+g}^{(1), {\rm lp}}(p_1,p_2;p_3;k) 
    &\xrightarrow[]{p_3\parallel p_1}
    \includegraphics[scale=0.8,page=7,valign=c]{figures_standalone/tikz_RV_xsec_pol.pdf} + 
    \includegraphics[scale=0.8,page=11,valign=c]{figures_standalone/tikz_RV_xsec_pol.pdf}+ \text{c.c.}
    \label{eq:xsec_pol_limp1p3}
\end{align}
Note that the current, before and after modification via either method 1 (symmetrization in the transverse plane) or method 2 (tensor reduction), fulfills the property in eq.~\eqref{eq:Jperp_property2}. Thus the first cut diagram exhibits a numerator suppression of the otherwise power-like denominator structure. Further details about the power counting of these cut diagrams is given in appendix~\ref{appendix-power_counting}. Due to the numerator suppression, other cut diagrams where the $\widetilde J_\perp$ rests on the $p_1$ line, become finite in this limit. The second cut diagram in eq.~\eqref{eq:xsec_pol_limp1p3} is zero due to the transverse property of $\widetilde J_\perp $ in eqs.~\eqref{eq:Jtransverse_property} and \eqref{eq:J_perp_split_finitepart}. Hence we can simplify the loop-polarization cross section in the limit as 
\begin{align}
    &\sigma_{q\bar q \to {\rm ew}+g}^{(1), {\rm lp}}(p_1,p_2;p_3;k) \xrightarrow[]{p_3\parallel p_1} 
     \includegraphics[scale=0.8,page=7,valign=c]{figures_standalone/tikz_RV_xsec_pol.pdf} + \text{c.c.} \,,
\end{align}
which is locally factorized and can be removed by subtracting
\begin{align}
     \frac{\sigma_{q\bar q \to {\rm H}+g}^{(1), {\rm lp}}(p_1,p_2;p_3;k)}{\sigma_{q\bar q \to {\rm H}}^{(0)}(k_1,k_2)} \sigma_{q\bar q \to {\rm ew}}^{(0)}(k_1,k_2) \, ,
\end{align}
as suggested in eq.~\eqref{eq:xsec_IRfinite_pol}. This subtraction moreover removes the double collinear singularities where both the real and virtual momenta are collinear to the same initial state parton.

\section{Checks of infrared limits}
\label{sec:numcheck}

We have developed a framework that renders both the one-loop amplitude and the real-virtual correction cross section for electroweak boson production in association with a jet infrared finite at the integrand level. This is achieved by subtracting universal counterterms derived from single-Higgs production processes and lower-order contributions.

We validated our construction by explicitly obtaining locally finite expressions for the cross section and amplitude in Higgs-pair production in association with either a gluon or a quark in the corresponding channels. The generation and manipulation of diagrams for the relevant amplitudes, cross sections, and counterterms were performed using QGRAF \cite{Nogueira:1991ex}, together with custom Maple \cite{maple} and FORM \cite{Vermaseren:2000nd, Tentyukov:2007mu, Kuipers:2012rf} code. The evaluation of the integrands in the single-virtual, single-real, and double real–virtual limits was carried out using Maple. Following an analogous procedure to section 7 of ref.~\cite{Anastasiou:2022eym}, we tested semi-numerically that all infrared singularities of the one-loop amplitude and cross section at order $\alpha_s^2$ in di-Higgs production indeed cancel locally. 

Explicitly, we verified that the difference of the two sides of eq.~\eqref{eq:M1IRapprox} for the one-loop amplitude in the quark-antiquark channel is infrared finite in all the virtual limits for ew=HH. Similarly the difference of the two sides in eq.~\eqref{eq:M1Lqg_IRapprox} for the one-loop amplitude in the quark-gluon channel for ew=HH is free of all virtual infrared singularities. We further verified the local finiteness of the corresponding cross sections in the quark–antiquark and quark–gluon channels, defined in eqs.~\eqref{eq:xsec_IRfinite_fermion_loop}, \eqref{eq:xsec_IRfinite_SLC}, \eqref{eq:xsec_IRfinite_LC}, \eqref{eq:xsec_IRfinite_pol}, and in eqs.~\eqref{eq:qgxsec_IRfinite_floops}, \eqref{eq:qgxsec_IRfinite_SLCLC}, \eqref{eq:qgxsec_IRfinite_pol}, respectively, for ew=HH. These checks validate our method in an example process of high complexity, involving a large number of diagrams.

As a final remark, we note that achieving a fully finite cross-section integrand also requires the local subtraction of ultraviolet singularities. This can be accomplished straightforwardly by employing the local, Ward-identity-preserving ultraviolet counterterms constructed in refs.~\cite{Anastasiou:2020sdt,Anastasiou:2022eym,Anastasiou:2024xvk}, the details of which we do not elaborate here.

\section{Conclusions and outlook}
\label{section-conclusions_outlook}

In this paper, we introduced a framework to construct infrared counterterms that render the real-virtual correction integrand for colorless multi-particle production at hadron colliders fully infrared finite, both in the phase space and loop momentum integrals.
Our starting point was the analysis of the one-loop amplitude for colorless multi-particle production in association with an external outgoing gluon through quark-antiquark annihilation in section~\ref{sec:RVamplitude}. We developed a refined treatment of loop polarizations arising from vertex corrections next to the incoming quark and antiquark legs, as well as power-like singularities originating from self-energies. These effects were previously addressed at two loops in refs.~\cite{Anastasiou:2020sdt,Anastasiou:2022eym} and a first attempt was made at three loops in ref.~\cite{Haindl:2025jte}.
This approach enables a systematic separation of loop-polarization contributions from the rest of the diagrams. Furthermore, the remaining diagrams are decomposed according to their color structure into leading-color, subleading-color and fermion-loop contributions. This decomposition permits an independent loop momentum routing for each part separately, leading to a local factorization in all infrared limits. As a result, we are able to formulate infrared counterterms that render the amplitude integrand fully infrared finite. In section~\ref{sec:IRanalyze} we demonstrated the local factorization of the singularities graphically.

In section~\ref{sec:section-finite_cross_section}, we defined the notion of cross-section integrands, local at the level of phase-space and loop-momentum integration. In section~\ref{sec:NLOxsec_localfact}, we constructed counterterms to render the order $\alpha_s$ corrections to electroweak boson production infrared finite. The infrared singularities of both the real and the virtual contributions exhibit factorization in a similar fashion, allowing their subtraction through counterterms derived from single-Higgs production processes. This observation is central to our approach: the infrared structure is governed locally by the same universal kernels that appear in simpler processes. 

In section~\ref{sec:RVxsec_localfact}, we lifted the amplitude construction of section~\ref{sec:RVamplitude} to the level of cross sections for quark-antiquark annihilation. At this stage, additional infrared singularities arise when the real parton becomes collinear to one of the incoming partons or when it becomes soft, leading to a richer infrared structure. The real–virtual correction cross section naturally inherits the decomposition of the amplitude into fermion-loop, leading-color, subleading-color, and loop-polarization contributions. The decomposition allows the implementation of local counterterms for each contribution separately, ensuring that the full real–virtual cross section remains locally integrable in all single-real, single-virtual and real-virtual infrared limits. The results are presented in section~\ref{sec:RVxsec_localfact} for the quark-antiquark annihilation channel, and in section~\ref{section-finite_RV_xsec_integrand_qg_channel} for the crossing into the quark-gluon scattering channel. In section~\ref{sec:IRanalyze}, we illustrate the local factorization of the real-virtual amplitude and real-virtual correction cross section in the quark-antiquark channel for all infrared singularities graphically. Additionally, the local factorization of the lower-order real correction cross section for the process of colorless multi-particle production via quark-antiquark annihilation is demonstrated.

% Outlook
The derivation of local momentum-space subtractions for real-virtual electroweak cross-section corrections presented in this article is an essential step toward developing a fully numerical method for NNLO cross sections. In particular, one could combine these real-virtual subtractions with local subtractions for two-loop amplitudes~\cite{Anastasiou:2022eym,Anastasiou:2025cvy} and real-real corrections under a common integrand. This can be achieved within the framework of Time-Ordered Perturbation Theory and its recent improvements~\cite{Sterman:2023xdj,Capatti:2022mly,Catani:2008xa,Capatti:2025khs}. We believe this combination can be arranged so that the full integrand of a hadronic cross section is free of singularities, factorizing initial-state singularities and ensuring that final-state singularities in infrared-safe observables cancel locally. We look forward to pursuing this research in a future publication.

\acknowledgments
We would like to thank Andrea Pelloni, Dario Kermanschah and Matilde Vicini for useful discussions. CA, JK, and YM are supported by the Swiss National Science Foundation under the project funding scheme, grant number 10001706.
The work of GS was supported in part by U.S.\ National Science Foundation award PHY-2210533.

\appendix
\section{Power counting examples}
\label{appendix-power_counting}

In this appendix, we explain some statements made in the main text regarding the power-counting analysis for the single-real limit $p_3\parallel p_1$. We shall explain:
\begin{enumerate}
    \item[\emph{1}.] The first diagram in the second line of eq.~\eqref{eq:xsec_0L_Sg_p3p1limit} is logarithmically divergent, with a suppression factor from the numerator.
    \item[\emph{2}.] The first three diagrams in eq.~\eqref{eq:xsec_SLC_limitp3p1} are all logarithmically divergent, with suppression factors from their numerators.
    \item[\emph{3}.] Original diagrams featuring loop polarizations contain power-like divergences as $p_3\parallel p_1$. The modification of such diagrams detailed in section ~\ref{sec:LP} leads to a numerator suppression and therefore to a logarithmic divergence. For example, the first diagram in eq.~\eqref{eq:xsec_pol_limp1p3} is logarithmically divergent.
\end{enumerate}
To this end, we parameterize the momentum $p_3$ as follows:
\begin{align}
    p_3^\mu = x p_1^\mu + \delta p_2^\mu + \delta^{\frac{1}{2}}p_{3\perp}^\mu
    \label{eq:collapprox_l}
\end{align}
where we have taken the center-of-mass frame such that $p_1$ and $p_2$ are back to back. The parameters $x\in (0,1)$ and $\delta\to 0$ characterize the $p_3\parallel p_1$ limit, in which the integration measure of $p_3$ scales as
\begin{align}
    \int \dmom{p_3}\sim \delta\,.
    \label{eq:measure_p3}
\end{align}
Below we examine the scaling behavior of the integrand for the cases \emph{1}--\emph{3} above.

First, the first diagram in the second line of eq.~\eqref{eq:xsec_0L_Sg_p3p1limit} has the same $p_3\parallel p_1$ behavior as $\sigma_{q\overline{q}\to {\rm H}+g}^{(0)}(p_1,p_2;p_3)$ where the real gluon attaches on the $p_1$ quark on both sides of the unphysical cut, which is proportional to
\begin{align}
    \includegraphics[scale=0.8,page=1,valign=c]{figures_standalone/tikz_RV_xsces_LO.pdf} \propto \frac{1}{\left(-2 p_1\cdot p_3\right)^2}\Tr{\slashed{p}_1 \gamma^\mu (\slashed{p}_1 -\slashed{p}_3) \slashed{p}_2 (\slashed{p}_1 -\slashed{p}_3) \gamma_\mu} \, .
\end{align}
The $\frac{1}{\left(-2 p_1\cdot p_3\right)^2}$ factor scales as $\delta^{-2}$, while the trace numerator scales as $\delta$ from a direct evaluation. Therefore, it is straightforward to see that including the scaling of the measure in eq.~\eqref{eq:measure_p3} leads to an overall logarithmic singularity. We have thus verified \emph{1}.

The analysis for the first two diagrams in eq.~\eqref{eq:xsec_SLC_limitp3p1} follows identically, because the loop momentum $k$ is hard thus not participating in the infrared power counting. As for the third diagram, note from eq.~(\ref{eq:p3Vqed}) that $V_{\rm qed}^{a \, \mu}$ has only nonzero components that are along $p_3^\mu$ or transverse to $p_3^\mu$, both of which lead to another $\delta$ suppression from the numerator. The rest of the reasoning follows. We have thus verified Statement \emph{2}.

Finally, we consider the power-counting analysis regarding loop polarizations. For example, the integrand of the following diagram scales as:
\begin{align}
    \includegraphics[scale=0.8,page=13,valign=c]{figures_standalone/tikz_RV_xsec_pol.pdf} &\xrightarrow[]{p_3 \parallel p_1} \frac{1}{\delta^2}\,.
\end{align}
The $\delta^2$ comes from the two propagators with the same momentum $p_1-p_3$. Meanwhile due to the loop polarization (which vanishes after integration but persists in the integrand) the numerator receives no suppression.
However, after the modification of loop polarizations as described in section~\ref{sec:LP}, the numerator obtains a suppression factor $\delta$ in the collinear limit.  This can be shown by inspection of the integrand $\widetilde{J}_\perp^{\mu c}$ for either of its alternate expressions, given in eq.\ \eqref{eq:Jcurrent_perp_tilde} and  \eqref{eq:Jcurrent_perp_tilde_alt}, using \eqref{eq:Jcurrent_perp}. We thus have
\begin{align}
    \includegraphics[scale=0.8,page=7,valign=c]{figures_standalone/tikz_RV_xsec_pol.pdf} & \xrightarrow[]{p_3 \parallel p_1} \frac{\delta}{\delta^2} \, .
\end{align}
Hence the first diagram in eq.~\eqref{eq:xsec_pol_limp1p3} is logarithmically divergent (including the scaling of the measure in eq.~\eqref{eq:measure_p3}), and Statement \emph{3} is verified.

\section{Ward identity cancellations in the single-real limits}
\label{appendix-Ward_identity_cancellations_single_real}

In section~\ref{sec:IRdemo_RVcrsec_singleR}, we investigated how $\sigma_{q\bar q \to {\rm ew}+g}^{(1), {\rm SLC}}$ and $\sigma_{q\bar q \to {\rm ew}+g}^{(1), {\rm LC}}$ locally factorize in the single-real limits. A crucial step is that the last line in eq.~(\ref{eq:xsec_SLC_limitp3p1}) and eq.~(\ref{eq:xsec_LC_limitp3p1}) factorize at the integrand level, even though they involve a (partial) sum over attachments of $p_3$ to a loop subdiagram which seems to generate shift mismatches. We briefly argued in the main text why this must be the case, and here we provide a detailed proof. The following subsections, \ref{appendix-Ward_identity_cancellations_single_real_SLC} and \ref{appendix-Ward_identity_cancellations_single_real_LC}, focus on the subleading- and leading-color cases, respectively.

\subsection{Subleading-color real-virtual cross section}
\label{appendix-Ward_identity_cancellations_single_real_SLC}

The last line in eq.~(\ref{eq:xsec_SLC_limitp3p1}) namely,
\begin{align}
    \includegraphics[scale = 0.75,page=26,valign=c]{figures_standalone/tikz_RV_xsecs_NNLO_SLC.pdf}
\end{align}
involves all cut diagrams where $p_3$ attaches to the hard subleading-color one-loop amplitude, defined in eq. \eqref{eq:M1SLC}, on the left-hand side of
the cut. In the following we depict only the left-hand side of the cut to investigate their cancellation. The diagrams are drawn to indicate both the position of the virtual gluon loop and the location of the electroweak final state ``ew". For example, the second diagram below implies that one element of the electroweak multi-particle state ``ew" must be emitted from a vertex closer to $p_1$ (and closer to $p_2$) than the gluon $p_3$ or the endpoints of the virtual gluon. This convention automatically excludes the loop-polarization and self-energy topologies stemming from diagrams shown in figure~\ref{fig:LoopPolarizationDiagrams}. From the Ward identity, we have
\begin{align}
    &\raisebox{1mm}{\includegraphics[scale=0.75,page=1,valign=c]{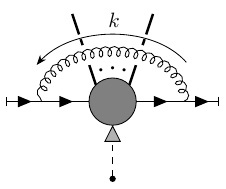}}
    + \includegraphics[scale=0.75,page=2,valign=c]{figures_standalone/tikz_RV_blob_appendix_SLC.pdf} + \includegraphics[scale=0.75,page=3,valign=c]{figures_standalone/tikz_RV_blob_appendix_SLC.pdf} + \includegraphics[scale=0.75,page=4,valign=c]{figures_standalone/tikz_RV_blob_appendix_SLC.pdf} \nn\\
    &+ \left( 1-\frac{C_A}{2C_F} \right)  
    \left(
    \includegraphics[scale=0.75,page=5,valign=c]{figures_standalone/tikz_RV_blob_appendix_SLC.pdf} + \includegraphics[scale=0.75,page=6,valign=c]{figures_standalone/tikz_RV_blob_appendix_SLC.pdf} + \includegraphics[scale=0.75,page=7,valign=c]{figures_standalone/tikz_RV_blob_appendix_SLC.pdf} + \includegraphics[scale=0.75,page=8,valign=c]{figures_standalone/tikz_RV_blob_appendix_SLC.pdf} \right)\nn\\
    =& \raisebox{1.5mm}{\includegraphics[scale=0.75,page=9,valign=c]{figures_standalone/tikz_RV_blob_appendix_SLC.pdf}} 
    + \raisebox{1.5mm}{\includegraphics[scale=0.75,page=10,valign=c]{figures_standalone/tikz_RV_blob_appendix_SLC.pdf}} 
    + \includegraphics[scale=0.75,page=11,valign=c]{figures_standalone/tikz_RV_blob_appendix_SLC.pdf} 
    + \includegraphics[scale=0.75,page=12,valign=c]{figures_standalone/tikz_RV_blob_appendix_SLC.pdf} \nn\\
    &+\includegraphics[scale=0.75,page=13,valign=c]{figures_standalone/tikz_RV_blob_appendix_SLC.pdf} + \includegraphics[scale=0.75,page=14,valign=c]{figures_standalone/tikz_RV_blob_appendix_SLC.pdf} + \includegraphics[scale=0.75,page=15,valign=c]{figures_standalone/tikz_RV_blob_appendix_SLC.pdf} + \includegraphics[scale=0.75,page=16,valign=c]{figures_standalone/tikz_RV_blob_appendix_SLC.pdf}  \nn\\
    &+ \left( 1-\frac{C_A}{2C_F} \right)  \left(
    \includegraphics[scale=0.75,page=17,valign=c]{figures_standalone/tikz_RV_blob_appendix_SLC.pdf} + \includegraphics[scale=0.75,page=18,valign=c]{figures_standalone/tikz_RV_blob_appendix_SLC.pdf} + \includegraphics[scale=0.75,page=19,valign=c]{figures_standalone/tikz_RV_blob_appendix_SLC.pdf} \right. \nn\\
    & \left.\hspace{3cm} 
    + \includegraphics[scale=0.75,page=20,valign=c]{figures_standalone/tikz_RV_blob_appendix_SLC.pdf} + \includegraphics[scale=0.75,page=21,valign=c]{figures_standalone/tikz_RV_blob_appendix_SLC.pdf} + \includegraphics[scale=0.75,page=22,valign=c]{figures_standalone/tikz_RV_blob_appendix_SLC.pdf}  \right) \,,    \label{eq:SLCline_p3p1}
\end{align}
where we have rewritten the sum over attachments into 14 diagrams on the right-hand side. All diagrams with two blobs represent the sum over all the partitions and permutations of the colorless final states attached to the blobs. We now show that the 3rd, 5th, and 13th diagrams cancel in their sum:
\begin{align}
    \includegraphics[scale=0.75,page=11,valign=c]{figures_standalone/tikz_RV_blob_appendix_SLC.pdf} +\includegraphics[scale=0.75,page=13,valign=c]{figures_standalone/tikz_RV_blob_appendix_SLC.pdf} + \left( 1-\frac{C_A}{2C_F} \right) 
    \includegraphics[scale=0.75,page=21,valign=c]{figures_standalone/tikz_RV_blob_appendix_SLC.pdf}  = 0\, .
\end{align}
By definition, the sum of the first two terms yields a diagram very similar to the third diagram, except for two differences. First, the vertex where $p_3$ is emitted from the blob lies inside, rather than outside, the virtual gluon loop. Second, its overall coefficient is $1$, rather than $\left( 1-\frac{C_A}{2C_F} \right)$. By the graphical notation of eq.~\eqref{eq:WI_qqg_vertdef}, we see that the first difference implies that the sign of its kinematic part is opposite to that of the third term, while the second shows that it carries the same color factor as the third term in the full diagram. Consequently, the sum of the first two terms cancels the third term, thereby establishing the identity above.

Similarly, the following combinations of terms in (\ref{eq:SLCline_p3p1}) cancel in their sum:
\begin{align}
    &\text{(2nd term) + (6th term) + (12th term)= 0}\,;\nn\\
    &\text{(4th term) + (8th term) + (9th term)= 0}\,;\nn\\
    &\text{(10th term) + (11th term) = 0}\,.\nn
\end{align}
Hence the only remaining diagrams are 
\begin{align}
    \raisebox{1.5mm}{\includegraphics[scale=0.75,page=9,valign=c]{figures_standalone/tikz_RV_blob_appendix_SLC.pdf}} +\includegraphics[scale=0.75,page=15,valign=c]{figures_standalone/tikz_RV_blob_appendix_SLC.pdf} + \left( 1-\frac{C_A}{2C_F} \right) \includegraphics[scale=0.75,page=22,valign=c]{figures_standalone/tikz_RV_blob_appendix_SLC.pdf} \, ,
    \label{eq:SLCline_p3p1_leftover}
\end{align}
which are all factorized. As cut diagrams, they can be displayed as 
\begin{align}
    \includegraphics[scale = 0.75,page=20,valign=c]{figures_standalone/tikz_RV_xsecs_NNLO_SLC.pdf}
    +\left( 1-\frac{C_A}{2C_F} \right) \includegraphics[scale = 0.75,page=25,valign=c]{figures_standalone/tikz_RV_xsecs_NNLO_SLC.pdf}
 \,,
\end{align}
where we have combined the first two diagrams in eq.~\eqref{eq:SLCline_p3p1_leftover} into the first one above. This is exactly the right-hand side of eq.~(\ref{eq:M1L_SLC_p3lim_general_factorized}). Note that for double or single electroweak boson production one or two of the heavy dashed lines in the diagrams may be absent.

\subsection{Leading-color real-virtual cross section}
\label{appendix-Ward_identity_cancellations_single_real_LC}
In this subsection we demonstrate the local factorization of the last cut diagram in eq.~(\ref{eq:xsec_LC_limitp3p1}) namely,
\begin{align}
    \includegraphics[scale = 0.75,page=25,valign=c]{figures_standalone/tikz_RV_xsecs_NNLO_LC.pdf}\, ,
\end{align}
where $p_3$ attaches to the hard leading-color one-loop amplitude, defined in eq. \eqref{eq:M1LC}, on the left-hand side of the cut. In the following we divide the analysis of the cut diagrams into two sets. The first contains all the $p_3$-ghost terms appearing when the longitudinal gluon $p_3$ attaches to a triple gluon vertex, i.e. the last two diagrams in eq.~\eqref{eq:scalar_to_ghost}. The second set contains the remaining diagrams of the leading-color amplitude contracted with a longitudinal gluon $p_3$. As in the subleading-color case, we draw all diagrams to indicate the positions of the virtual gluon loop and the electroweak final state ``ew", thereby excluding the loop-polarization topologies in figure~\ref{fig:LoopPolarizationDiagrams}.

We first demonstrate that the $p_3$-ghost contributions locally factorize into a form involving a ghost self-energy correction (see section 5.3 of ref.~\cite{Anastasiou:2022eym}). For simplicity, let us show only the left-hand side of the cut, and obtain:
\begin{align}
    &\includegraphics[scale=0.75,page=1,valign=c]{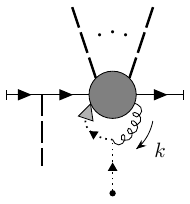} + \includegraphics[scale=0.75,page=2,valign=c]{figures_standalone/tikz_RV_blob_appendix_LC.pdf} + \includegraphics[scale=0.75,page=3,valign=c]{figures_standalone/tikz_RV_blob_appendix_LC.pdf} + \includegraphics[scale=0.75,page=4,valign=c]{figures_standalone/tikz_RV_blob_appendix_LC.pdf}  \nn\\
    =&  \includegraphics[scale=0.75,page=5,valign=c]{figures_standalone/tikz_RV_blob_appendix_LC.pdf} + \includegraphics[scale=0.75,page=6,valign=c]{figures_standalone/tikz_RV_blob_appendix_LC.pdf}+ \includegraphics[scale=0.75,page=8,valign=c]{figures_standalone/tikz_RV_blob_appendix_LC.pdf} \nn\\
    &+\includegraphics[scale=0.75,page=7,valign=c]{figures_standalone/tikz_RV_blob_appendix_LC.pdf}
    + \includegraphics[scale=0.75,page=9,valign=c]{figures_standalone/tikz_RV_blob_appendix_LC.pdf} \, ,
\end{align}
where again all diagrams exhibiting two blobs are a sum over all partitions and permutations of the colorless final states attached to the blobs.
On the right-hand side above, the 1st and 4th diagrams cancel in the sum, and the last diagram is automatically factorized. The two remaining terms combine into longitudinal ghosts with self-energy corrections as explained in detail in section 5.3 of ref.~\cite{Anastasiou:2022eym}. Diagrammatically, we find
\begin{align}
    \raisebox{-0.8mm}{\includegraphics[scale=0.75,page=10,valign=t]{figures_standalone/tikz_RV_blob_appendix_LC.pdf}}+
    \raisebox{-0.8mm}{\includegraphics[scale=0.75,page=11,valign=t]{figures_standalone/tikz_RV_blob_appendix_LC.pdf}} = \frac{1}{2}\raisebox{-1mm}{\includegraphics[scale=0.75,page=12,valign=t]{figures_standalone/tikz_RV_blob_appendix_LC.pdf}}
    + \frac{1}{2}\raisebox{-1mm}{\includegraphics[scale=0.75,page=13,valign=t]{figures_standalone/tikz_RV_blob_appendix_LC.pdf}}\,.
\end{align}
These point into the fermion line, leading to Ward-identity cancellations and, ultimately, factorization. The factorized $p_3$-ghost terms as cut diagrams are
\begin{align}
    \frac{1}{2}\left(\includegraphics[scale = 0.75,page=22,valign=c]{figures_standalone/tikz_RV_xsecs_NNLO_LC.pdf}
    +\includegraphics[scale = 0.75,page=23,valign=c]{figures_standalone/tikz_RV_xsecs_NNLO_LC.pdf}
     \right) + \includegraphics[scale = 0.75,page=24,valign=c]{figures_standalone/tikz_RV_xsecs_NNLO_LC.pdf} \, ,
\end{align}
and correspond to the second line of eq.~\eqref{eq:M1L_cutgrpah_p3p1_limit_summary}.

At last we investigate the local cancellations of the remaining cut diagrams in this limit. Again, let us only depict the subdiagram corresponding to the left-hand side of the cut, which is:
\begin{align}
    & \raisebox{-1mm}{\includegraphics[scale=0.75,page=14,valign=c]{figures_standalone/tikz_RV_blob_appendix_LC.pdf} } 
    + \raisebox{-1mm}{\includegraphics[scale=0.75,page=15,valign=c]{figures_standalone/tikz_RV_blob_appendix_LC.pdf} } 
    + \raisebox{-1mm}{\includegraphics[scale=0.75,page=16,valign=c]{figures_standalone/tikz_RV_blob_appendix_LC.pdf} } 
    + \raisebox{-1mm}{\includegraphics[scale=0.75,page=17,valign=c]{figures_standalone/tikz_RV_blob_appendix_LC.pdf}}  \nn\\
    & +\frac{C_A}{2C_F}\left( 
    \includegraphics[scale=0.75,page=18,valign=c]{figures_standalone/tikz_RV_blob_appendix_LC.pdf}  + \includegraphics[scale=0.75,page=19,valign=c]{figures_standalone/tikz_RV_blob_appendix_LC.pdf}  + \includegraphics[scale=0.75,page=20,valign=c]{figures_standalone/tikz_RV_blob_appendix_LC.pdf}  + \includegraphics[scale=0.75,page=21,valign=c]{figures_standalone/tikz_RV_blob_appendix_LC.pdf}  \right) \nn\\
    &= \includegraphics[scale=0.75,page=22,valign=c]{figures_standalone/tikz_RV_blob_appendix_LC.pdf}   +\includegraphics[scale=0.75,page=23,valign=c]{figures_standalone/tikz_RV_blob_appendix_LC.pdf}   +\includegraphics[scale=0.75,page=24,valign=c]{figures_standalone/tikz_RV_blob_appendix_LC.pdf}   +\includegraphics[scale=0.75,page=25,valign=c]{figures_standalone/tikz_RV_blob_appendix_LC.pdf}   \nn\\
    &+ \includegraphics[scale=0.75,page=26,valign=c]{figures_standalone/tikz_RV_blob_appendix_LC.pdf}   +\includegraphics[scale=0.75,page=27,valign=c]{figures_standalone/tikz_RV_blob_appendix_LC.pdf}   +\includegraphics[scale=0.75,page=29,valign=c]{figures_standalone/tikz_RV_blob_appendix_LC.pdf}   +\includegraphics[scale=0.75,page=28,valign=c]{figures_standalone/tikz_RV_blob_appendix_LC.pdf}    \nn\\
    &+ \frac{C_A}{2C_F}\left(
    \includegraphics[scale=0.75,page=30,valign=c]{figures_standalone/tikz_RV_blob_appendix_LC.pdf}   +\includegraphics[scale=0.75,page=31,valign=c]{figures_standalone/tikz_RV_blob_appendix_LC.pdf}   +\includegraphics[scale=0.75,page=32,valign=c]{figures_standalone/tikz_RV_blob_appendix_LC.pdf}   +\includegraphics[scale=0.75,page=33,valign=c]{figures_standalone/tikz_RV_blob_appendix_LC.pdf}   \right. \nn\\
    &\left.\hspace{2cm} 
    +\includegraphics[scale=0.75,page=34,valign=c]{figures_standalone/tikz_RV_blob_appendix_LC.pdf}   +\includegraphics[scale=0.75,page=35,valign=c]{figures_standalone/tikz_RV_blob_appendix_LC.pdf}   + \right) \,,
    \label{eq:LC_lines_Wids}
\end{align}
where we have repeatedly applied the gluon-quark lowest-order Ward identity eq.\ \eqref{eq:Wardid_qqg_eq} and gluon-scalar 
identities of \eqref{eq:scalar_to_ghost} to obtain the 14 diagrams of on the right-hand side.
Most of these diagrams cancel in the sum. For example, the sum over the 3rd, 5th, and 13th diagrams vanishes:
\begin{align}
     \includegraphics[scale=0.75,page=24,valign=c]{figures_standalone/tikz_RV_blob_appendix_LC.pdf}   +\includegraphics[scale=0.75,page=26,valign=c]{figures_standalone/tikz_RV_blob_appendix_LC.pdf}   + \frac{C_A}{2C_F} \includegraphics[scale=0.75,page=34,valign=c]{figures_standalone/tikz_RV_blob_appendix_LC.pdf}    = 0\, . 
\end{align}
As in the subleading-color case, the sum of the first two diagrams above yields a diagram that differs from the third diagram in only two aspects. First, the vertex where $p_3$ is emitted from the blob lies inside (rather than outside) the virtual gluon loop. Second, its overall coefficient is $1$ (rather than $\tfrac{C_A}{2C_F}$). The first difference implies by eqs.~\eqref{eq:WI_qqg_vertdef} and \eqref{eq:scalar_to_ghost} that the sign of its kinematic part is opposite to that of the third term, while the second shows that it carries the same color factor as the third term in the full diagram. Consequently, the sum of the first two terms cancels the third term, establishing the identity above.

For the same reason, all other diagrams cancel except the 1st, 7th, and 14th diagrams, which remain in the final factorized form:
\begin{align}
    \includegraphics[scale=0.75,page=22,valign=c]{figures_standalone/tikz_RV_blob_appendix_LC.pdf}   + \includegraphics[scale=0.75,page=29,valign=c]{figures_standalone/tikz_RV_blob_appendix_LC.pdf}   + \frac{C_A}{2C_F} \includegraphics[scale=0.75,page=35,valign=c]{figures_standalone/tikz_RV_blob_appendix_LC.pdf}   \, .
\end{align}
The expression above corresponds exactly to the left-hand side of the cut in diagrams on the first line of eq.~\eqref{eq:M1L_cutgrpah_p3p1_limit_summary}, namely
\begin{align}
    \includegraphics[scale = 0.65,page=19,valign=c]{figures_standalone/tikz_RV_xsecs_NNLO_LC.pdf}+ \frac{C_A}{2C_F} \includegraphics[scale = 0.65,page=25,valign=c]{figures_standalone/tikz_RV_xsecs_NNLO_SLC.pdf}\, .
\end{align}

\bibliographystyle{JHEP}
\bibliography{biblio}

\end{document}